\documentclass[11pt]{article}
\usepackage{jheppub}
\usepackage{amsfonts,amsmath,amssymb,epsf}
\usepackage{amsmath,braket}
\usepackage{mathtools}
\usepackage{physics}
\usepackage{graphicx,hyperref,color}
\usepackage[dvipsnames]{xcolor}
\usepackage{float}
\hypersetup{
    bookmarks=true,         
    unicode=false,          
    pdftoolbar=true,        
    pdfmenubar=true,        
    linktocpage=true,       
    pdffitwindow=false,     
    pdfstartview={FitH},    
    pdfnewwindow=true,      
    colorlinks=true,       
    linkcolor=blue,          
    citecolor=blue,        
    filecolor=blue,      
    urlcolor=blue           
}
\usepackage[subrefformat=parens]{subcaption}

\numberwithin{equation}{section}									

\newcommand{\de}{\partial}
\newcommand{\be}{\begin{equation}}
\newcommand{\ba}{\begin{eqnarray}}
\newcommand{\ea}{\end{eqnarray}}
\newcommand{\ee}{\end{equation}}

\newcommand{\s}{\sqrt}
\newcommand{\vp}{\varphi}

\newcommand{\ti}{\tilde}
\newcommand{\ap}{\alpha}

\newcommand{\no}{\nonumber \\}
\newcommand{\la}{\langle}
\newcommand{\lb}{\rangle}
\newcommand{\bea}{\begin{eqnarray}}
\newcommand{\eea}{\end{eqnarray}}
\newcommand{\bes}{\begin{equation*}}
\newcommand{\beas}{\begin{eqnarray*}}
\newcommand{\eeas}{\end{eqnarray*}}
\newcommand{\bas}{\begin{array*}}
\newcommand{\eas}{\end{array*}}
\newcommand{\ees}{\end{equation*}}

\newcommand{\p}{\partial}
\newcommand{\ep}{\epsilon}

 \let\p=\phi  
   \let\vp=\varphi

\DeclareMathOperator{\arccosh}{arccosh}

\newcommand{\tee}[1]{S^{(\text{T})}_{#1}}
\newcommand{\pe}[1]{S^{(\text{P})}_{#1}}
\newcommand{\itee}[1]{I^{(\text{T})}_{#1}}
\newcommand{\jtee}[1]{J^{(\text{T})}_{#1}}

\newcommand\myeq{\mathrel{\stackrel{\makebox[0pt]{\mbox{\normalfont\tiny reg}}}{=}}}

\usepackage{tikz}
\usetikzlibrary{arrows,arrows.meta,intersections, calc,positioning,decorations.pathreplacing,decorations.pathmorphing,shapes}
\usetikzlibrary{patterns}
\usetikzlibrary{decorations.markings}
\usetikzlibrary{knots}

\preprint{YITP-23-22}

\emailAdd{kazuki.doi@yukawa.kyoto-u.ac.jp}
\emailAdd{jonathan.harper@yukawa.kyoto-u.ac.jp}
\emailAdd{ali.mollabashi@yukawa.kyoto-u.ac.jp}
\emailAdd{takayana@yukawa.kyoto-u.ac.jp}
\emailAdd{yusuke.taki@yukawa.kyoto-u.ac.jp}

\title{\boldmath Timelike entanglement entropy}

\author[a]{Kazuki Doi,}
\author[a]{Jonathan Harper,}
\author[a]{Ali Mollabashi,}
\author[a,b,c]{Tadashi Takayanagi}
\author[a]{and Yusuke Taki}

\affiliation[a]{Center for Gravitational Physics, Yukawa Institute for Theoretical Physics, Kyoto University,\\
Kitashirakawa Oiwakecho, Sakyo-ku, Kyoto 606-8502, Japan}

\affiliation[b]{Inamori Research Institute for Science,\\
620 Suiginya-cho, Shimogyo-ku,
Kyoto 600-8411, Japan}

\affiliation[c]{Kavli Institute for the Physics and Mathematics
 of the Universe (WPI),\\
University of Tokyo, Kashiwa, Chiba 277-8582, Japan}

\abstract{We define a new complex-valued measure of information called the timelike entanglement entropy (EE) which in the boundary theory can be viewed as a Wick rotation that changes a spacelike boundary subregion to a timelike one. An explicit definition of the timelike EE in 2d field theories is provided followed by numerical computations which agree with the analytic continuation of the replica method for CFTs. We argue that timelike EE should be correctly interpreted as another measure previously considered, the pseudo entropy, which is the von Neumann entropy of a reduced transition matrix. Our results strongly imply that the imaginary part of the pseudo entropy describes an emergent time which generalizes the notion of an emergent space from quantum entanglement. For holographic systems we define the timelike EE as the total complex valued area of a particular stationary combination of both space and timelike extremal surfaces which are homologous to the boundary region. For the examples considered we find explicit matching of our optimization procedure and the careful implementation of the Wick rotation in the boundary CFT. We also make progress on higher dimensional generalizations and relations to holographic pseudo entropy in de Sitter space.}
\begin{document}

\maketitle

\section{Introduction}

The AdS/CFT correspondence \cite{Maldacena:1997re} tells us that a space coordinate in an anti-de Sitter (AdS) spacetime can emerge from a conformal field theory (CFT). The mechanism of this emergent space can be described quantitatively by considering the holographic entanglement entropy 
(HEE) \cite{Ryu:2006bv,Ryu:2006ef,Hubeny:2007xt}. Specifically the entanglement entropy (EE) in a CFT is computed from the area of an extremal surface in AdS. The entanglement entropy $S_A$ for a subsystem $A$ is defined by 
\ba
S_A=-\mbox{Tr}_A[\rho_A\log\rho_A].
\ea
$\rho_A$ is the reduced density matrix obtained by tracing out the complement $B$ of $A$:
\ba
\rho_A=\mbox{Tr}_B\rho_{\text{tot}},
\ea
where $\rho_{\text{tot}}$ is the density matrix for the total system and the total Hilbert space is assumed to factorize as 
${\cal H}_{\text{tot}}={\cal H}_A\otimes {\cal H}_B$.
The holographic entanglement entropy computes $S_A$ from the area of an extremeal surface by
\ba
S_A=\frac{\mbox{Area}(\Gamma_A)}{4G_N}. \label{HEEF}
\ea
This relation leads to the remarkable idea that the space coordinate in an AdS emerges from quantum entanglement \cite{Swingle:2009bg,VanRaamsdonk:2010pw}.
\begin{figure}[H]
    \centering
    \includegraphics[width=.6\textwidth]{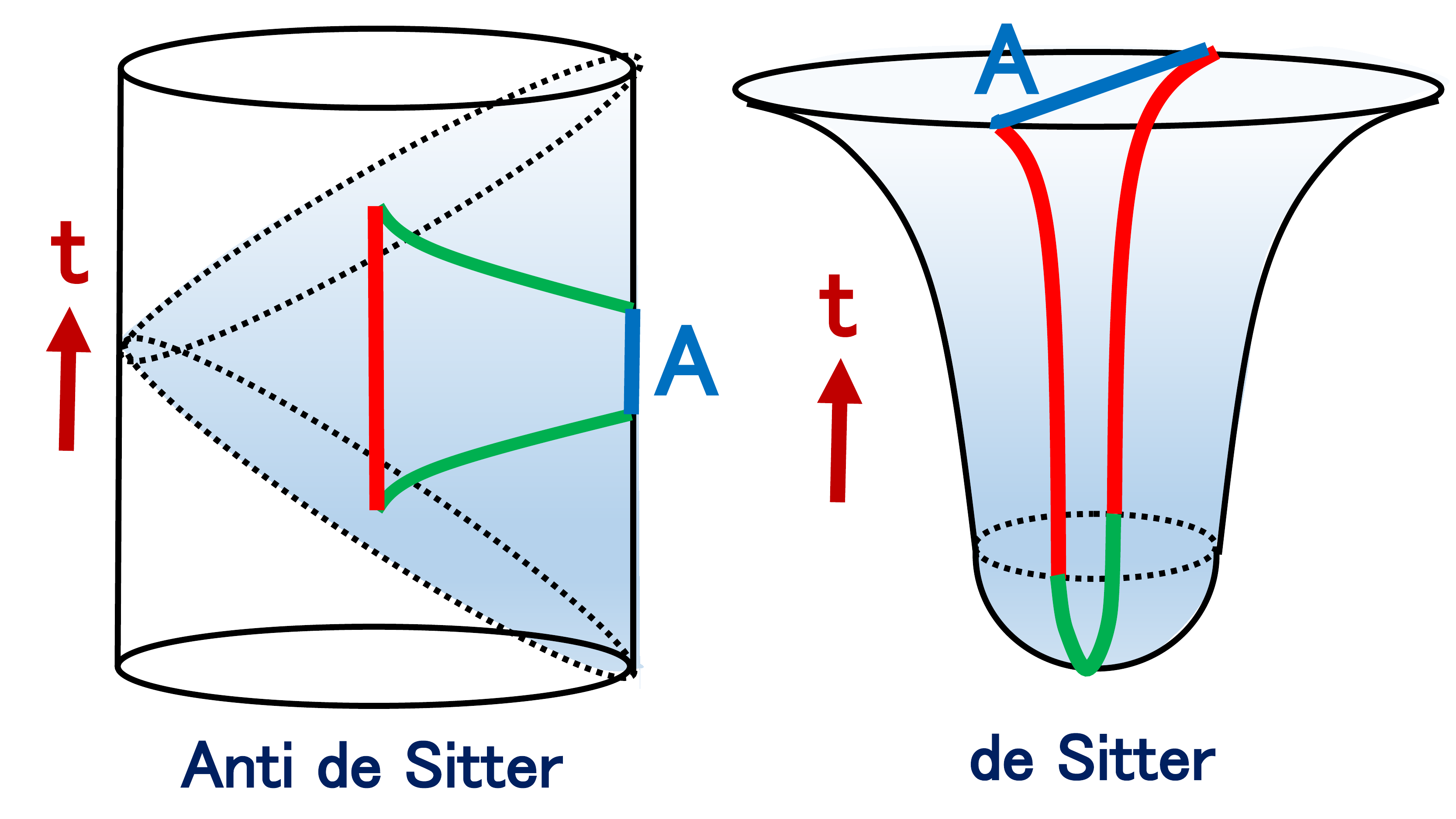}
    \caption{Holographic Timelike Entanglement Entropy in AdS$_3/$CFT$_2$ (left) and Holographic Pseudo entropy in dS$_3/$CFT$_2$ (right). The green curves and red curves describe the spacelike and timelike geodesic, whose lengths give the real and imaginary part of the entropy. Each of blue intervals is the subsystem A.}
    \label{fig:skt}
\end{figure}
This raises a natural question: can the time coordinate also emerge from some quantum information theoretic property?
To make progress in this problem, we first need to find a quantity which is directly related to the emergence of the time coordinate. Motivated by this, the main purpose of this paper is to introduce a new quantity called timelike entanglement entropy (timelike EE) and to study its properties\footnote{ Refer to
\cite{PhysRevLett.54.857,Fitzsimons:2013gga,Olson:2011bq,Cotler:2017anu,Cotler:2018sbu,Lerose:2021sag,Giudice:2021smd,Parzygnat:2022pax}, for other ideas on how to study quantum entanglement in timelike setups.}. The timelike EE $\tee{A}$ is defined by analytically continuing the standard EE to the case where the subsystem $A$ is a timelike region.
Indeed, in the AdS/CFT correspondence, the imaginary part of this quantity is related to the area of a timelike extremal surface, providing a generalization of (\ref{HEEF}), as depicted in left panel of figure \ref{fig:skt}. A part of these new results was briefly reported in the letter article \cite{Doi:2022iyj}, focusing on a few simple setups of AdS$_3/$CFT$_2$. Refer to \cite{Liu:2022ugc,Narayan:2022afv} for independent works on timelike EE and see also \cite{Diaz:2021snw,Reddy:2022zgu,Li:2022tsv}. In this full paper, we will present more general results of timelike EE in various setups of AdS/CFT.

Moreover, in the context of dS/CFT \cite{Strominger:2001pn}, we can also see that the holographic entanglement entropy has contributions from timelike extremal surfaces in addition to a spacelike surface as depicted in the right panel of figure \ref{fig:skt}. Holography in de Sitter space (dS), so called the dS/CFT correspondence \cite{Strominger:2001pn}
argues that gravity on a de Sitter space is dual to a Euclidean CFT on its future infinity. In this holography, as opposed to AdS/CFT, the time coordinate emerges from the Euclidean CFT. Such a Euclidean CFT is expected to be exotic and non-unitary. A limited number of examples of CFTs dual to de Sitter spaces have been found in four dimensional higher spin gravity \cite{Anninos:2011ui}, in three dimensional Einstein gravity \cite{Hikida:2021ese,Hikida:2022ltr} and in two dimensions \cite{Maldacena:2019cbz,Cotler:2019nbi}. Since there is no spacelike geodesic between two distinct points on the dS boundary at future infinity, the holographic entropy becomes complex-valued \cite{Narayan:2015vda,Sato:2015tta,Miyaji:2015yva,Narayan:2017xca,Narayan:2020nsc,Hikida:2022ltr}.

 We point out that both the timelike entanglement entropy and the complex-valued holographic entropy in dS/CFT, can properly be interpreted as pseudo entropy, introduced in \cite{Nakata:2021ubr}. 
 Pseudo entropy is defined as follows. Decomposing the total Hilbert space into those of subsystems $A$ and $B$, we introduce the reduced transition matrix for two pure states $|\psi\lb$ and $|\vp\lb$, by
\ba
\tau_A=\mbox{Tr}_B\left[\frac{|\psi\lb \la \vp|}{\la \vp|\psi\lb}\right].
\ea 
Finally, pseudo entropy $\pe{A}$ is defined by
\ba
\pe{A}=-\mbox{Tr}[\tau_A\log\tau_A].\label{PEdef}
\ea
Remarkably, in  Euclidean time-dependent asymptotically AdS spaces, the pseudo entropy is related to the minimal surface area in the same formula as (\ref{HEEF}).
See \cite{Mollabashi:2020yie,Camilo:2021dtt,Mollabashi:2021xsd,Nishioka:2021cxe,Goto:2021kln,Miyaji:2021lcq,
Akal:2021dqt,Berkooz:2022fso,Akal:2022qei,Mori:2022xec,Mukherjee:2022jac,Guo:2022sfl,Ishiyama:2022odv,
Miyaji:2022cma,Bhattacharya:2022wlp,Guo:2022jzs,Diaz:2021snw,Reddy:2022zgu,Li:2022tsv,He:2023eap} for further related developments. Indeed, in  both the timelike EE and the EE in dS/CFT, the reduced density matrices are not Hermitian and thus are not standard von Neumann entropies, but rather pseudo entropies instead. 

This paper is organized as follows. In section \ref{sec:2}, we define the timelike entanglement entropy and analyize it in quantum field theories. In section \ref{sec:3}, we will show how to calculate the timelike entanglement entropy in AdS/CFT and evaluate it in various examples.
In section \ref{sec:4}, we study the holographic calculation of timelike entanglement entropy in higher dimensional AdS/CFT. In section \ref{sec:5}, we discuss holographic pseudo entropy in dS/CFT. Finally in section \ref{sec:6}, we summarize our conclusions and discuss future problems. Appendix \ref{apend:tmi} contains a derivation of the thermofield mutual information for 2d holographic CFTs.

\section{Timelike entanglement entropy in QFT}\label{sec:2}

In this section we explain the definition of timelike entanglement entropy and present calculations of this quantity in conformal field theories.

\subsection{Definition via the replica method}
\begin{figure}[H]
    \centering
    \includegraphics[width=.6\textwidth]{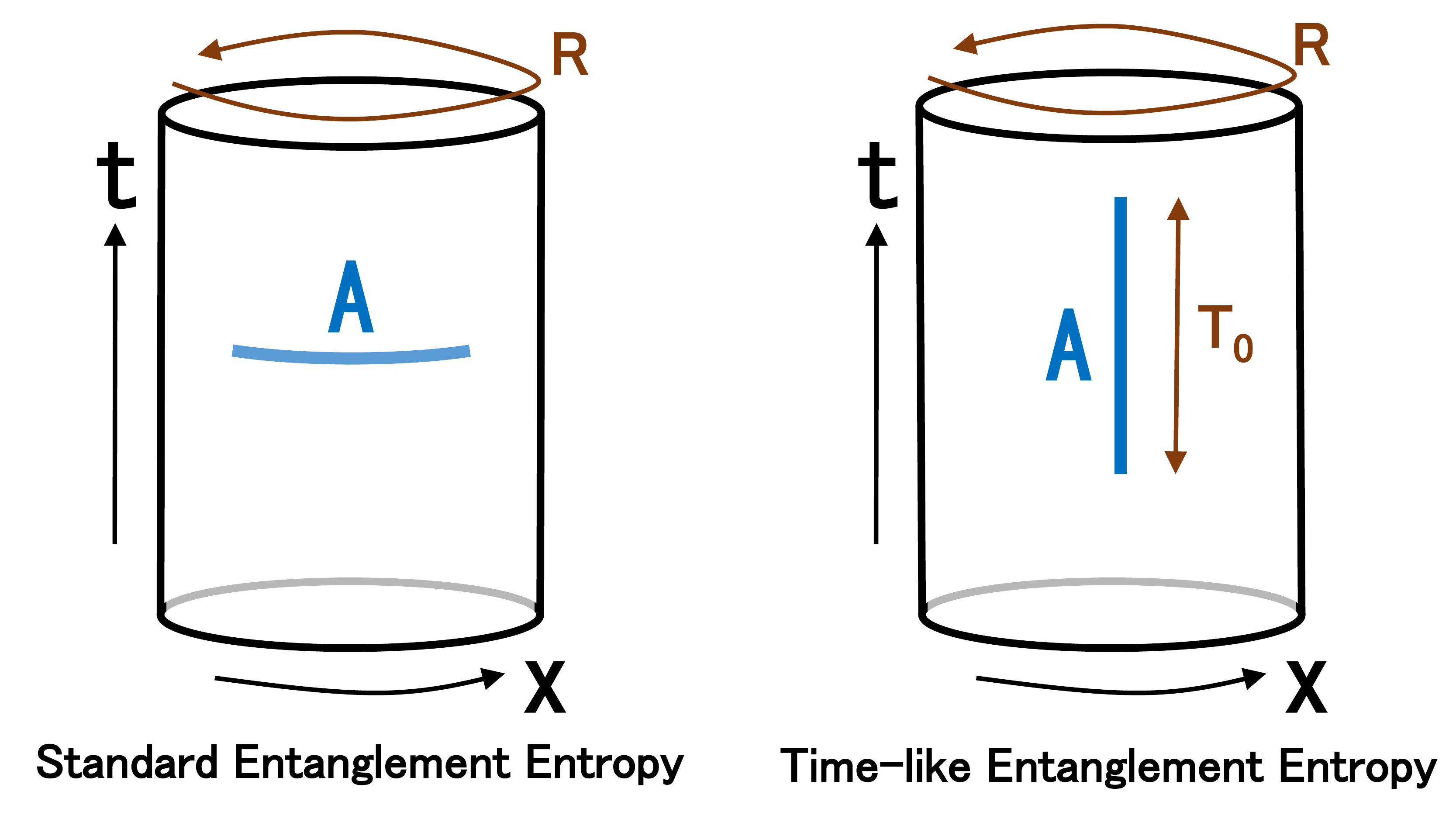}
    \caption{Definition of standard entanglement entropy (left) and timelike entanglement entropy (right) in two dimensional field theories.}
    \label{fig:TEE}
\end{figure}
 Consider a two dimensional quantum field theory in a flat spacetime whose time and space coordinate are denoted by $(t,x)$. We first define the timelike entanglement entropy in this two dimensional case and will later generalize to higher dimensions. In the standard definition of entanglement entropy $S_A$ we choose the subsystem $A$ to be a spacelike region in order to specify the Hilbert space ${\cal H}_A$ as in the left panel of figure \ref{fig:TEE}. The timelike entanglement entropy is defined by analytically continuing the entanglement entropy to a timelike subsystem $A$ as in the right panel of figure \ref{fig:TEE}, which we write as $\tee{A}$.

To see this procedure explicitly, consider the replica method computation of entanglement entropy \cite{Calabrese:2004eu}. We assume subsystem $A$ is given by a spacelike interval $A=I_{PQ}$, where the two end points are situated at $P=(t_P,x_P)$ and $Q=(t_Q,x_Q)$. 
The Renyi entanglement entropy $S^{(n)}_A=\frac{1}{1-n}\log\mbox{Tr}(\rho_A)^n$ is computed from the two point function of twist operators
$\sigma_n$ and $\bar{\sigma}_n$:
\ba
S^{(n)}_A=\frac{1}{1-n}\log\la\sigma_n(P)\bar{\sigma}_n(Q)\lb
=\frac{1}{1-n}\log\left[\left(\frac{\ep}{\s{(x_P-x_Q)^2-(t_P-t_Q)^2}}\right)^{2\Delta_n}
\right].
\ea
In this equation, $\Delta_n=\frac{c}{12}\left(n-\frac{1}{n}\right)$, where $c$ is the central charge of the two dimensional CFT, is the conformal dimension of the twist operator and $\ep$ is the UV cutoff. By taking the limit $n\to 1$ of the von Neumann entropy, the entanglement entropy is evaluated to be
\ba
S_A=S^{(1)}_A=\frac{c}{3}\log\left[\frac{\s{(x_P-x_Q)^2-(t_P-t_Q)^2}}{\ep}\right]. \label{EErep}
\ea

Now the timelike entanglement entropy $\tee{A}$ is defined by continuing the expression (\ref{EErep}) to the case where the interval $A$ is timelike i.e. $(x_P-x_Q)^2-(t_P-t_Q)^2<0$. This is evaluated as follows:
\ba
\tee{A}=\frac{c}{3}\log\left[\frac{\s{(t_P-t_Q)^2-(x_P-x_Q)^2}}{\ep}\right]+\frac{c\pi }{6}i. \label{TEErep}
\ea
In particular, when the subsystem $A$ is purely timelike i.e. $x_Q-x_P=0$
and $t_Q-t_P=T_0$, we find 
\ba
\tee{A}=\frac{c}{3}\log\frac{T_0}{\ep}+\frac{c\pi }{6}i. \label{TEEarep}
\ea

It is straightforward to generalize this to $\tee{A}$ in a finite size CFT and $\tee{A}$ in a finite temperature CFT. For a CFT on a circle with a circumference $R$ (at zero temperature), the entanglement entropy for a boosted subsystem $A$ is given by
\begin{align}
    S_A=\frac{c}{6}\log\left[\frac{R^2}{\pi^2\epsilon^2}\sin\left(\frac{\pi}{R}(\Delta\phi+\Delta t)\right)\sin\left(\frac{\pi}{R}(\Delta\phi-\Delta t)\right)\right].
\end{align}
In a similar fashion, the timelike entanglement entropy is defined by continuing $A$ to be timelike $\Delta\phi-\Delta t<0$, then
\begin{align}
    \tee{A}=\frac{c}{6}\log\left[\frac{R^2}{\pi^2\epsilon^2}\sin\left(\frac{\pi}{R}(\Delta t+\Delta \phi)\right)\sin\left(\frac{\pi}{R}(\Delta t-\Delta \phi)\right)\right]+\frac{i\pi c}{6}.
\end{align}
In particular, when $A$ is purely timelike $\Delta \phi=0$, $\Delta t=T_0$, we have
\begin{align}
    \tee{A}=\frac{c}{3}\log\left[\frac{R}{\pi\epsilon}\sin\frac{\pi T_0}{R}\right]+\frac{i\pi c}{6}.
    \label{TEEsize}
\end{align}
On the other hand, for a CFT at temperature $1/\beta$ (on an infinite line), the entanglement entropy is given by 
\begin{align}
    S_A=\frac{c}{6} \log\left[\frac{\beta^2}{\pi^2\epsilon^2}\sinh\left(\frac{\pi}{\beta}(\Delta x+\Delta t)\right)\sinh\left(\frac{\pi}{\beta}(\Delta x-\Delta t)\right)\right].
\end{align}
Therefore the timelike entanglement entropy is
\begin{align}
    \tee{A}=\frac{c}{6} \log\left[\frac{\beta^2}{\pi^2\epsilon^2}\sinh\left(\frac{\pi}{\beta}(\Delta t+\Delta x)\right)\sinh\left(\frac{\pi}{\beta}(\Delta t-\Delta x)\right)\right]+\frac{i\pi c}{6},
\end{align}
and particularly
\begin{align}
    \tee{A}=\frac{c}{3}\log\left[\frac{\beta}{\pi\epsilon}\sinh\frac{\pi T_0}{\beta}\right]+\frac{i\pi c}{6}.
     \label{TEEtemp}
\end{align}
when $A$ is purely timelike. Note that these basically take the form of the standard entanglement entropy plus $\frac{i\pi}{6}$.

We can also define timelike EE in higher dimensions in a similar way: We start from the EE for a spacelike subsystem and boosting the subsystem with an analytic continuation until it becomes timelike.

\subsection{Another equivalent definition via Wick rotation of coordinates}\label{subsec:formulation}

We can define the timelike entanglement entropy in another way, which is useful for numerical calculations. For simplicity, consider a free scalar field theory with mass $m$ on a cylinder, where the space and time coordinate are again denoted by $x$ and $t$. The spacelike direction is compactified as $x\sim x+R$.
The action of the scalar field reads
\ba
S=\frac{1}{2}\int dtdx\left[(\de_t\phi)^2-(\de_x\phi)^2-m^2\phi^2\right].\label{Laciona}
\ea
The total partition function on the Lorentzian spacetime looks like
\ba
Z_{\phi}=\int D\phi e^{iS}.  \label{totpar}
\ea
\begin{figure}[H]
    \centering
    \includegraphics[width=.6\textwidth]{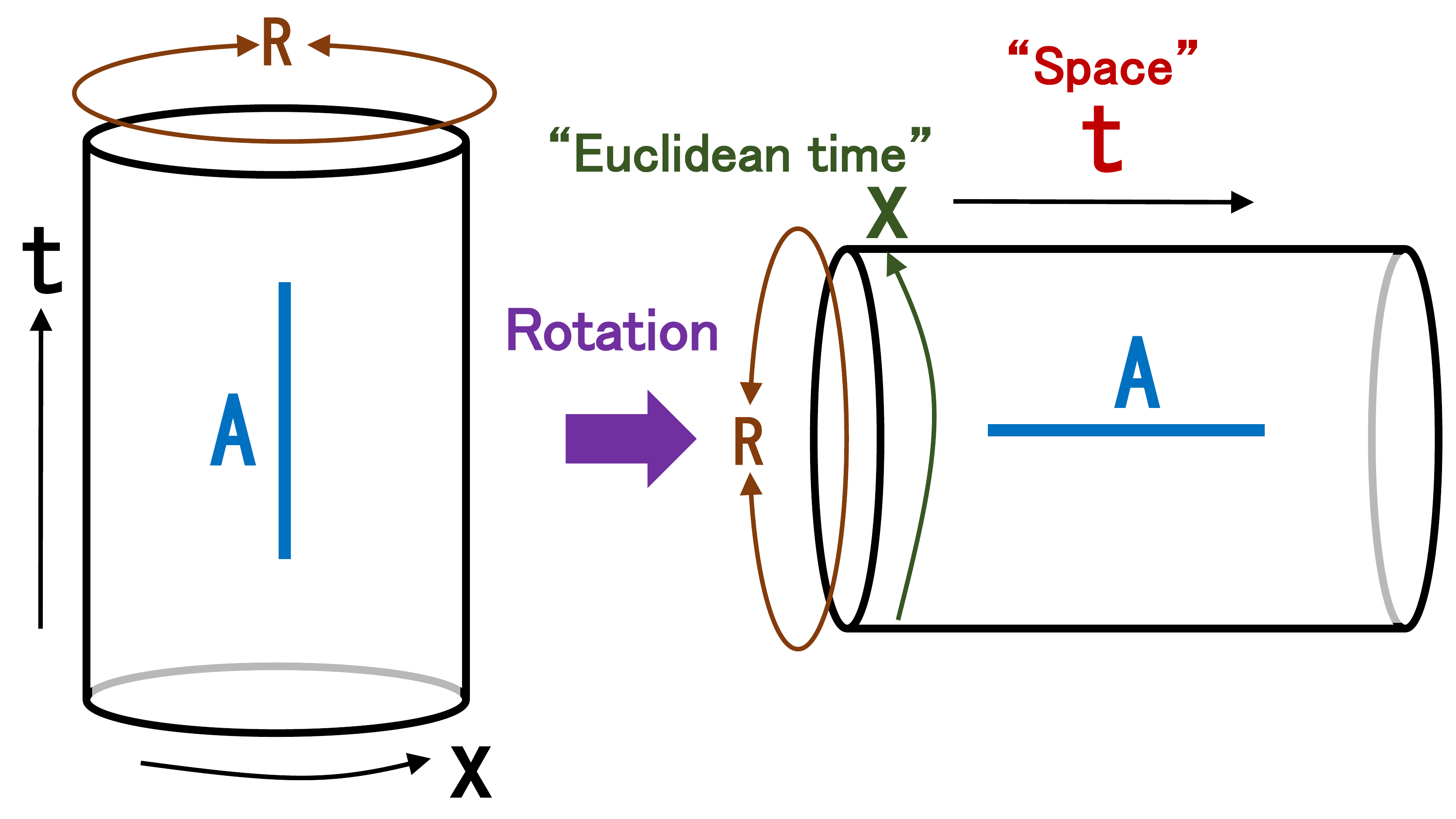}
    \caption{Timelike entanglement entropy via a Wick rotation of coordinates.}
    \label{fig:QFTPE}
\end{figure}
However, let us regard $t$ as the ``space'' direction and $x$ as the Euclidean time, such that 
$ix(\equiv-T)$ is the real time, by rotating the spacetime by ninety degree, as depicted in figure \ref{fig:QFTPE}.
The ``Hamiltonian" $H$ reads 
\ba\label{eq:dHd}
H=-\frac{i}{2}\int dt\left[\pi^2+(\de_t\phi)^2-m^2\phi^2\right],  \label{hamtim}
\ea
where 
\ba
\pi=-\de_x\phi, \label{monw}
\ea
is the canonical momentum such that the commutation relation is
\ba
[\phi(t),\pi(t')]=i\delta(t-t').
\ea
To see this, by introducing the real time $T$ related to the Euclidean time $x$ via
$x=iT$, we can rewrite the action (\ref{Laciona}) as follows
\ba
S=\frac{i}{2}\int dTdt\left[(\de_T\phi)^2+(\de_t\phi)^2-m^2\phi^2\right].\label{hampp}
\ea
Then the momentum conjugate to $\phi$, defined by $\pi=\frac{\delta S}{\delta (\de_T\phi)}$, is found to be (\ref{monw}).
The Hamiltonian is also canonically defined as $H=\frac{\delta L}{\delta (\de_T\phi)}(\de_T\phi)-L$, where
$L=\frac{i}{2}\int dt[(\de_T\phi)^2+(\de_t\phi)^2-m^2\phi^2]$ is the Lagrangian of $S$. This leads to (\ref{hampp}).

In this formulation, regarding $t$ as a spatial coordinate and $x$ as the Euclidean time with a periodicity $R$, the partition function 
\ba
Z_{\phi}=\Tr [e^{-R H}],
\ea
where $H$ is the timelike Hamiltonian \eqref{hamtim}, can be rewritten as 
\begin{align}
    Z_{\phi}=\mbox{Tr}\left[e^{iR \ti{H}}\right],
\end{align}
using the rescaled Hamiltonian $\tilde{H}=iH$, which takes the conventional form with a minus sign for the mass term:
\ba
\ti{H}=\frac{1}{2}\int dt\left[\pi^2+(\de_t\phi)^2-m^2\phi^2\right].
\ea
By tracing out the region $B$, the reduced density matrix $\rho_A$ is now given by 
\ba\label{TDM}
\rho_A=\mbox{Tr}_B[e^{iR \ti{H}}].
\ea
Therefore, we can obtain the timelike EE from the ordinary EE of a spacelike subsystem at finite temperature $1/\beta_{S}$ by the transformation:
\ba
\beta_{S}\to -iR, \ \ \ m\to-im.   \label{transmb}
\ea
In the massless limit $m=0$, the transformation above acts purely on $\beta_{S}$ and thus we expect this is universal for any two dimensional CFTs. 

The entanglement entropy for the thermal state at temperature $1/\beta_{S}$ is known to be given by \cite{Calabrese:2004eu}:
\begin{align}\label{EES1}
    S_A=\frac{c}{3}\log\left[\frac{\beta_{S}}{\pi\tilde{\epsilon}}\sinh\frac{\pi X}{\beta_{S}}\right],
\end{align}
where $X$ is the length of the spacelike interval $A$. 
Note that the energy cutoff $\tilde{\epsilon}$ is defined for $\tilde{H}$, so it is related as $\tilde{\epsilon}=-i\epsilon$ to $\epsilon$ defined for the Hamiltonian $H$ we are interested in. 
Thus by applying (\ref{transmb}) and setting $X=T_0$, we obtain the timelike EE:
\begin{align}\label{eq:tee}
    \tee{A}=\frac{c}{3}\log\left[\frac{R}{\pi\epsilon}\sin\frac{\pi T_0}{R}\right]+\frac{i\pi c}{6}.
\end{align}
This indeed reproduces our previous result for a CFT on a circle from the replica method (\ref{TEEsize}).
By taking the limit $R\to \infty$ we can also derive the decompactified result (\ref{TEEarep}).

We can also consider a compactification in $t$ direction to compute the timelike EE at a finite temperature, where we do not compactify $x$ coordinate. For the temperature $1/\beta$ we compactify $t$ as $t\sim t-i\beta$ so that the periodicity of the Euclidean time $it$ is $\beta$. Remember that the ordinary EE for a spacelike interval $A$ on a circle with the circumference $R_S$ reads 
\ba\label{eq:EEcompact}
S_A=\frac{c}{3}\log\left[\frac{R_S}{\pi \ti{\ep}}\sin\left(\frac{\pi X}{R_S}\right)\right].
\ea
To obtain the timelike EE at finite temperature, we identify $R_S=-i\beta$ and 
$\tilde{\epsilon}=-i\epsilon$ as we explained. This perfectly reproduce our previous result (\ref{TEEtemp}) based on the replica method.

It is important to note that the reduced density matrix (\ref{TDM}) for our timelike EE is not hermitian as opposed to the case of standard EE defined for a spacelike subsystem $A$. This shows that the timelike entanglement entropy can properly be regarded as a special example of pseudo entropy \cite{Nakata:2021ubr}. More explicitly, we can see this by a purification procedure as follows. Consider a thermofield double-like description by doubling the total Hilbert space and by introducing the following two states 
\ba
&& |\Psi\lb=\frac{1}{\s{Z(\delta)}}\sum_n e^{i(R+i\delta) E_n/2}|n\lb_1|n\lb_2,\no
&& |\Psi^*\lb=\frac{1}{\s{Z(\delta)}}\sum_n e^{-i(R-i\delta) E_n/2}|n\lb_1|n\lb_2,
\ea
where $\delta$ is an infinitesimally small regularization parameter;  $E_n$ and $|n\lb$ are the eigenvalues and eigenstates of $\ti{H}$. They satisfy 
\ba
\mbox{Tr}_2|\Psi\lb \la\Psi^{*}|
=e^{i(R+i\delta)\ti{H}}.
\ea
By tracing out $B$ further, this is identical to the reduced density matrix $\rho_A$  in (\ref{TDM}). In this way, we can rewrite the timelike EE in terms of a transition matrix which shows that the timelike EE is an an example of pseudo entropy.
It is curious to note that such a thermofield double state at an imaginary temperature was introduced in \cite{Cotler:2023xku}
to find a holographic interpretation of de Sitter space in terms of the CFT on the future infinity and the other one on the past one. 

As we have seen in this subsection, we have two prescriptions to calculate the timelike EE: the Wick rotations  of $(i)$ the interval $A$ and of $(ii)$ the coordinates $(t,x)$. In the remainder of this section, we will investigate the timelike entanglement entropy in the various models by adopting a suitable prescription. It is also useful to note that the higher dimensional generalization is straightforward just by adding extra coordinates $y^i$ to the arguments above.

\subsection{Timelike entanglement entropy of Dirac fermion on a torus}

As our next example, we would like to analyze a finite temperature state of a two dimensional CFT on a circle. In this case we cannot use the conformal mapping method to obtain a universal result for the entanglement entropy. Thus, we focus on the $c=1$ Dirac fermion CFT, whose entanglement entropy was first computed in \cite{Azeyanagi:2007bj}.
We consider this CFT on a torus and choose the subsystem $A$ to be an interval $A=PQ$, where $P=(0,0)$ and $Q=(z,\bar{z})$. The periodicity of the Euclidean time Im$z$ and the spatial coordinate 
Re$z$ are set to be $\beta_{S}$ and $R_S=1$, respectively. We assume the NS-sector spin structure. The two point function of twist operator reads 
\ba
\la \sigma_k(z,\bar{z})\sigma_{-k}(0,0)\lb=\left(\frac{4\pi^2\eta(\tau)^2}{\theta_1(z|\tau)\theta_1(\bar{z}|\bar{\tau})}\right)^{2\Delta_k}\cdot 
\frac{\theta_3\left(\frac{k}{N}z|\tau \right)\theta_3\left(\frac{k}{N}\bar{z}|\bar{\tau}\right)}{\theta_3(\tau)^2},
\ea
where $\theta_3(\nu|\tau), \theta_1(\nu|\tau)$ and $\eta(\tau)$ are standard theta and eta functions.
We take $z=L+T$ and $\bar{z}=L-T$.  In the spacelike limit $T=0$, i.e. $(z,\bar{z})=(L,L)$, the entanglement entropy in the high temperature expansion reads \cite{Azeyanagi:2007bj}
\ba
S_A&=&\frac{1}{3}\log\left[\frac{\beta_{S}}{\pi\ep}\sinh\left(\frac{\pi L}{\beta_{S}}\right)\right]
+\frac{1}{3}\sum_{m=1}^\infty\log\left[\frac{(1-e^{\frac{2\pi L}{\beta_{S}}}e^{-2\pi R_S\frac{m}{\beta_{S}}})(1-e^{-\frac{2\pi L}{\beta_{S}}}e^{-2\pi R_S\frac{m}{\beta_{S}}})}
{(1-e^{-2\pi R_S\frac{m}{\beta_{S}}})^2}\right]\no
&&+2\sum^\infty_{l=1}\frac{(-1)^l}{l}\cdot \frac{\frac{\pi Ll}{\beta_{S}}\coth\left(\frac{\pi Ll}{\beta_{S}}\right)-1}{\sinh\frac{\pi R_Sl}{\beta_{S}}},
\label{highex}
\ea
where $R_S=1$ is the periodicity of the $\Re z$ direction. In the low temperature expansion, this is equivalently written as follows \cite{Azeyanagi:2007bj}
\ba
S_A&=&\frac{1}{3}\log\left[\frac{1}{\pi\ep}\sin\left(\pi L\right)\right]
+\frac{1}{3}\sum_{m=1}^\infty\log\left[\frac{(1-e^{2\pi iL}e^{-2\pi\beta_{S} m})(1-e^{-2\pi iL}e^{-2\pi\beta_{S} m})}
{(1-e^{-2\pi\beta_{S} m})^2}\right]\no
&&+2\sum^\infty_{l=1}\frac{(-1)^{l}}{l}\cdot \frac{\pi Ll{\beta_{S}}\cot\left(\pi Ll\right)-1}{\sinh\pi l\beta_S}.\label{lowex}
\ea

In the timelike limit, i.e. $(z,\bar{z})=(T,-T)$, we find
\ba
\tee{A}&=&\frac{1}{3}\log\left[\frac{\beta}{\pi\ep}\sinh\left(\frac{\pi T}{\beta}\right)\right]+\frac{\pi}{6}i\no
&&+\frac{1}{3}\sum_{m=1}^\infty\log\left[\frac{(1-e^{\frac{2\pi T}{\beta}}e^{-2\pi\frac{m}{\beta}})(1-e^{-\frac{2\pi T}{\beta}}e^{-2\pi\frac{m}{\beta}})}
{(1-e^{-2\pi\frac{m}{\beta}})^2}\right]
+2\sum^\infty_{l=1}\frac{(-1)^l}{l}\cdot \frac{\frac{\pi Tl}{\beta}\coth\left(\frac{\pi Tl}{\beta}\right)-1}{\sinh\frac{\pi l}{\beta}},\no
\label{timetorus}
\ea
where we set $\beta_{S}=\beta$.

Thus the only difference from the standard EE is again just the imaginary 
part $\frac{\pi c}{6}i$ in this torus calculation.
To see this note that $\theta_1(z|\tau)$ and $\theta_3(z|\tau)$ are odd and even function w.r.t. $z$, respectively. Note also that this timelike EE is oscillating with the periodicity $2\pi$ under the $T$ evolution.

The other definition of timelike EE, which Wick rotates the whole spacetime, is argued to give the identical result (\ref{timetorus}). We can confirm this as follows.
As we learned in section \ref{subsec:formulation},
the timelike EE is related to the standard spacelike EE via the double Wick rotation:
\ba
\tee{A}(R_,\beta,T)=S_A(R_S,\beta_{S},L)+\frac{\pi c}{6}i,
\label{eeweqax}
\ea
with the identification
\ba
(R_S,\beta_{S},L)= (-i\beta,-iR,T),  \label{identig}
\ea
where $\beta$ and $R(=1)$ are the inverse temperature and the periodicity in $x$ for the timelike EE, respectively. By applying this relation (\ref{eeweqax}) to (\ref{highex}) we obtain the following expression of timelike EE:
\ba\label{eq:timetorusHT}
\tee{A}&=&\frac{1}{3}\log\left[\frac{1}{\pi\ep}\sin\left(\pi T\right)\right]
+\frac{1}{3}\sum_{m=1}^\infty\log\left[\frac{(1-e^{2\pi iT}e^{-2\pi\beta m})(1-e^{-2\pi iT}e^{-2\pi\beta m})}
{(1-e^{-2\pi\beta m})^2}\right]\no
&&+2\sum^\infty_{l=1}\frac{(-1)^{l}}{l}\cdot \frac{\pi Tl{\beta}\cot\left(\pi Tl\right)-1}{\sinh\pi l\beta}+\frac{\pi c}{6}i.\label{eetimegqweel}
\ea
As the low temperature expansion of (\ref{highex}) is given by (\ref{lowex}),  we can show that the result above (\ref{eetimegqweel}) indeed agrees with the timelike EE from the Wick rotation of the interval itself (\ref{timetorus}).

\subsection{Numerical Method}
\subsection*{Free Scalar Theory}
In this part we adapt the correlator method \cite{Casini:2009sr, Peschel} to the Hamiltonian we have introduced in equation \eqref{eq:dHd} in order to compute timelike EE numerically. To do so we need to be more precise about how we have defined equation \eqref{eq:dHd}. In the ordinary free scalar theory (the vertical cylinder in figure \ref{fig:QFTPE}), the mode expansion of the field is given by
\ba
\phi(\mathbf{x},t)=\int \frac{d^dk}{(2\pi)^{d/2}}\frac{1}{\sqrt{2\omega_{\mathbf{k}}}}
\left(a_{\mathbf{k}}\,e^{-i\omega_{\mathbf{k}} t}+a^{\dagger}_{-\mathbf{k}}\,e^{i\omega_{\mathbf{k}}t}\right)e^{i{\mathbf{k}}\cdot{\mathbf{x}}}\,,
\ea
where $\omega^2_{\mathbf{k}}=|\mathbf{k}|^2+m^2$. In our conventions we denote the original spatial coordinate $x$ to play the role of Euclidean time and the rest $(d-1)$ spatial coordinates with $\mathbf{y}$. 
Using this expansion one can find the correlation functions which lead to the expected results for timelike entanglement, but we will not proceed in this way to use the field expansion corresponding to the vertical cylinder.
We rather considered the viewpoint of the right horizontal cylinder in figure \ref{fig:QFTPE}, where $T=-ix$ plays the role of Lorentzian time. Within this point of view, we may define the scalar theory defined via \eqref{hampp} or equivalently in the Hamiltonian formalism. The Hamiltonian in $(d+1)$ dimensions is given by
\ba\label{eq:dHdD}
H=-\frac{i}{2}\int dtd\mathbf{y}\left[\pi^2+(\de_t\phi)^2-(\nabla_y \phi)^2-m^2\phi^2\right]\,.
\ea
Using the Hamiltonian equations of motion
\ba
\partial_T\p=\frac{\delta H}{\delta \pi}
\;\;\;\;\;\;\;,\;\;\;\;\;\;\;
-\partial_T\pi=\frac{\delta H}{\delta \p}
\ea
we find that  $\pi=-\partial_x\p$ and 
\ba
(\partial_t^2-\partial_x^2-\nabla_y^2+m^2)\p=0  \label{EOMsctim}
\ea

Now considering a decomposed notation for the momenta as $\mathbf{k}=(k,\mathbf{k}_y)$ where $\mathbf{k}$ corresponds to the momenta of $(t,\mathbf{y})$ and we consider
\ba
\p(t,x,\mathbf{y})=
\int \frac{d^{d-1}k_y\,dk}{(2\pi)^{d/2}}\p(k,x,\mathbf{k}_y) e^{-ik t+i{\mathbf{k}}_y\cdot{\mathbf{y}}}
\ea
which leads to
\ba
\left(\partial_x^2+\Omega_\mathbf{k}^2\right)\p(k,x,\mathbf{k}_y)=0\;,
\ea
where $\Omega_\mathbf{k}=\sqrt{k^2-|\mathbf{k}_y|^2-m^2}$.
Following the standard procedure in Klein-Gordon theory based on the similarity between this equation and harmonic oscillator equation of motion, we consider the following Fourier expansions 
\begin{align}
\begin{split}
    \phi(t,\mathbf{y})&=\int \frac{d^{d-1}k_y\,dk}{(2\pi)^{d/2}} \frac{1}{\sqrt{2\Omega_{\mathbf{k}}}}\left(a_{\mathbf{k}}+a^{\dagger}_{-\mathbf{k}}\right) e^{-ik t+i{\mathbf{k}}_y\cdot{\mathbf{y}}}\,,\\
    \pi(t,\mathbf{y})&=-i\int \frac{d^{d-1}k_y\,dk}{(2\pi)^{d/2}} \sqrt{\frac{\Omega_{\mathbf{k}}}{2}}\left(a_{\mathbf{k}}-a^{\dagger}_{-\mathbf{k}}\right) e^{-ik t+i{\mathbf{k}}_y\cdot{\mathbf{y}}}\,.
\end{split}
\end{align}
With these expressions one can check that assuming the standard commutation relations between the creation and annihilation operators, the canonical commutation relations are preserved
and the Hamiltonian is diagonalized up to a constant as
\ba
H=-i\int d^{d-1}k_y\,dk
\;\Omega_{\mathbf{k}}\,a^{\dagger}_{\mathbf{k}}a_{\mathbf{k}}
\ea
where the structure of the dispersion relation is as expected from the form of equation \eqref{eq:dHdD}, namely the sign of the $y$-direction derivatives and the mass term are opposite to the Hamiltonian in standard Lorentzian theory.

\begin{figure}[t]
  \centering
  \includegraphics[scale=.31]{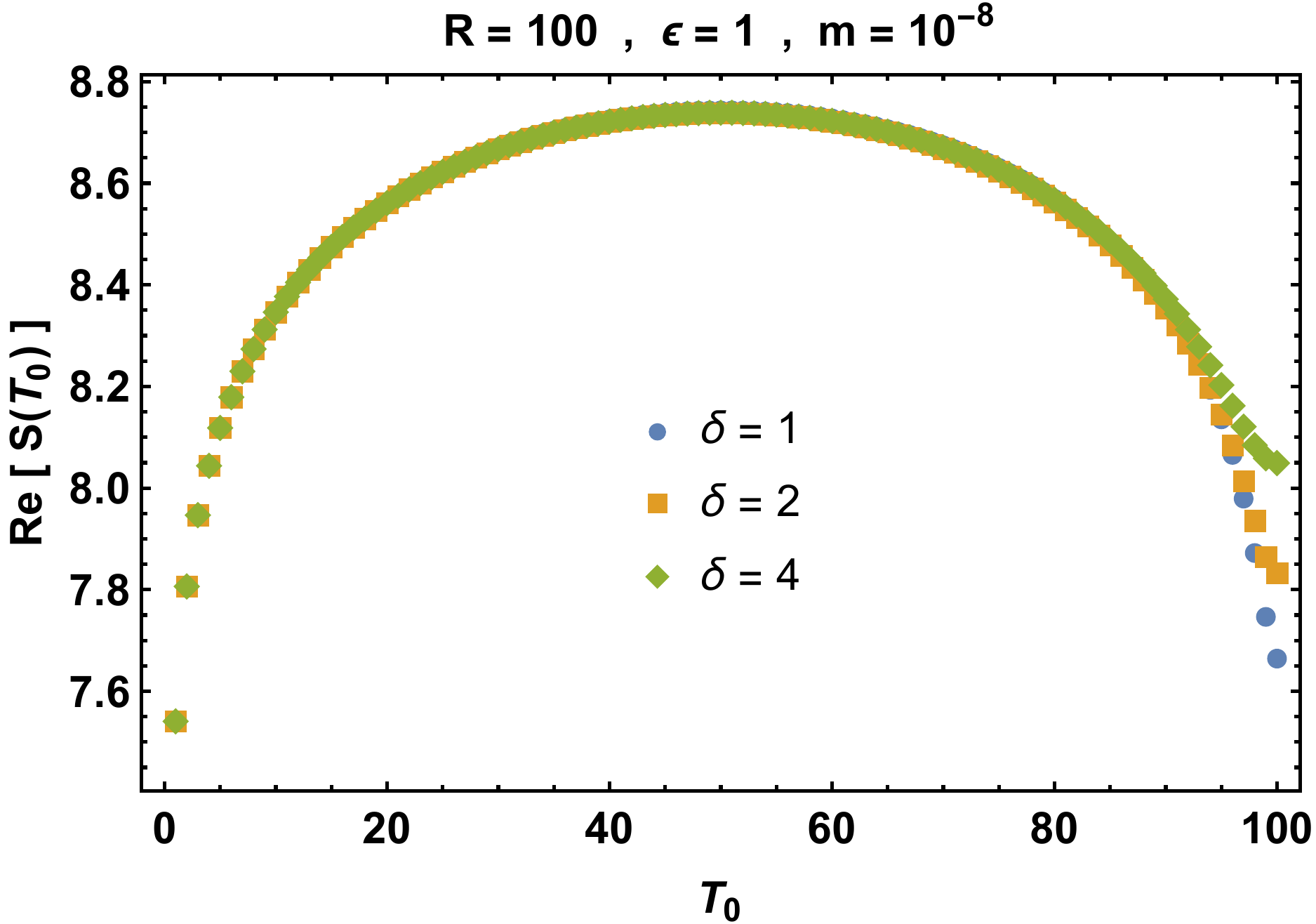}
  \includegraphics[scale=.29]{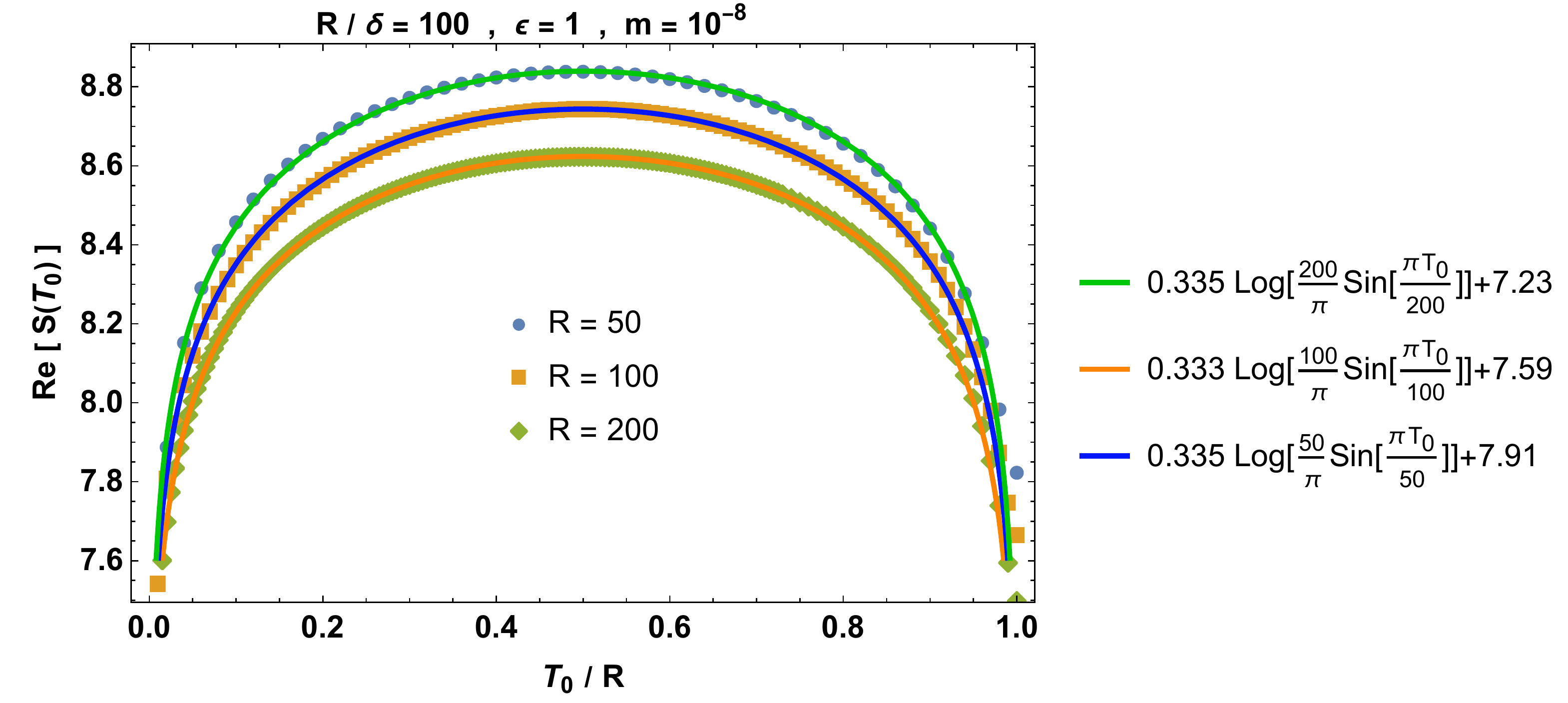}
  \caption{Timelike entanglement for free scalar theory for a connected region. The $t$-direction is infinite and the $x$-direction is compactified as $x\sim x+R$ and we also put the regulator $\delta$. The left panel where the data of different color lie on the top of each other corresponds to fixed $R$ and different $\delta$. As expected modulo some numerical defects for large sugberions the timelike EE does not depend on the ratio of $R / \delta$. The right panel is showing results for different $R$ with fixed $R/\delta$. The structure of the fit function perfectly matches with our analytic expectation. As $R$ is increasing larger subregions are considered and the difference of data sets is encoded in different UV cut-offs which is reflected in the constant term in our fit functions.} 
\label{fig:NumD2scalar1}
\end{figure}

Now we are equipped with what we need to calculate timelike entanglement entropy. We consider the $d=1$ case in this paper. In order to find the correct prescription of the correlator method to compute timelike EE, it is enough to focus on a single mode contribution to the correlations functions. Tracing over the $x$ direction is the nontrivial part where we have to introduce a cut-off to sum the geometric series of the form
$$\sum_n e^{iR\,\Omega_k\,n}f(n)
\;\;\; \longrightarrow \;\;\;  \sum_n e^{i(R+i\delta)\,\Omega_k\,n}f(n)=\sum_n e^{-(R+i\delta)\,\Omega_{ik}\,n}f(n).$$
In the case that the $t$-direction is an infinite line, namely $\beta\to\infty$, we consider a straight line along $t$ as the entangling region. All we need to find is the correlation matrix that which its non-trivial elements are given by
\footnote{It is worth reminding the reader the correlators relevant to calculate standard EE in finite temperature case
\begin{align}\label{eq:SEEcor}
\begin{split}
    \Phi_{xx'}&\equiv
    \mathrm{Tr}\left[e^{-\beta_S H}\phi(x)\phi(x')\right]=
    \int \frac{dk}{2\pi} \frac{1}{2\omega_{k}}\coth \left(\frac{\beta_S\, \omega_{k}}{2}\right)\,e^{ik(x-x')}\,,
    \\
    \Pi_{xx'}&\equiv
    \mathrm{Tr}\left[e^{-\beta_S H}\pi(x)\pi(x')\right]=
    \int \frac{dk}{2\pi} \frac{\omega_{k}}{2}\coth \left(\frac{\beta_S\, \omega_{k}}{2}\right)\,e^{ik(x-x')}\,.
\end{split}
\end{align}
}
\begin{align}\label{eq:TEEcorR}
\begin{split}
    \Phi_{tt'}&\equiv
    \mathrm{Tr}\left[e^{i(R+i\delta)\tilde{H}}\phi(t)\phi(t')\right]=
    \int \frac{dk}{2\pi} \frac{i}{2\Omega_{ik}}\coth \left(\frac{(R+i\delta)\, \Omega_{ik}}{2}\right)\,e^{ik(t-t')}\,,
    \\
    \Pi_{tt'}&\equiv
        \mathrm{Tr}\left[e^{i(R+i\delta)\tilde{H}}\pi(t)\pi(t')\right]=\int \frac{dk}{2\pi} \frac{\Omega_{ik}}{2i}\coth \left(\frac{(R+i\delta)\,\Omega_{ik}}{2}\right)\,e^{ik(t-t')}\,.
\end{split}
\end{align}
For CFT case in the continuum, one can simply verify that $\Omega_{ik}=i\omega_k$. While this is not true on the lattice, we introduce the aforementioned expressions for the correlators as the correct prescription to calculate timelike EE. 
To numerically work out the timelike EE we consider the discrete version on this infinite lattice, i.e. we integrate over $-\pi<k<\pi$ where $t$ takes integer values and the dispersion relation is $\epsilon^2\Omega_k^2=4\sin^2(\frac{k}{2})-(m\,\epsilon)^2$. The timelike EE is related to the spectrum of the operator $\sqrt{\Phi\cdot\Pi}$ shown by $\{\mu\}$ via
\be \label{eq:EEcorr}
\tee{A}=\sum_i\left[\left(\mu_i+\frac{1}{2}\right)\log\left(\mu_i+\frac{1}{2}\right)-\left(\mu_i-\frac{1}{2}\right)\log\left(\mu_i-\frac{1}{2}\right)\right]
\ee
In figure \ref{fig:NumD2scalar1} we show the numerical results corresponding to the case where $\beta\to\infty$. Our results are capturing the expected periodic behavior for timelike EE. It is worth noting that although the real part of the timelike EE is similar to the standard EE on a compact space \eqref{TEEsize}, the structure of the periodicity of the standard EE and timelike EE on the compactified space direction are of completely different origin. In standard EE the periodicity of entropy is inherited from the periodicity of the fields which is reflected in the corresponding correlators as on a compact space we have to consider the replacement $e^{ik(x-x')}\to \cos\frac{2\pi k(x-x')}{R}$ in \eqref{eq:SEEcor}. The periodicity in the case of timelike EE rather originates in the regulated summation leading to the hyperbolic cotangent in $\coth$ the integrands of \eqref{eq:TEEcorR}.  

\begin{figure}[t]
  \centering
  \includegraphics[scale=.4]{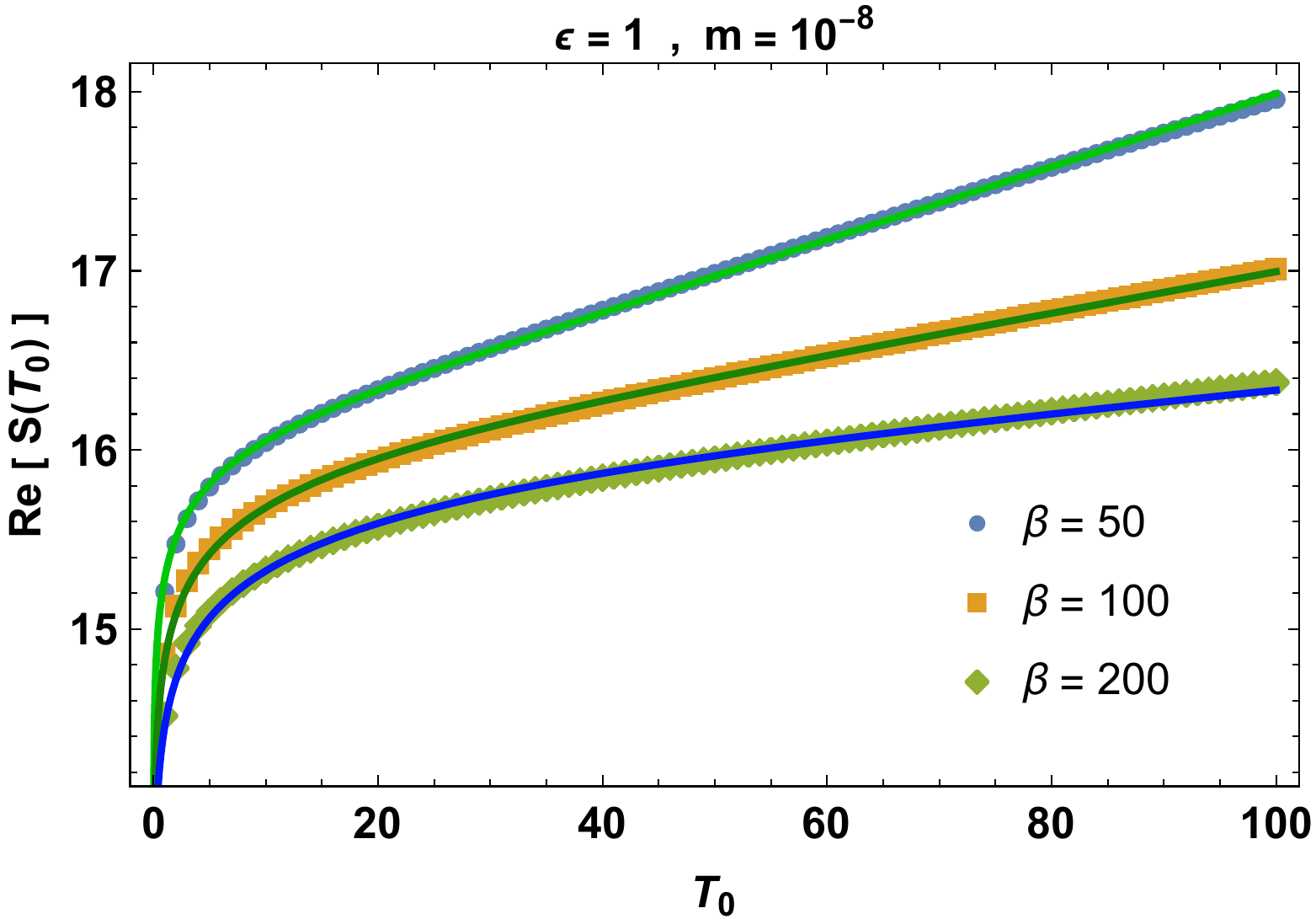}
  \caption{Timelike EE for free scalar theory for a connected region. We consider $t$ to be compactified as $t\sim t-i\beta$ and $R=\infty$. The data points correspond to different values for $\beta$ and the solid curves are fitting functions up to an irrelevant constant of the form  $a\log\left(\frac{\beta}{\pi\epsilon}\sinh\frac{\pi T_0}{\beta}\right)$ which lead to $a=0.324$, $a=0.367$, and $a=0.376$ decreasing the temperature. 
  } 
\label{fig:NumD2scalar2}
\end{figure}

The other case which we are interested in is the finite temperature case, namely when we consider Euclidean time to be compactified as $t\sim t-i\beta$. In this case one should be careful about the generalization of the \eqref{eq:TEEcorR} to the finite temperature case due to imaginary time periodicity. To do so we consider the standard method in thermal field theory, i.e. rather than working with imaginary time in real space, we sum over all mirror images in the Fourier space (considering the momentum $k$ to be held as a continuous variable) 
\begin{align}\label{eq:TEEcorRBeta}
\begin{split}
    \Phi_{tt'}&=
    \sum_{n=-\infty}^{\infty}\int_{-\pi}^{\pi} \frac{dk}{2\pi} \frac{i}{2\Omega_{ik}}\coth \left(\frac{(R+i\delta)\, \Omega_{ik}}{2}\right)\,e^{ik(t-t'-in\beta)}\,,
\end{split}
\end{align}
and similarly for $\Pi_{tt'}$. When $n$ and $k$ have the same sign, this integral diverges and we need to regularize it. Here we regularize these correlators by simply excluding the aforementioned cases that cause the divergence, namely those cases where $n$ and $k$ have the same sign. Doing so one arrives at
\begin{align}\label{eq:TEEcorRBetaReg}
\begin{split}
    \Phi_{tt'}&\myeq
    \int_{-\pi}^{\pi} \frac{dk}{2\pi} \frac{i}{2\Omega_{ik}}\coth \left(\frac{(R+i\delta)\, \Omega_{ik}}{2}\right)\frac{1}{1-e^{-|k|\beta}}\,e^{ik(t-t')}\,,
\end{split}
\end{align}
with a similar expression for conjugate momenta correlations. In figure \ref{fig:NumD2scalar2} we have presented the result of using this regularization for the case that $R\gg\beta$. The real part perfectly agrees with our expectation from analytical results.\footnote{Other Dirichlet regularization schemes, such as acting with $e^{-n^2}$ damping function leads to similar results with lower numerical precision.} One can see the expected behavior of \eqref{TEEtemp} with reasonable values for the prefactor reported in the caption. In the $R\gg\beta$ regime one can even do better after approximating the $\coth$ part in the integrand by unity and analytically performing the integrals which again need a regularization prescription. We avoid to present these results since they are not more illuminating than what is presented in figure \ref{fig:NumD2scalar2}.


\begin{figure}[t]
  \centering
  \includegraphics[scale=.36]{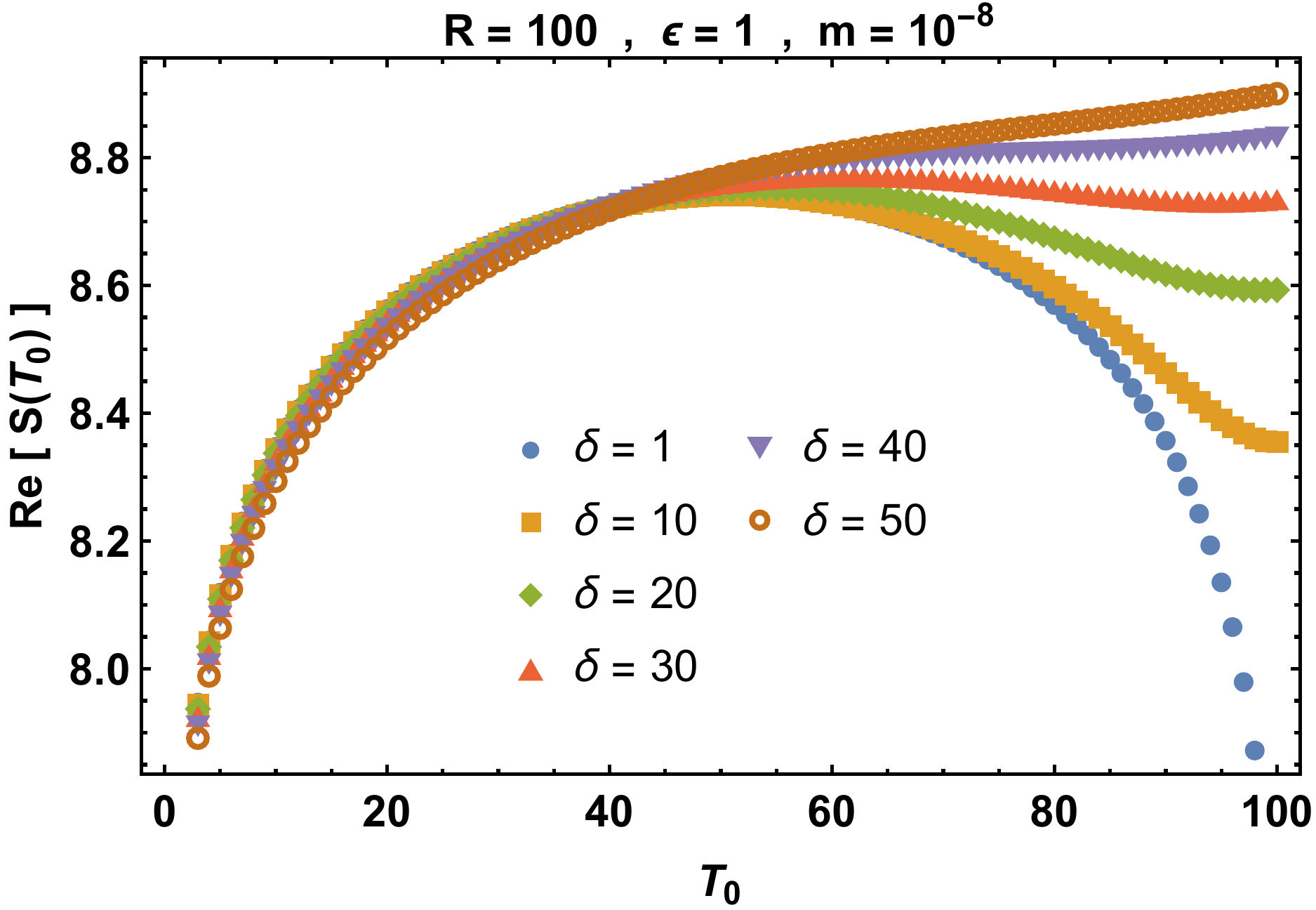}
  \includegraphics[scale=.45]{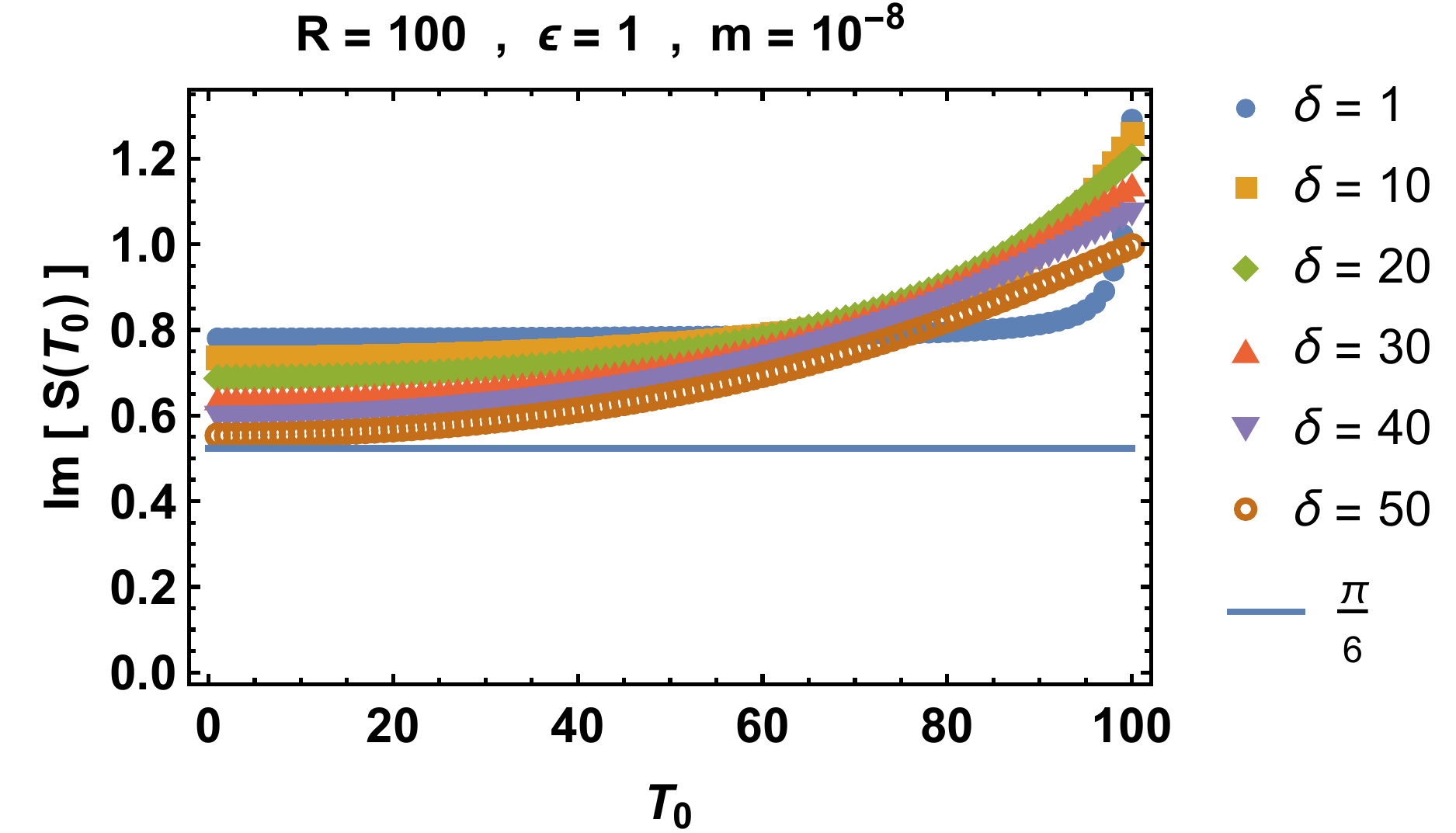}
  \caption{Here we show the challenge to find the numerical value of the imaginary part of timelike EE. Focusing on the region that the numerical results are stable, namely on smaller subregions in this figure, the data presented in the right panel is strongly suggesting that the imaginary part takes a constant value. As we increase the value of $\delta$, this constant approaches to the expected value $\pi/6$ while in this direction the results become less stable for larger subregions. The stability of the results can be seen from the left panel where we show how the real part depart from the \eqref{eq:tee} behavior. Numerics confirm our analytical expectation value for the imaginary part in the small subregion regime.} 
\label{fig:NumD2scalar3}
\end{figure}

As figure \ref{fig:NumD2scalar1} and figure \ref{fig:NumD2scalar2} are showing the real part of timelike EE, a reasonable question is whether our numerical method can also confirm the constant imaginary part as well. Numerically finding the imaginary part is much more challenging than the real part. This is technically due to the difference of the origin of the imaginary part and the real part. The real part, similar to the case of standard EE, gets its dominant contribution from a small number of eigenvalues in \eqref{eq:EEcorr}. In other words, the corresponding spectrum contains a huge gap where a very few large eigenvalues contribute dominantly and the rest contribute very slightly. The case is very different for the imaginary part which almost all eigenvalues contribute to this part democratically. So relatively a very high-precision is needed to find the small eigenvalues as precise as possible. Since the integrand of the correlators are rapidly oscillating, this means that we need to increase the cut-off $\delta$ to improve the convergence. But physically we need to take $\delta/R\to 0$, as we confirmed this in our results for the real part. So there is a trade-off between getting reasonable physical results and considering larger values of $\delta$ to improve the convergence and finding a reliable value for the imaginary part. In figure \ref{fig:NumD2scalar3} we have shown how this works. From the numerical data one can verify that even in the $\delta\ll R$ regime, which the real part is reliable on the whole $R$-cycle, the imaginary part takes a constant value. We have shown that by increasing $\delta$, we can numerically approach a regime where although $\delta\lesssim R$, the real part is still reliable for small enough $T_0$ and the imaginary part in this regime approaches the expected value of $\pi/6$ for the $c=1$ scalar theory. 

\subsection*{Free Dirac Fermion Theory}
Following a similar procedure to what we did for the scalar theory, we show that the other tractable example with our numerical method Dirac fermion theory also leads to similar results. Using conventions similar to our scalar theory analysis we have
\ba
S=\int dtdxd\mathbf{y}\,\bar{\psi}\left(i\gamma^0\partial_t+i\gamma^1\partial_x+i\mathbb{\gamma}^y\cdot\mathbb{\nabla}_y-m\right)\psi\,,
\ea
where $\bar{\psi}=\psi^\dagger\gamma^1$. With the same definition for $x=iT$ we have
\ba
S=i\int dtdTd\mathbf{y}\,\bar{\psi}\left(i\gamma^0\partial_t+\gamma^1\partial_T+i\mathbb{\gamma}^y\cdot\mathbb{\nabla}_y-m\right)\psi\,,
\ea
where the conjugate momentum is defined as $\pi=i\bar{\psi}\gamma^1$ which leads to the following Hamiltonian
\ba\label{eq:HDirac}
H=-i\int dtd\mathbf{y}\,\bar{\psi}\left(i\gamma^0\partial_t+i\mathbb{\gamma}^y\cdot\mathbb{\nabla}_y-m\right)\psi\,.
\ea
Now similar to the scalar theory formulation, the partition function is given by
\ba
Z_{\psi}=\Tr [e^{-\beta H}]=\Tr [e^{i\beta \tilde{H}}],
\ea
where
\ba
\tilde{H}=\int dtd\mathbf{y}\,\bar{\psi}\left(i\gamma^0\partial_t+i\mathbb{\gamma}^y\cdot\mathbb{\nabla}_y-m\right)\psi\,.
\ea
Considering the 2$d$ case, using the algebra of gamma matrices and the reduced density matrix defined as $\rho_A=\Tr_A [e^{i\beta \tilde{H}}]$, one can verify that timelike EE is related to ordinary EE via the same analytic continuations considered in equation \eqref{transmb}
\ba
\beta\to -i\beta, \ \ \ m\to-im.   
\ea
In the rest we focus on the 2$d$ case. We first consider the case where $t$ is defined on an infinite line to calculate timelike EE. In order to explicitly diagonalize the Hamiltonian we chose $\gamma^0=\sigma_2$ and $\gamma^1=i\sigma_1$ and we consider the $\psi_k^T=(u_k\;\;d_k)$ where $\psi_k$ is the Fourier transform of the Dirac field defined as
\be\label{eq:Fourierfer}
\psi(t)=\int\frac{dk}{\sqrt{2\pi}}\,\psi_k\,e^{ikt}\,.
\ee
Using the following Bogoliubov transformations
\ba
u_k=\cos\theta_k b_k+ i \sin\theta_k b^\dagger_{-k}
\;\;\;\;\;,\;\;\;\;\;\;
d_k=\sin\theta_k b_k- i \cos\theta_k b^\dagger_{-k}
\ea
one can easily check that by choosing 
\ba\label{eq:dd}
\cos 2\theta_k=\frac{k}{\Omega_k}
\;\;\;\;\;,\;\;\;\;\;\;
\sin 2\theta_k=\frac{im}{\Omega_k}
\;\;\;\;\;,\;\;\;\;\;\;
\Omega_k^2=k^2-m^2
\ea
the Hamiltonian \eqref{eq:HDirac} in 2$d$ is diagonalized as
\ba\label{eq:HDiracdiag}
H=i\int dt\,\Omega_k\,\left(b_k^\dagger b_k+b_{-k}^\dagger b_{-k}\right)\,.
\ea
\begin{figure}[t]
  \centering
  \includegraphics[scale=.37]{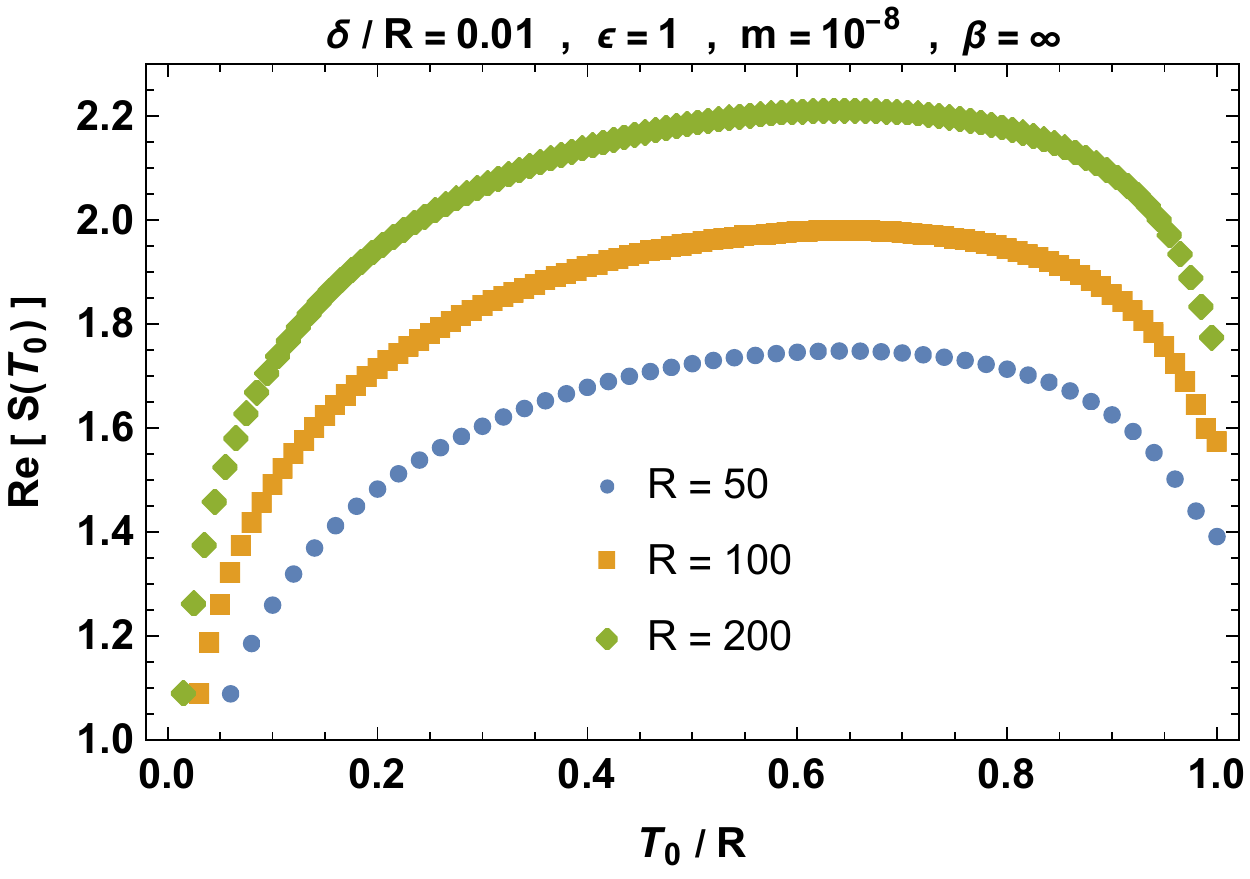}
  \hspace{2mm}
  \includegraphics[scale=.37]{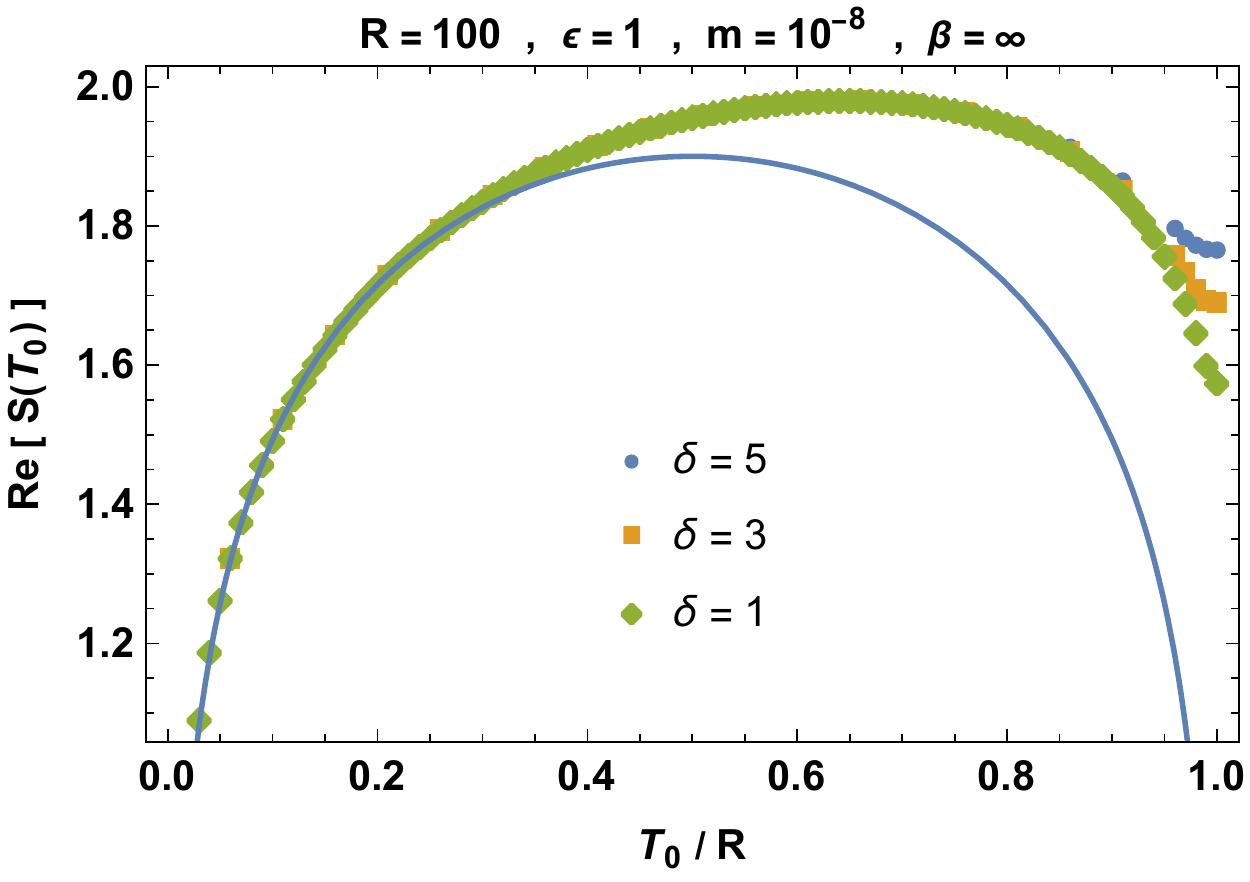}
  \hspace{2mm}
  \includegraphics[scale=.31]{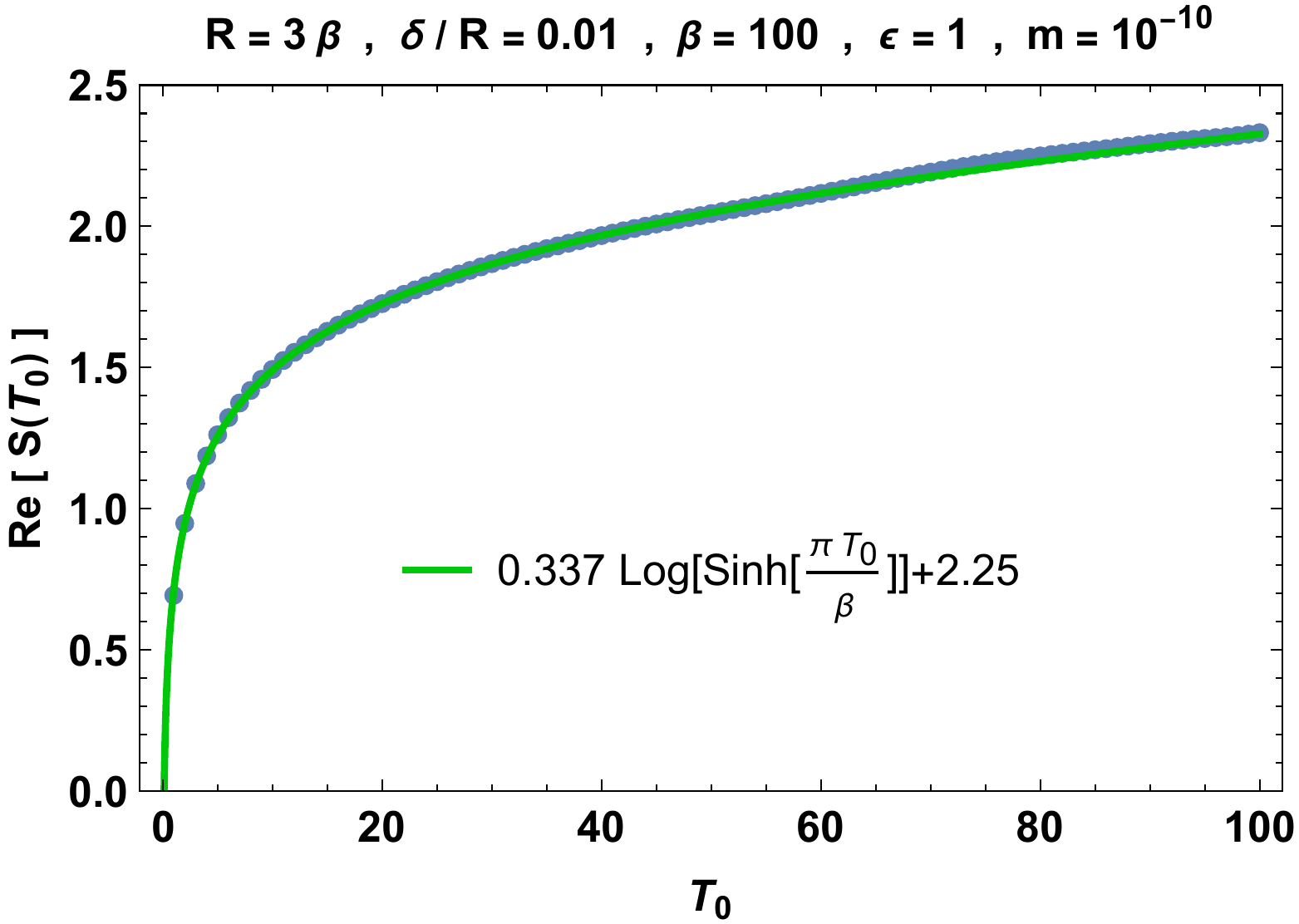}
  \caption{Timelike EE of a connected region for the free Dirac fermion theory. 
The left and the middle panels correspond to the infinite lattice $\beta=\infty$ case. In this case, our numerical results are matching the expected behavior except for the regime where the subregion size approaches $R$. The numerics in this regime are less stable compared to the scalar case so we cannot probe the $\delta\ll R$ limit stronger than the presented results. The middle panel is showing that for a fixed $R$ as we decrease $\delta$ we get closer to the expected result, namely the first term of \eqref{eq:timetorusHT}.
The right panel corresponds to the imaginary time compactified case $\beta$ and finite $R$. We show that the expected behavior as like \eqref{TEEtemp} is reproduced for the fermion case as well.} 
\label{fig:NumD2fer1}
\end{figure}
Again we adopt the correlator method to calculate timelike entanglement entropy in Dirac theory. Applying the same prescription introduced for the scalar field, we need the following correlation functions
\begin{align}
    \begin{split}
\mathbf{C}_{tt'}&\equiv
\langle\psi^\dagger(t)\psi(t')\rangle=    \mathrm{Tr}\left[e^{i(R+i\delta)\tilde{H}}\psi^\dagger(t)\psi(t')\right]
\\&
=\frac{1}{4\pi}\int dk
\tanh\left(\frac{(R+i\delta)\Omega_{ik}}{2}\right)
\begin{pmatrix}
1-\cos2\theta_k & \sin 2\theta_k
\\
\sin 2\theta_k & 1+\cos2\theta_k
\end{pmatrix}
e^{i(t-t')k}\,.
\end{split}
\end{align}
In order to calculate timelike EE we consider the regularized version of the Hamiltonian on a lattice where the time coordinate takes integer values, except we should replace $k\to\sin k$ in \eqref{eq:dd} and the integration runs over $-\pi<k<\pi$. With these in hand the timelike EE can be read from the spectrum of $\mathbf{C}$ denoted by $\{\nu\}$ as
\ba
\tee{A}=-\sum_i\left[\left(1-\nu_i\right)\log\left(1-\nu_i\right)+\nu_i\log\nu_i\right]\,.
\ea
The numerical results of the case of finite $R$ 
and $\beta\to\infty$ are presented in the middle and right panels of figure \ref{fig:NumD2fer1}. We have also considered the case of $R\to\infty$ 
and finite $\beta$. In this case we have applied the same procedure discussed in the scalar case for the imaginary time compactified case. To produce the numerical results presented in the left panel of figure \ref{fig:NumD2fer1}, we have used a regularization similar to 
\eqref{eq:TEEcorRBetaReg}. 

We have shown that with a careful adaptation of the correlator method to our second definition of timelike EE via Wick rotation of the coordinates, one can numerically find the same results which we have found via the analytic continuation of the the replica method in CFT.

\section{Holographic timelike entanglement entropy in \texorpdfstring{AdS$_3$/CFT$_2$}{AdS3/CFT2}}\label{sec:3}
In the previous section, we introduced the timelike entanglement entropy as a generalization of entanglement entropy. In this section, we will present a holographic calculation of the timelike EE in the framework of the AdS/CFT correspondence. This leads to a generalization of the holographic entanglement entropy formula \cite{Ryu:2006bv,Ryu:2006ef,Hubeny:2007xt}. 

Below we will study holographic timelike EE in the pure AdS$_3$, BTZ black holes and shock wave geometries. 

\begin{figure}[H]
  \centering
  \includegraphics[width=12cm]{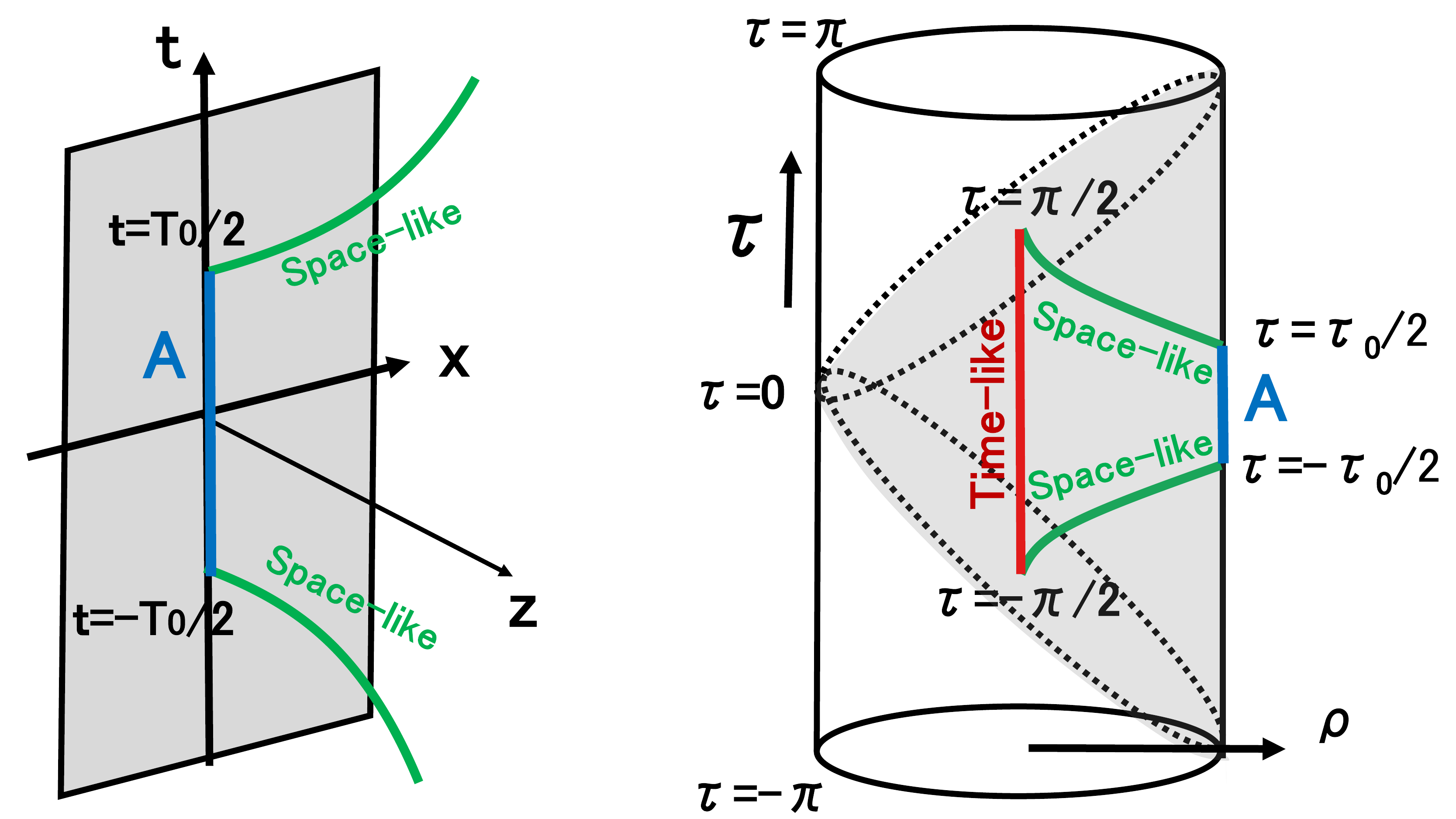}
  \caption{Sketches of calculation of HEE for a timelike subsystem $A$.} 
\label{fig:HEET}
\end{figure}

\subsection{Pure \texorpdfstring{AdS$_3$}{AdS3}}\label{subsec:pureads}
Consider a two dimensional CFT and choose the timelike subsystem to be an interval $A$ which connects two points: $t=-T$ and $t=T$ at $x=0$.
From the result (\ref{TEEarep}), we expect that this timelike EE is given by
\be
S_A=\frac{c}{3}\log\frac{2T}{\ep}+\frac{c}{6}\pi i. \label{HEET}
\ee
In the dual AdS$_3$, we can interpret this by generalizing the holographic entanglement entropy (HEE) as follows. In Poincare AdS$_3$ coordinates,
\be
ds^2=\frac{dz^2-dt^2+dx^2}{z^2},
\ee
the geodesic which computes the timelike EE is identified with
\ba
t=\s{z^2+T^2},
\ea
by Wick rotating the familiar semi-circle geodesic as 
depicted in the left panel of figure \ref{fig:HEET}.
Indeed the length of this spacelike geodesic reads
\be
S_A=\frac{1}{4G_N}\int^{\infty}_{\ep}dz \frac{2T}{z\s{z^2+T^2}}=\frac{c}{3}\log\frac{2T}{\ep},
\ee
which explains the real part of (\ref{HEET}).
The imaginary part of (\ref{HEET}) can be found by going beyond the Poincare patch to the global coordinate 
\be\label{globalc}
ds^2=-\cosh^2\rho dt+d\rho^2+\sinh^2\rho d\theta^2,
\ee
as sketched in the right panel of figure \ref{fig:HEET}. Since Poincare coordinates only cover the gray region, we need to connect the two endpoints at $\rho=0$ and $t=\pm\frac{\pi}{2}$ by a timelike geodesic. 
Since the length is $\pi$, this leads to a contribution to the HEE of $\frac{c}{6}\pi i$, which agrees with the imaginary part.

The same result can be found by the geodesic calculation directly in the global AdS$_3$. The geodesic distance between the two points $(\rho_\infty,T_0/2,\phi_0)$ and $(\rho_\infty,-T_0/2,\phi_0)$ is found to be
\begin{align*}
    D&=\cosh^{-1}\left(\cosh^2\rho_\infty\cos T_0-\sinh^2\rho_\infty\right) \\
    &\simeq\log\left(-e^{2\rho_\infty}\sin^2\frac{T_0}{2}\right)\\
    &=\pi i+\log\left(\frac{\sin^2\frac{T_0}{2}}{\epsilon^2}\right),
\end{align*}
where we have set $\epsilon=e^{-\rho_\infty}$. Thus the entanglement entropy is calculated as
\begin{align}\label{eq:global_TEE}
    \tee{A}=\frac{c}{3}\log\left(\frac{2\sin\frac{T_0}{2}}{\epsilon}\right)+\frac{c}{6}\pi i,
\end{align}
which reproduces the correct timelike EE in the CFT side \eqref{TEEsize}
at $R=2\pi$.

\subsection{Comments on extreme surfaces in \texorpdfstring{AdS$_3$}{AdS3}}\label{subsec:extremalsurface}

The surface which gives the holographic timelike entanglement entropy consists of the union of spacelike and timelike geodesic and it does not look like an extremal surface at first sight. To see this we consider the AdS$_2$ slice of global AdS$_3$:
\ba
ds^2=d\eta^2+\cosh^2\eta\left(-\cosh^2 r dt^2+dr^2\right),
\ea
where the new coordinate $(\eta,r)$ is related to $(\rho,\theta)$ in (\ref{globalc}) via
\ba
\cosh\eta\cosh r=\cosh\rho,\ \ \ \sinh\eta=\sinh\rho\cos\phi.
\ea
\begin{figure}[H]
    \centering
    \includegraphics[width=.75\textwidth]{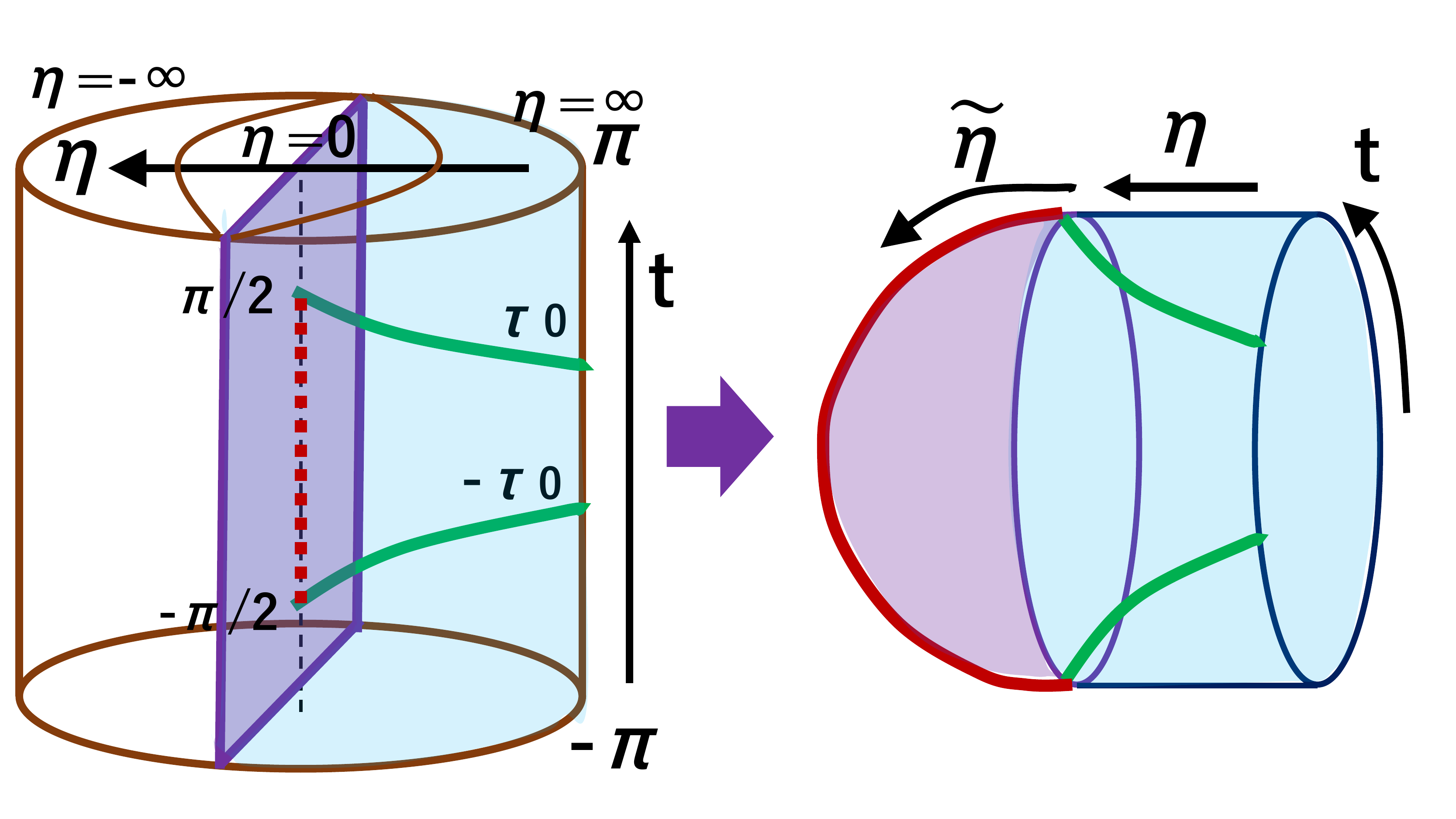}
    \caption{The holographic timelike entanglement entropy in global AdS$_3$ (left) and that in a ``Hartle-Hawking'' like geometry (right). The length of the red curve and that of the green curve give the imaginal and real part of the holographic timelike entanglement entropy, respectively. }
    \label{fig:global_HH}
\end{figure}

As depicted in figure \ref{fig:global_HH}, the boundary at $\eta=\infty$ covers a half of the original boundary cylinder of the global AdS$_3$, where the timelike interval $A$ is situated. $\eta=-\infty$ corresponds the other boundary. Now, starting from $\eta=\infty$,  at $\eta=0$ we continue to the 'Wick-rotated' geometry defined by $\eta=i\ti{\eta}$, which is given by 
\ba\label{capmetric}
ds^2=-d\ti{\eta}^2+\cos^2\ti{\eta}\left(-\cosh^2 r dt^2+dr^2\right),\quad (0<\ti{\eta}<\pi/2).
\ea
This geometry has AdS$_2$ as its boundary at $\tilde{\eta}=0$. We expect that the dual extremal surface $\Gamma_A$ is situated on a hyperplane $r=0$, whose metric reads 
\ba
ds^2=-(d\ti{\eta}^2+\cos^2\ti{\eta}dt^2),\ \ \ (0<\ti{\eta}<\pi/2).
\ea
If we compactify $t$ such that $t\sim t+2\pi$, then this metric is identical to that of minus times the sphere metric. The minus sign means that the length is given by $i$ times the length in the semisphere as depicted in the right of figure \ref{fig:global_HH}.
This glued geometry helps us to find the correct geodesics which compute the holographic timelike EE.
This is similar to the Hartle-Hawking prescription which is employed to calculate the holographic pseudo entropy in dS/CFT as we will see in section \ref{sec:dscftpe}.

Let us check that the timelike entanglement entropy is indeed derived from an ``extremization'' condition analogous to the ordinary geodesic condition. Since the geodesic length we are considering is complex-valued, this condition amounts to imposing the extremality to both the real and imaginary parts, which come from spacelike and timelike surfaces respectively. A geodesic emanating from the boundary at $t=T_0/2$ will attach to some point on the surface $\eta=0$. We fix the $t$-coordinate of the end point of the geodesic by $t_1$, and that of the other geodesic emanating from $t=-T_0/2$ by $t_2$. 

First we consider the imaginary part. The geodesic connecting $t_1$ and $t_2$ in the Euclidean region is determined by varying the following area functional
\begin{align}\label{TEEimarea}
    i\int d\tilde{\eta} \sqrt{1+\cos^2\tilde{\eta}\,\left(\frac{dt(\tilde{\eta})}{d\tilde{\eta}}\right)^2}.
\end{align}
As described above, the overall factor $i$ comes from the minus sign in \eqref{capmetric}. From the Euler-Lagrange equation, we have
\begin{align}\label{ELE}
    t'(\tilde{\eta})=\frac{\cos \tilde{\eta}_*}{\cos\tilde{\eta}\sqrt{\cos^2\tilde{\eta}-\cos^2\tilde{\eta}_*}},
\end{align}
where $\tilde{\eta}_*$ satisfies $d\tilde{\eta}/dt=0$. The condition that the geodesic ends on the $\tilde{\eta}=0$ surface at $t=t_1,t_2$ leads to 
\begin{align}
    t(0)=t_2,\quad t(\tilde{\eta}_*)=\frac{t_1+t_2}{2}.
\end{align}
By integrating \eqref{ELE} with these boundary conditions, we obtain 
\begin{align}\label{Euccond}
    t_1-t_2=\pi.
\end{align}
The fact that this condition does not depend on $\tilde{\eta}_*$ means that the geodesic can exist only when the endpoints $t_1,t_2$ satisfy this condition, i.e. the two points on $\mathbb{S}^2$ should be at the antipodal points to each other. The length of this geodesic is 
\begin{align}
    2i\int_0^{\tilde{\eta}_*}d \tilde{\eta}\frac{\cos\tilde{\eta}}{\sqrt{\cos^2\tilde{\eta}-\cos^2\tilde{\eta}_*}}=i\pi.
\end{align}
Therefore the imaginary part always takes the value of $\pi$, which is identical to the length of geodesics that lie along the boundary $\mathbb{S}^2$. 

Next we consider the real part. We can straightforwardly evaluate the sum of the two spacelike geodesics with end points $t_1,t_2$ as 
\begin{align*}
    \cosh^{-1}&\left[\cosh\eta_\infty\cos\left(\frac{T_0}{2}-t_1\right)\right]+\cosh^{-1}\left[\cosh\eta_\infty\cos\left(\frac{T_0}{2}+t_2\right)\right] \\
    &\simeq\log\left[e^{2\eta_\infty}\cos\left(\frac{T_0}{2}-t_1\right)\cos\left(\frac{T_0}{2}+t_2\right)\right]\\
    &=\log\left[-\frac{1}{2}e^{2\eta_\infty}\left(\cos T_0+\cos 2t_1\right)\right],
\end{align*}
where $\eta_\infty$ is a large cutoff. In the second equality, we used the condition \eqref{Euccond}. The condition that the derivative with $t_1$ vanishes leads to $\sin 2t_1=0$. For $t_1$ and $t_2$ to be in the range $-\pi<t<\pi$, they should take 
\begin{align}
    t_1=\frac{\pi}{2},\quad t_2=-\frac{\pi}{2}.
\end{align}
Under this condition the sum of the geodesic lengths is
\begin{align}
    \log\left[e^{2\eta_\infty}\sin^2\frac{T_0}{2}\right].
\end{align}

Finally, we evaluate the timelike entanglement entropy
\begin{align}
\begin{aligned}
    \tee{A}&=\frac{1}{4G_N}\left(2\log\left[ e^{\eta_\infty}\sin\frac{T_0}{2}\right]+i\pi\right)\\
    &=\frac{c_{\text{AdS}}}{3}\log\left[\frac{2\sin\frac{T_0}{2}}{\epsilon}\right]+\frac{i\pi c_{\text{AdS}}}{6},\qquad (\epsilon\equiv 2e^{-\eta_\infty}),
\end{aligned}
\end{align}
which reproduces \eqref{eq:global_TEE}.

In this way, the extremality condition imposed in the geometry obtained after the Wick rotation as the originally spacelike geodesic turned into the timelike one, requires that the length of the timelike geodesic is $\pi$ and fixes the end-points of the spacelike geodesics. This uniquely determined the total shape of $\Gamma_A$, which is a union of spacelike and timelike geodesics.  

We can summarize this procedure as follows:

(i) First we construct candidates of $\Gamma_A$ from a union of spacelike and timelike geodesics such that $\de\Gamma_A=\de A$. 

(ii) Then we take variations of the joining points of spacelike and timelike geodesics and require that they are all stationary. Here we consider 
the Wick rotated geometry for the timelike geodesics.

\subsection{BTZ black hole}
The metric for the BTZ black hole is given by
\ba
ds^2=-\left(r^2-r_{+}^2\right)dt^2+\frac{dr^2}{\left(r^2-r_{+}^2\right)}+r^2d\phi^2
\ea
where $r_+$ is the horizon radius and we have set the AdS length $R_{\text{AdS}}=1$ in the remainer of this section. 
The inverse temperature can be written in terms of the horizon radius as $\beta=\frac{2\pi}{r_+}$.
The BTZ black hole and global AdS$_3$ are related by the following coordinate transformation
\be\label{eq:global_to_BTZ}
\begin{split}
 \chi_0&=\cosh(\rho)\sin(\tau)=\frac{\sqrt{r^2-r_+^2}}{r_+}\sinh(r_+t)\\
 \chi_1&=\cosh(\rho)\cos(\tau)=\frac{r}{r_+}\cosh(r_+\phi)\\
 \chi_2&=\sinh(\rho)\cos(\theta)=\frac{r}{r_+}\sinh(r_+\phi)\\
 \chi_3&=\sinh(\rho)\sin(\theta)=\frac{\sqrt{r^2-r_+^2}}{r_+}\cosh(r_+t)
\end{split}
\ee

\begin{figure}[t]
    \centering
    \includegraphics[width=.45\textwidth,page=1]{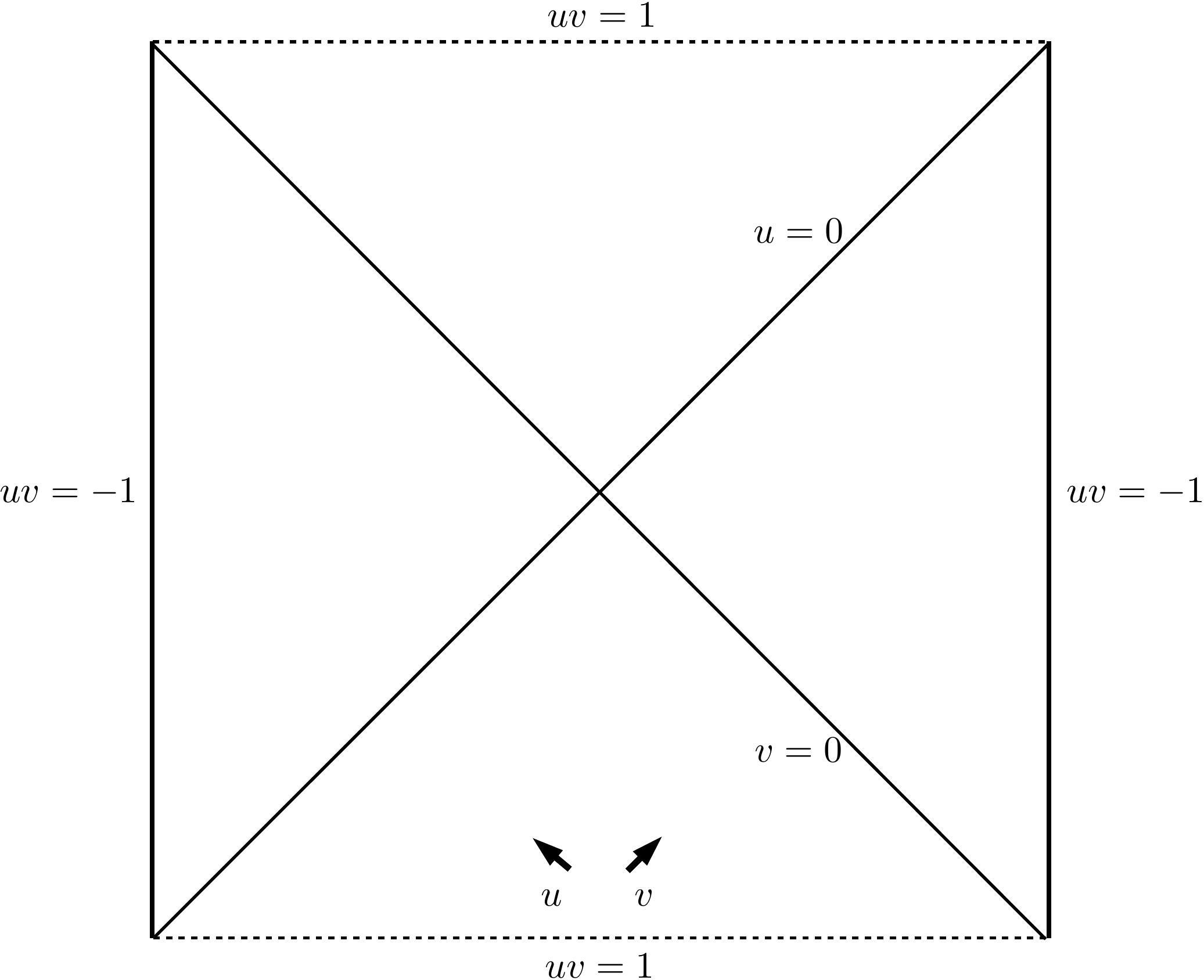}
    \caption{The BTZ black hole in Kruskal coordinates. The left and right boundary are determined by $uv=-1$ while the past and future singularity $uv=1$. the two horizons are at $u=0$ and $v=0$ respectively.}
    \label{fig:kruskal}
\end{figure}

Our first task will be to reproduce \eqref{TEEtemp} by considering the lengths of paths of geodesics in the bulk spacetime and applying our optimization procedure. To aid us we find it useful to utilize Kruskal coordinates defined by the transformation
\be\label{eq:global_to_KC}
\begin{split}
 \chi_0&=\frac{u+v}{1+uv}\\
 \chi_1&=\frac{1-uv}{1+uv}\cosh(r_+\phi)\\
 \chi_2&=\frac{1-uv}{1+uv}\sinh(r_+\phi)\\
 \chi_3&=\frac{v-u}{1+uv}.
\end{split}
\ee
The corresponding metric is given by
\be
ds^2=-4\frac{dudv}{(1+uv)^2}+\frac{(1-uv)^2}{(1+uv)^2}r_+^2d\phi^2
\ee
and the geometry is shown in figure \ref{fig:kruskal}.

For the calculation of the timelike entanglement entropy we will restrict to the constant $\phi=0,\pi$ slice. With this choice we can solve for the geodesics which are given by the equation
\be
u''(1+uv)+2u'(u-u'v)=0
\ee
and obtain the solution
\be\label{eq:kruskal_geo}
u(v)=\frac{c_1+c_2v}{1+c_1v}, \quad u'(v)=\frac{c_2-c_1^2}{(1+c_1v)^2}.
\ee
The causality of these geodesics is determined by the relative values of $c_1,c_2$
\be
\begin{split}
&c_1^2\geq c_2 \quad \text{spacelike}\\
&c_1^2\leq c_2 \quad \text{timelike}.\\
\end{split}
\ee
The length of geodesics can generally be directly evaluated by integration, however it is often easier to make use of the embedding distance $\Theta$ between points $X_1$ and $X_2$ in the embedding space $\mathbb{R}^{2,2}$. By making use of the coordinate transformation \eqref{eq:global_to_KC} this is given by
\be
\Theta(X_1,X_2)=g_{AB}X_1^AX_2^B=\frac{-2(v_1u_2+u_1v_2)-(1-v_1u_1)(1-v_2u_2)\cosh(r_+(\phi_1-\phi_2))}{(1+v_1u_1)(1+v_2u_2)}
\ee
In terms of the embedding distance the geodesic distance is then given by
\be\label{eq:geod_kc}
D(X_1,X_2)=\begin{cases}
        \arccosh(-\theta(X_1,X_2)), & \text{spacelike}\\
        i\arccos(-\theta(X_1,X_2)), & \text{timelike}.
    \end{cases}
\ee
To regulate the length of geodesics we introduce a UV cutoff $\omega$ such that $u_bv_b=-1+\omega$ for any point $(u_b,v_b)$ on the boundary. Examining $\chi_0+\chi_3$ using \eqref{eq:global_to_KC} and \eqref{eq:global_to_BTZ} we have
\be
\begin{split}
\frac{2v}{1+uv}&=\frac{\sqrt{r^2-r_+^2}}{r_+}\left(\sinh(r_+t)+\cosh(r_+t)\right)\\
\frac{2v_b}{\omega}&=\frac{r_\infty}{r_+}e^{r_+t}\\
\frac{2v_b}{\omega}&=\frac{1}{r_+\delta}e^{r_+t}.
\end{split}
\ee
which allows us to relate $\omega$ to the cutoff in BTZ coordinates $\delta$ and well as the boundary coordinates. As such we will work with the conventions
\be\label{eq:K_conv}
\frac{2}{\omega}=\frac{1}{r_+\delta},\quad v_{r,l}=\pm e^{r+t}(1-\frac{\omega}{2})\approx \pm e^{r+t},\quad u_{r,l}=\mp e^{-r_+t}(1-\frac{\omega}{2}) \approx \mp e^{-r+t}
\ee
where the upper (lower) signs are for the right (left) boundary respectively. In what follows we will always work at leading order $\frac{1}{\omega}$ dropping any subleading terms. 

We proceed by considering several examples. We shall start by determining the timelike entanglement entropy of a single region $A$. Its result can then be used to consider two regions $A$ and $B$ and define the timelike mutual information $\itee{AB}$ and thermofield mutual information $\jtee{AB}$ (between regions on different boundaries). Lastly we shall consider the BTZ black hole with a shock wave.

\subsubsection{Single interval}
We start by determining the timelike entanglement entropy of a single region $A=[T_1,T_2]$ on the right boundary where $T_1<T_2$. In Kruskal coordinates given by $(v_1,u_1) =(a_1,-\frac{1}{a_1})$ and $(v_2,u_2)=(a_2,-\frac{1}{a_2})$ where making use of \eqref{eq:K_conv}
\be\label{eq:krus_to_BH}
a_i=e^{r_+T_i}.
\ee
\begin{figure}[H]
    \centering
    \includegraphics[width=.45\textwidth,page=20]{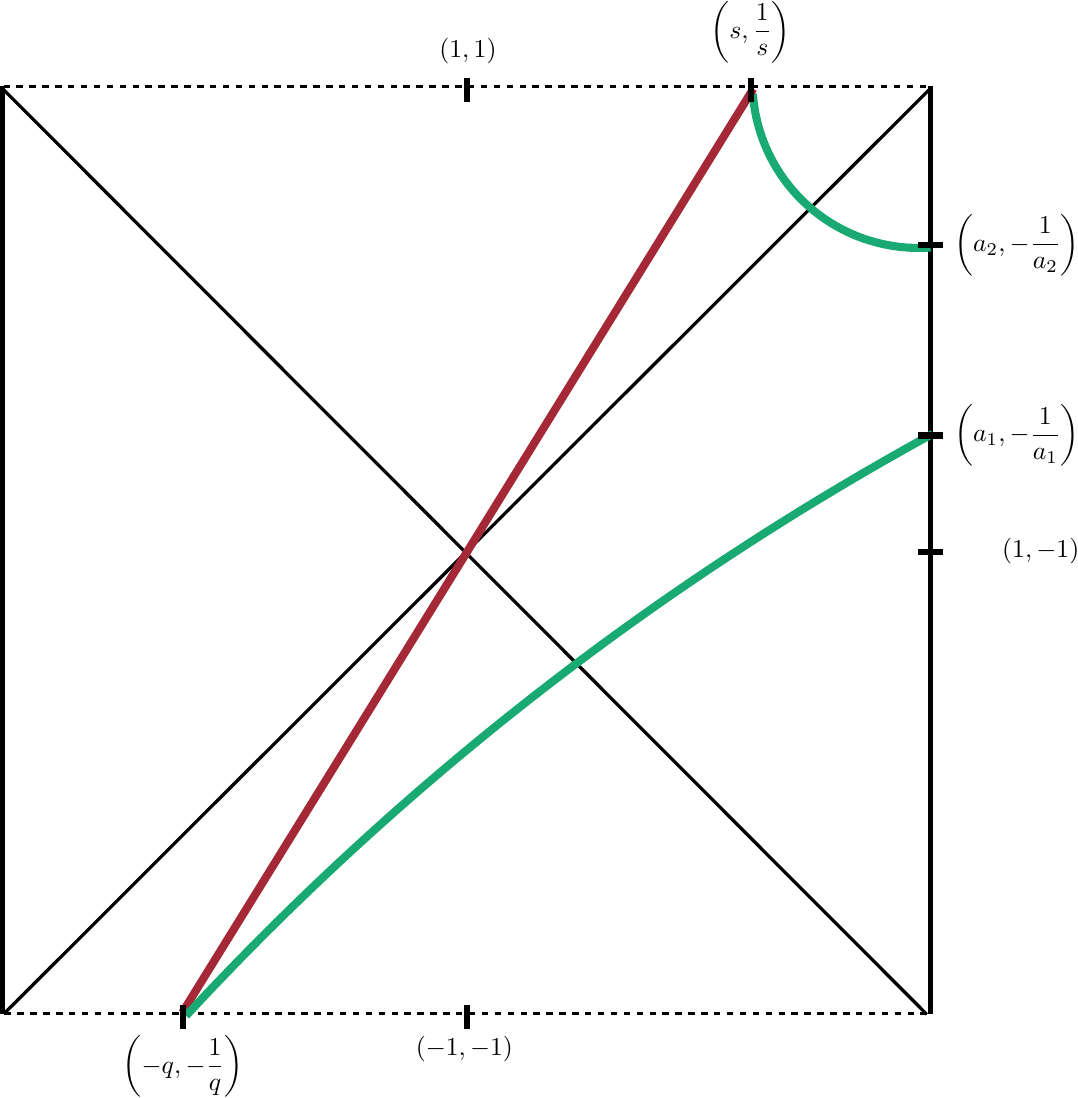}
    \caption{The system of geodesics used to determine the timelike entanglement entropy $\tee{A}$. In Kruskal coordinates $A$ is determined by the values $a_1,a_2$ which label the endpoints of $A$ on the right boundary. $s,q$ determines the intersection points of the geodesics on the singularities and are initially unconstrained.}
    \label{fig:kc_setup}
\end{figure}
Consider the collection of three geodesics shown in figure \ref{fig:kc_setup}. This consists of two spacelike geodesics one from the boundary at $v=a_2$ to the future singularity at $v=s$ and one from the boundary at $v=a_1$  to the past singularity at $v=q$. We also include a timelike geodesic between the past singularity at $v=q$ and future singularity at $v=s$. This forms a family of candidates $\Gamma_A$ where the values of $s,q$ are to be determined.

Examining the geodesic equation \eqref{eq:kruskal_geo} we require a timelike geodesic between the points on the past and future singularity. This is only possible if we take $s=q$ and provides a family of timelike geodesics $g_t$
\be
g_{t}:\quad u(v)=\frac{c_1+\frac{1}{s^2}v}{1+c_1v}, \quad c_1\in\left(-\frac{1}{s},\frac{1}{s}\right)
\ee
It can be confirmed using the geodesic distance \eqref{eq:geod_kc} that all such curves of this form are of maximal length
\be
l_{t}=i\pi.
\ee
For the choice $c_1=0$ the geodesic simplifies to
\be
u(v)=\frac{v}{s^2}.
\ee
The two spacelike geodesics $g_{su}$ and $g_{sl}$ and their lengths are found to be
\be
g_{su}:\quad u(v)=\frac{2va_2-(a_2^2+s^2)}{2a_2s^2-(a_2^2+s^2)v}, \quad a_2\geq s
\ee
\be
l_{su}=\log\left(\frac{1}{\omega}\frac{(a_2^2-s^2)}{a_2s}\right)
\ee
\be
g_{sl}:\quad u(v)=\frac{2va_1-(a_1^2+q^2)}{2a_1q^2-(a_1^2+q^2)v}, \quad q\geq a_1
\ee
\be
l_{sl}=\log\left(\frac{1}{\omega}\frac{(q^2-a_1^2)}{a_1q}\right).
\ee
Making use of the requirement $q=s$ the two spacelike geodesics have total length
\be
l_s=l_{su}+l_{sl}=\log\left(\frac{1}{\omega^2}\frac{(s^2-a_1^2)(a_2^2-s^2)}{s^2a_1a_2}\right).
\ee

\begin{figure}[H]
    \centering
    \begin{tabular}{cc}
    \includegraphics[width=.45\textwidth,page=3]{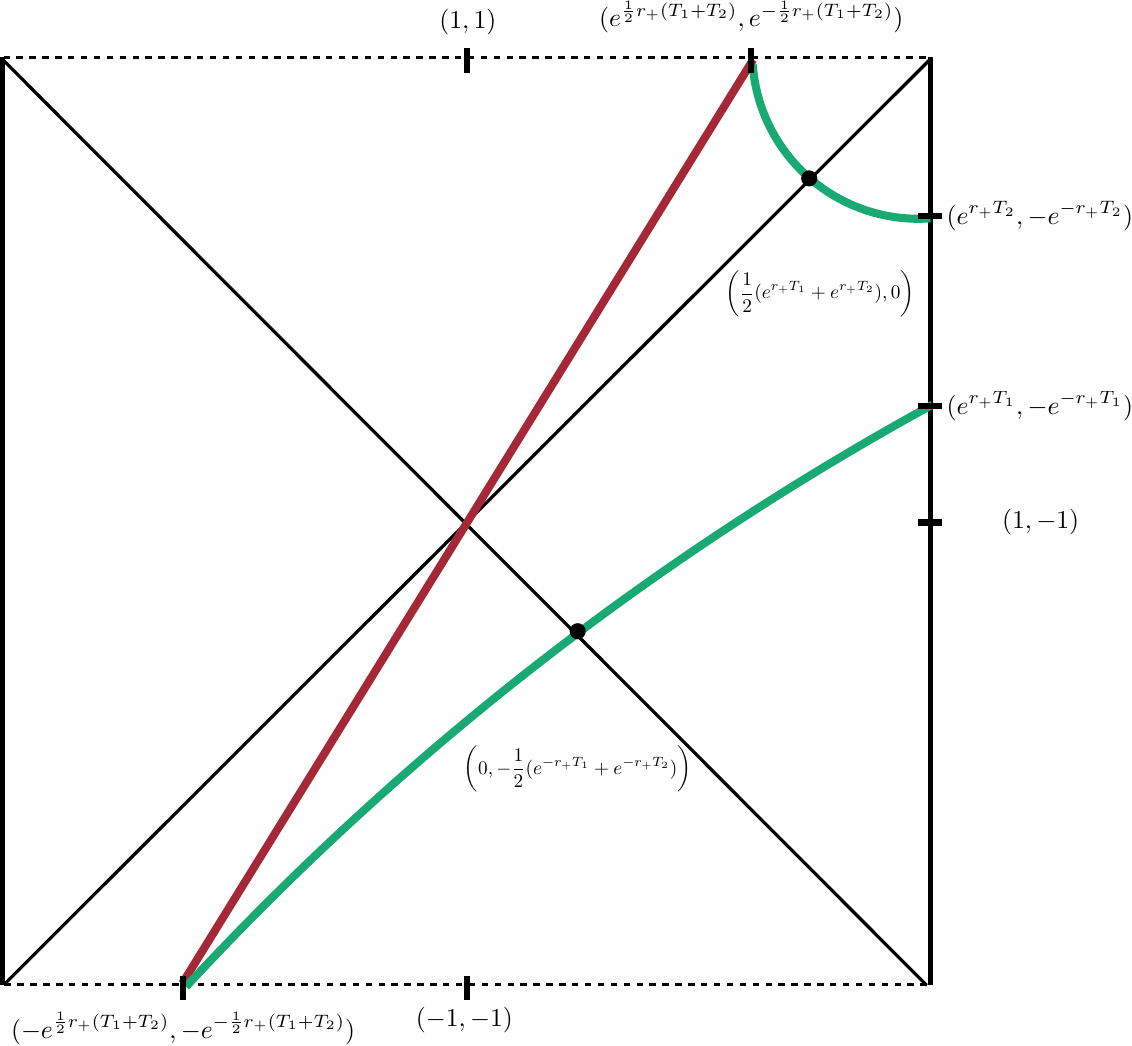}&
    \includegraphics[width=.45\textwidth,page=2]{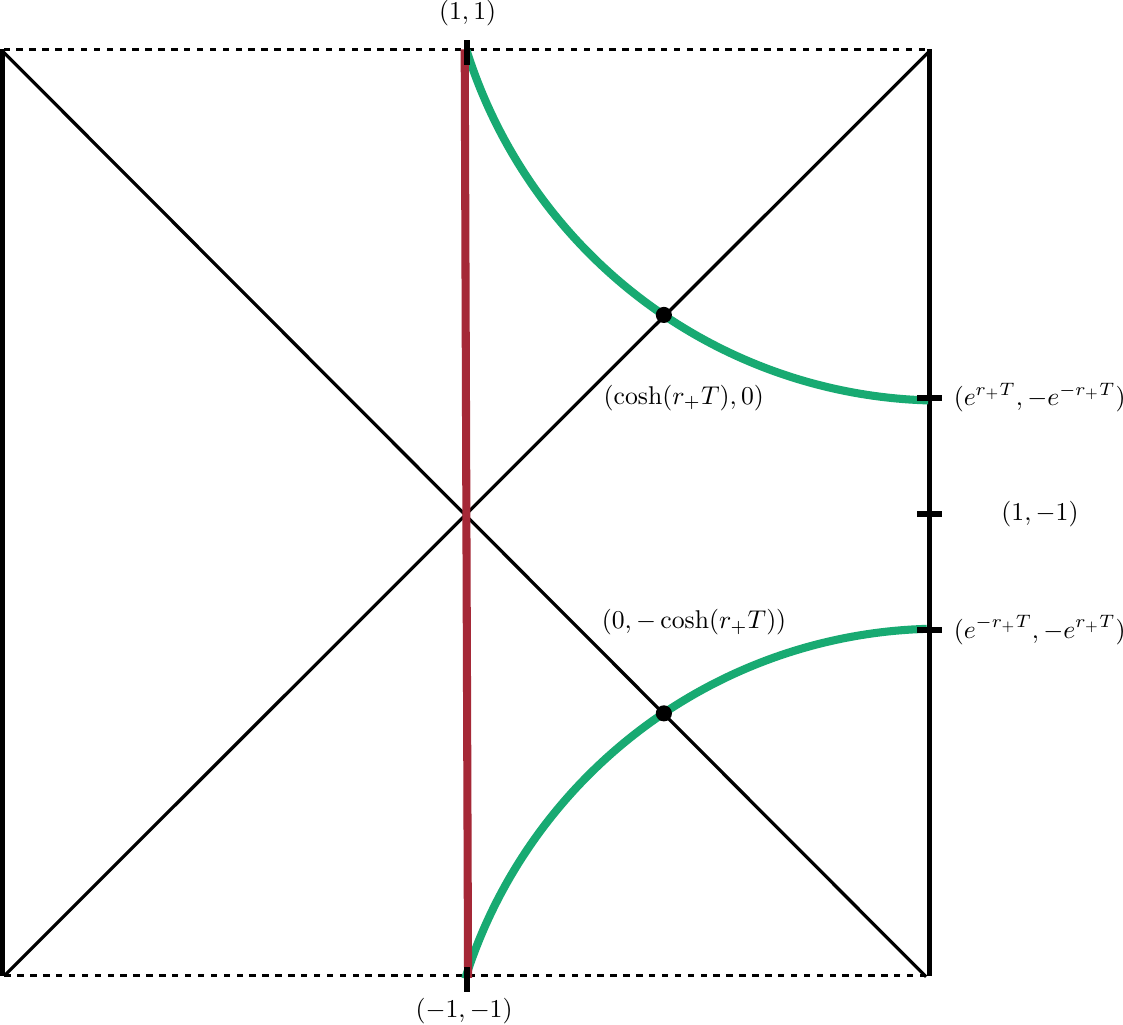}
    \end{tabular}
    \caption{Examples of stationary paths $\Gamma_A^*$ whose area calculates the the timelike entanglement entropy $\tee{A}$ in the BTZ geometry. For regions symmetric about $t=0$ the timelike curves extends from the center of the past singularity to the center of the future singularity.}
    \label{fig:KC_single}
\end{figure}
\noindent Only the parameter $s$ remains unfixed and labels for us many possible paths $\Gamma_A$ all of varying lengths. Among all these candidate paths we should choose the one which is stationary\footnote{As constructed each candidate path $\Gamma_A$ is the union of geodesics meaning each spacelike curve is minimal and each timelike curve maximal. It is important that the optimization over such paths is an extremization i.e. we look for paths which are stationary. In this example the stationary path $\Gamma_A^*$ is in fact a maximum \emph{among the possible paths $\Gamma_A$}.} which requires $\partial_s l_s=0$ and thus $s^2=a_1a_2$. The total length of the geodesics which make up the optimized path $\Gamma_A^*$ is
\be
l=\log\left(\frac{1}{\omega^2}\frac{(a_2-a_1)^2}{a_1a_2}\right)+i\pi.\\
\ee
Using \eqref{eq:krus_to_BH} the timelike entanglement entropy is found to be
\be\label{eq:BH_SI_tee}
\tee{A}=\frac{c}{3}\log\left(\frac{\beta}{\pi\delta}\sinh\left(\frac{\pi}{\beta}(T_2-T_1)\right)\right)+\frac{c}{6}i\pi
\ee
which is the same as \eqref{TEEtemp}. Examples of geodesics which calculate the timelike entanglement entropy are shown in figure \ref{fig:KC_single}.

\subsubsection{Timelike mutual information}

\begin{figure}[H]
    \centering
    \includegraphics[width=.45\textwidth,page=4]{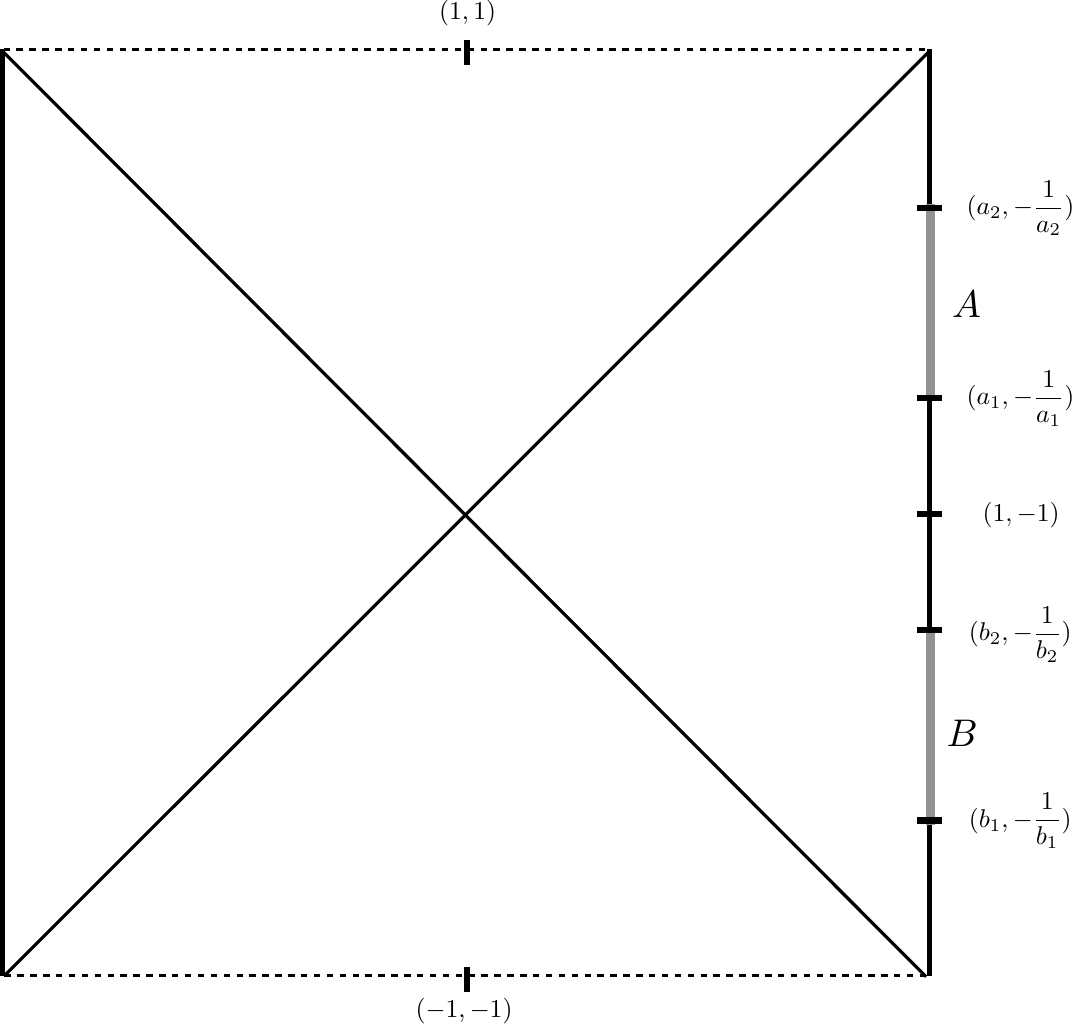}
    \caption{Two time like regions $A=[T_{A1},T_{A2}]$ and $B=[T_{B1},T_{B2}]$}
    \label{fig:KK_MI_setup}
\end{figure}

We now consider two timelike intervals $A=[T_{A1},T_{A2}]$ and $B=[T_{B1},T_{B2}]$ on the right boundary of the BTZ geometry. We assume the interval are separated with $A$ later than $B$. In Kruskal coordinates this can be written as
\be\label{eq:g_k_conv}
\begin{split}
&A: \quad v_{A1}=a_1, \quad v_{A2}=a_2,\quad a_i=e^{r_+T_{Ai}}\\
&B: \quad v_{B1}=b_1, \quad v_{B2}=b_2, \quad \quad b_i=e^{r_+T_{Bi}}
\end{split}
\ee
where because these points are on the boundary we have $v_ru_r=-1$. The set up is shown in figure \ref{fig:KK_MI_setup}.

\begin{figure}[H]
    \centering
    \begin{tabular}{cc}
    \includegraphics[width=.45\textwidth,page=6]{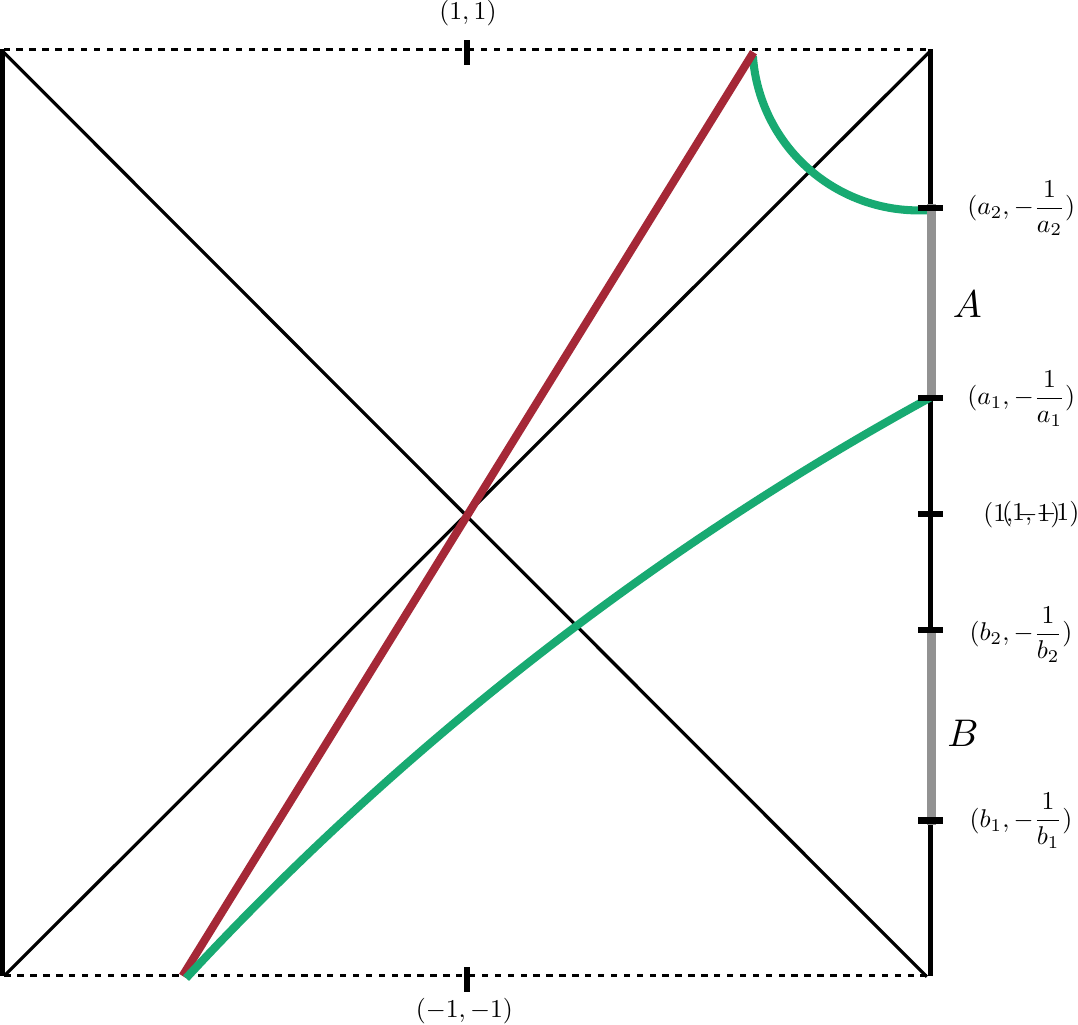}&
    \includegraphics[width=.45\textwidth,page=7]{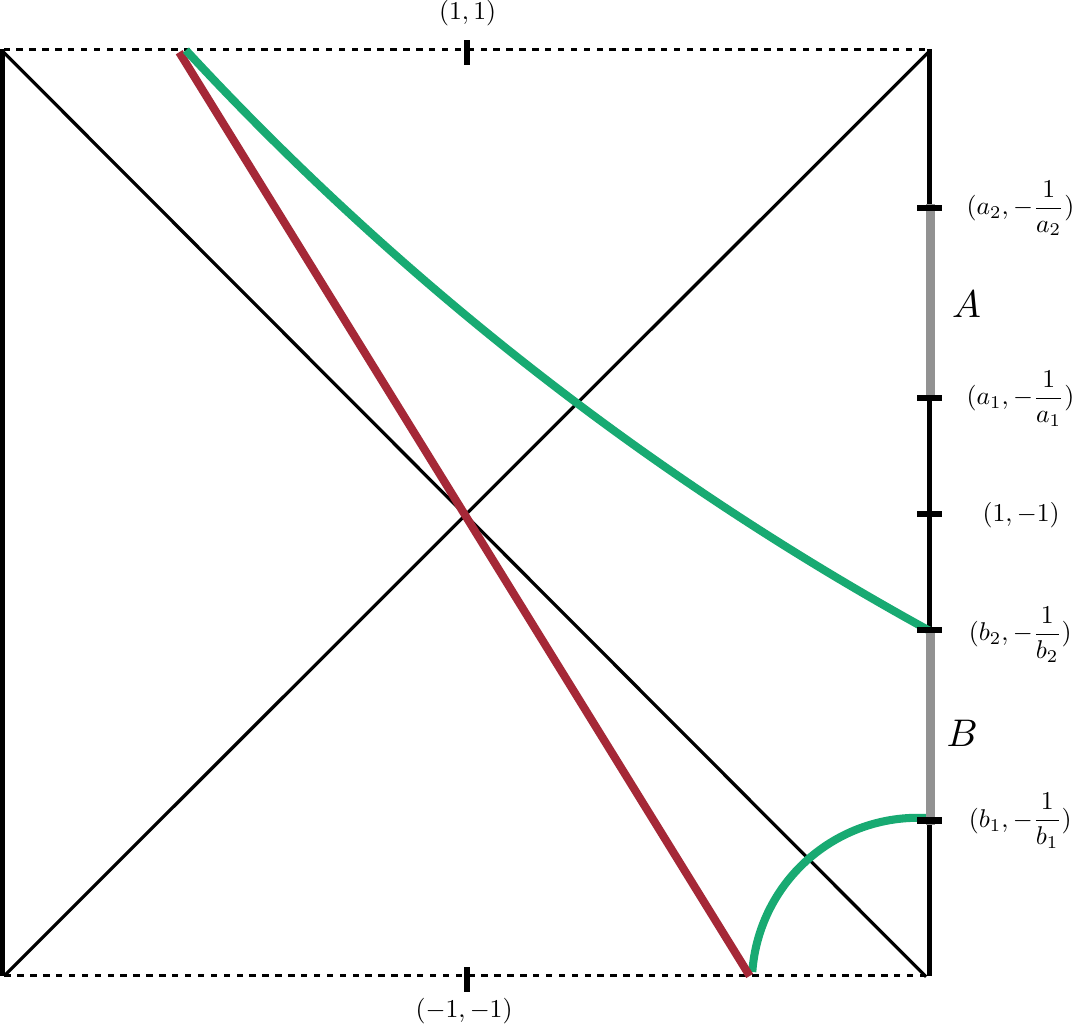}
    \end{tabular}
    \caption{Paths $\Gamma_{A}$ (left) and $\Gamma_B$ (right) whose area gives the timelike entanglement entropy $\tee{A}$ and $\tee{B}$.}
    \label{fig:KK_MI_SA_SB}
\end{figure}
\noindent Using our results \eqref{eq:BH_SI_tee} the single party timelike entanglement entropies $\tee{A}$ and $\tee{B}$ are given by
\be
\tee{A_i}=\frac{c}{3}\log\left(\frac{\beta}{\pi\delta}\sinh\left(\frac{\pi}{\beta}(T_{A_i2}-T_{A_i1})\right)\right)+\frac{c}{6}i\pi.
\ee
with the corresponding geodesics shown in figure \ref{fig:KK_MI_SA_SB}.

For the two party timelike entanglement entropy $\tee{AB}$ there are two configurations of geodesics we can consider. The first corresponds to the disconnect phase in which $\tee{AB}=\tee{A}+\tee{B}$ so that
\be
\tee{AB}=\frac{c}{3}\log\left(\frac{\beta^2}{\pi^2\delta^2}\sinh\left(\frac{\pi}{\beta}(T_{A2}-T_{A1})\right)\sinh\left(\frac{\pi}{\beta}(T_{B2}-T_{B1})\right)\right)+\frac{c}{3}i\pi.
\ee
The second corresponds to the connect phase where the geodesics correspond to the two intervals of length $T_{A2}-T_{B1}$ and $T_{A1}-T_{B2}$ so that
\be
\tee{AB}=\frac{c}{3}\log\left(\frac{\beta^2}{\pi^2\delta^2}\sinh\left(\frac{\pi}{\beta}(T_{A2}-T_{B1})\right)\sinh\left(\frac{\pi}{\beta}(T_{A1}-T_{B2})\right)\right)+\frac{c}{3}i\pi.
\ee
In this example we see that the two possible phases for $\tee{AB}$ are comprised of stationary paths but have the same imaginary value. To distinguish between these configurations, we choose, in analogy with the case of the usual entanglement entropy, to take the one with the smallest real value to be the optimal configuration\footnote{This is a definition and it is possible that other prescriptions should be considered. We comment briefly on this in the discussion section \ref{sec:6}.}.
\begin{figure}[H]
    \centering
    \begin{tabular}{cc}
    \includegraphics[width=.45\textwidth,page=8]{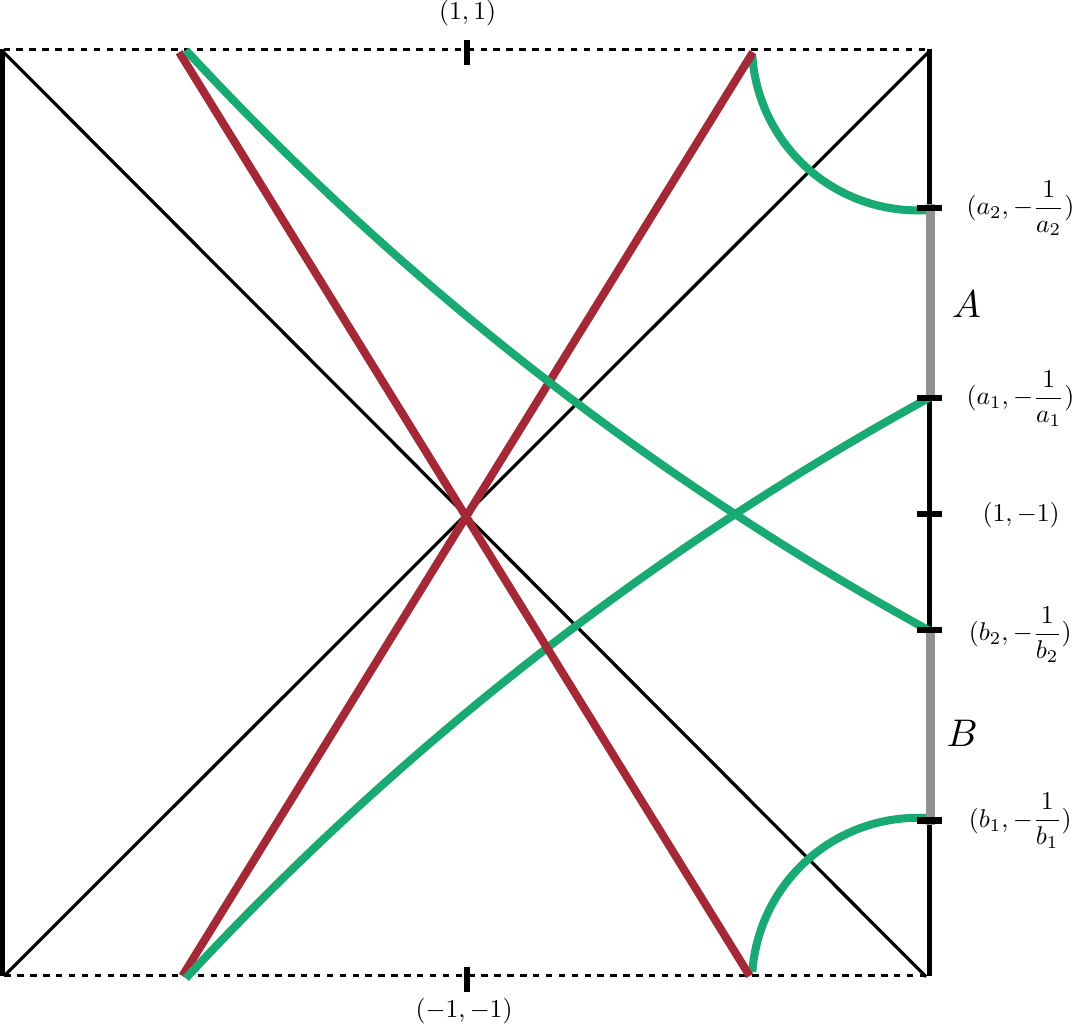}&
    \includegraphics[width=.45\textwidth,page=5]{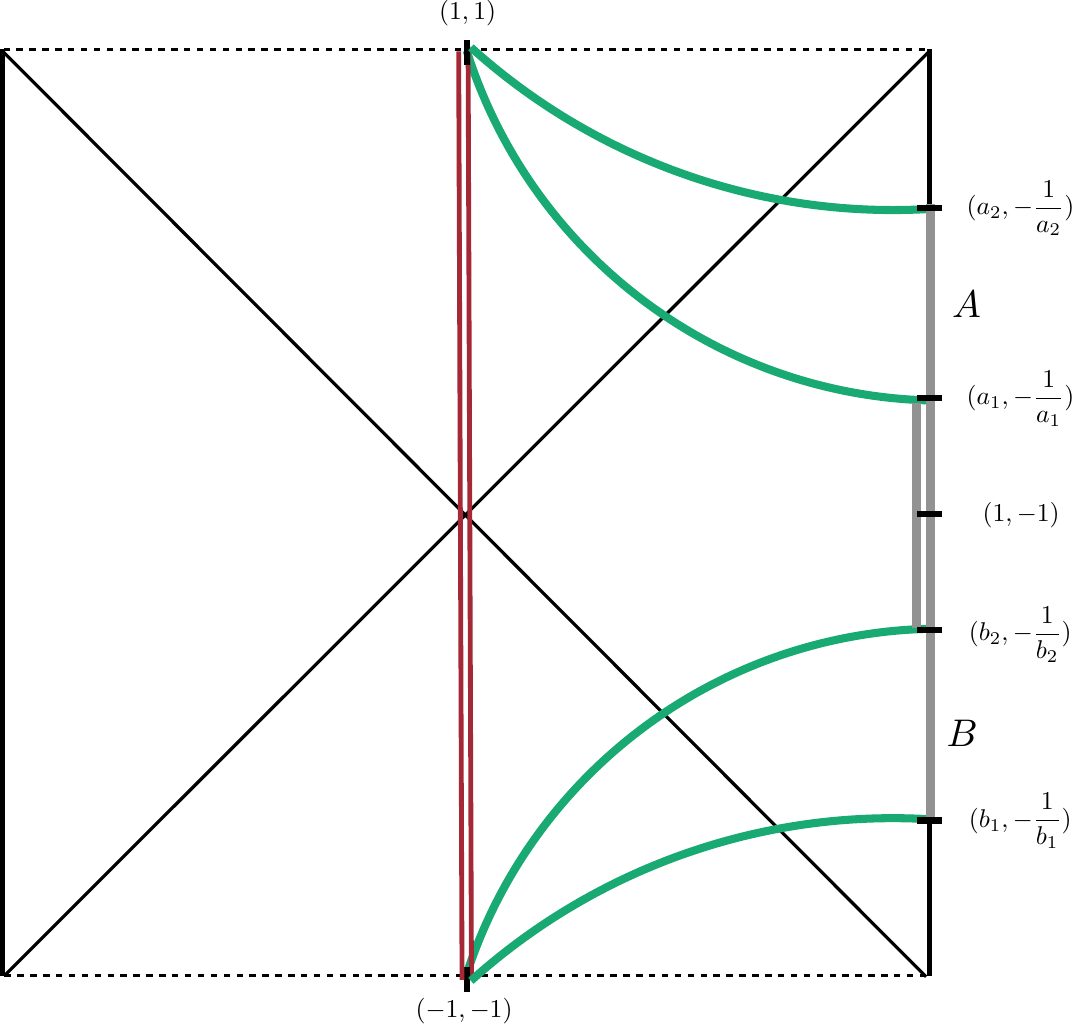}
    \end{tabular}
    \caption{Paths $\Gamma_{AB}$ whose area gives the timelike entanglement entropy $\tee{AB}$. There are two possible phases in analogy with the connected/disconnected phase transition for the entanglement entropy $S_{AB}$. L: Disconnected phase R: Connected phase}
    \label{fig:KK_MI_SAB}
\end{figure}
\noindent Both sets of geodesics are shown in figure \ref{fig:KK_MI_SAB}. We can now consider the timelike mutual information
\be
\itee{AB}=\tee{A}+\tee{B}-\tee{AB}
\ee
where we have taken the spacelike entanglement entropy and replaced each term with the respective timelike entanglement entropy. Since we choose the configuration of $\tee{AB}$ for which the real part is minimal the timelike mutual information should be maximized with respect to the two possible configurations. Combining our results we find
\be
\itee{AB}=\max\begin{cases}
     \frac{c}{3}\log\left(\frac{\sinh\left(\frac{\pi}{\beta}(T_{A2}-T_{A1})\right)\sinh\left(\frac{\pi}{\beta}(T_{B1}-T_{B2})\right)}{\sinh\left(\frac{\pi}{\beta}(T_{A2}-T_{B1})\right)\sinh\left(\frac{\pi}{\beta}(T_{A1}-T_{B2})\right)}\right)  \\
     0
\end{cases}
\ee
Note that here the imaginary part completely cancels so that the maximization is over an entirely real quantity. The timelike mutual information should be compared with the mutual information of two spacelike regions \cite{2010PhRvD..82l6010H, 2017LNP...931.....R} $A=[\theta_{A1},\theta_{A2}]$ and $B=[\theta_{B1},\theta_{B2}]$ which we take to be arranged such that there is no thermal contribution coming from the area of the horizon
\be
I_{AB}=\max\begin{cases}
     \frac{c}{3}\log\left(\frac{\sinh\left(\frac{\pi}{\beta}(\theta_{A2}-\theta_{A1})\right)\sinh\left(\frac{\pi}{\beta}(\theta_{B1}-\theta_{B2})\right)}{\sinh\left(\frac{\pi}{\beta}(\theta_{A2}-\theta_{B1})\right)\sinh\left(\frac{\pi}{\beta}(\theta_{A1}-\theta_{B2})\right)}\right)  \\
     0.
     \end{cases}
     \ee

\subsubsection{Timelike thermofield mutual information}
In the BTZ geometry it is also possible to consider entanglement entropies where the regions are placed on opposite boundaries. For example \cite{2013JHEP...07..081M} considered the mutual information of two spacelike regions $A=[\theta_{A1},\theta_{A2}]$ and $B=[\theta_{B1},\theta_{B2}]$ in such a set up. This quantity was defined to be the thermofield mutual information $J_{AB}$ and found by calculating the lengths of minimal geodesics\footnote{The authors of \cite{2013JHEP...07..081M} also provided derivations of the thermofield mutual information for the CFT of a 2d massless Dirac fermion, but were unable to extend the result to 2d holographic CFTs. This is essentially a historical artifact as \cite{2013JHEP...07..081M} appeared before \cite{2013arXiv1303.6955H}  which provided the necessary tools of vacuum block expansion. To our knowledge a complete derivation for 2d holographic CFTs has not appeared in the literature so we have provided one in Appendix \ref{apend:tmi}.} that it takes the form
\be
J_{AB}=\max \begin{cases}
     \frac{c}{3}\log\left(\frac{\sinh\left(\frac{\pi}{\beta}(\theta_{A2}-\theta_{A1})\right)\sinh\left(\frac{\pi}{\beta}(\theta_{2B}-\theta_{1B})\right)}{\cosh(\frac{\pi}{\beta}(\theta_{A1}+\theta_{B1}))\cosh(\frac{\pi}{\beta}(\theta_{A2}+\theta_{B2}))}\right) \\
     0.
\end{cases}
\ee
\begin{figure}[H]
    \centering
    \includegraphics[width=.45\textwidth,page=12]{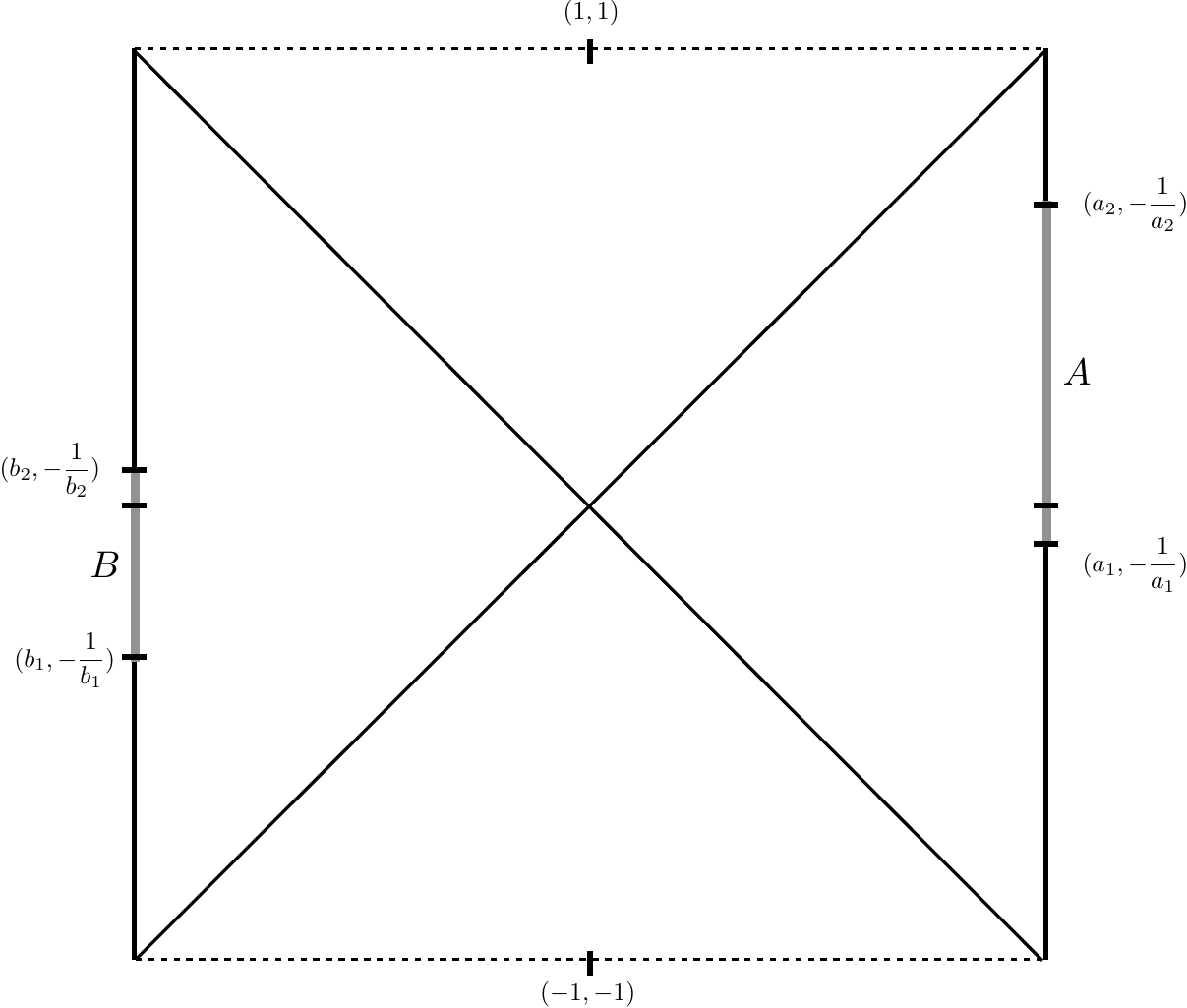}
    \caption{Two timelike regions $A=[T_{A1},T_{A2}]$ and $B=[T_{B1},T_{B2}]$. For the calculation of the timelike thermofield mutual information one region is paced on each of the left and right boundaries.}
    \label{fig:KK_TMI_setup}
\end{figure}
To determine the timelike thermofield mutual information we instead consider two time like intervals on different boundaries. The interval on the right boundary is given by $A=[T_{A1},T_{A2}]$ while the left by $B=[T_{B1},T_{B2}]$ both of which are defined with respect to the time coordinate on the right boundary. We use the same conventions \eqref{eq:g_k_conv} to define these intervals in Kruskal coordinates. The setup is shown in figure \ref{fig:KK_TMI_setup}.
\begin{figure}[H]
    \centering
    \begin{tabular}{cc}
    \includegraphics[width=.45\textwidth,page=15]{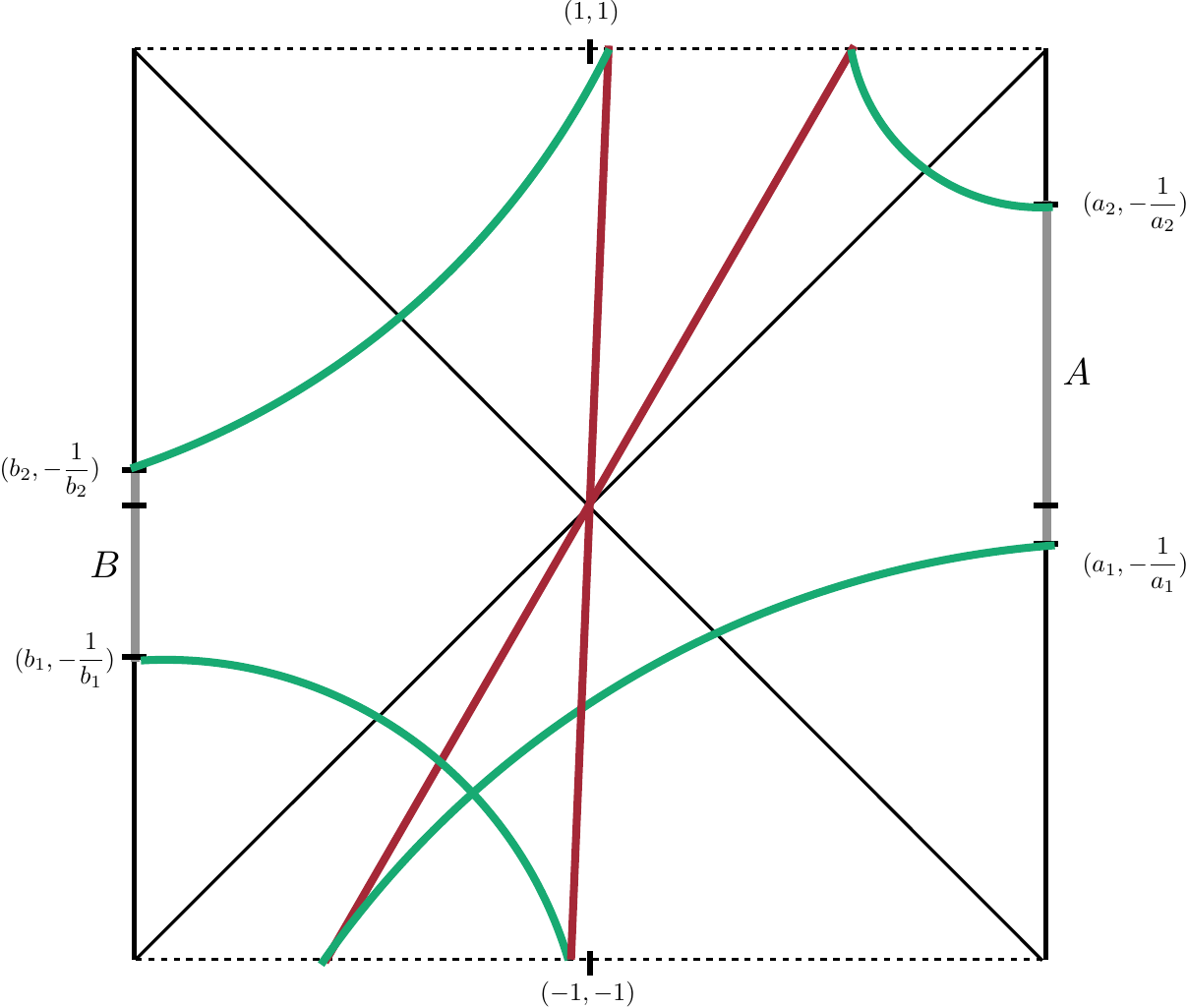}&
    \includegraphics[width=.45\textwidth,page=13]{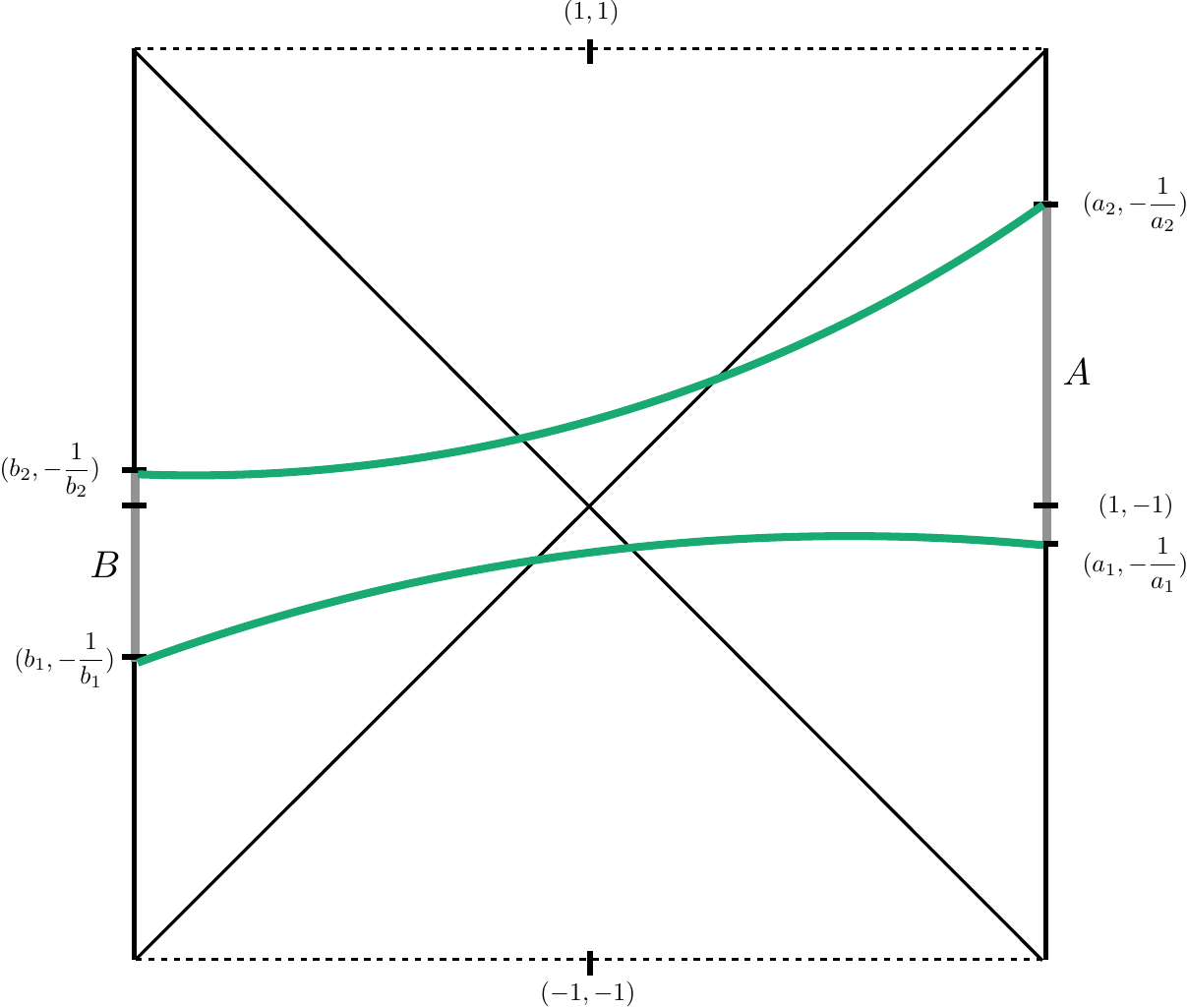}
    \end{tabular}
    \caption{Paths $\Gamma_{AB}$ whose area gives the timelike entanglement entropy $\tee{AB}$. There are two possible phases in analogy with the connected/disconnected phase transition of the thermofield mutual information. L: The disconnected phase is constructed from our result in the single interval case. Because time runs oppositely on the two boundaries the two imaginary geodesics should contribute with opposite sign. As such these areas cancel so that $\tee{AB}$ is entirely real. R: The connected phase consists of two spacelike geodesics extending through the geometry to opposite boundaries.}
    \label{fig:KK_TMI_SAB}
\end{figure}
\noindent There are once again two possible configurations of geodesics for the two party entropy $\tee{AB}$ (see figure \ref{fig:KK_TMI_SAB}). For the connected phase the geodesic are entirely spacelike and connects through the geometry between endpoints on either boundary. The geodesic connecting the left boundary at $v_L$ to the right boundary at $v_R$ is given by
\be
u(v)=\frac{v_Lv_R-1-2v_Lv}{(v_Lv_R-1)v+2v_R}
\ee
which has length
\be\label{eq:BTZ_across_geo}
\log\left(\frac{\beta}{\pi\delta}\cosh\left(\frac{\pi}{\beta}(T_L+T_R)\right)\right).
\ee
so that in this phase we have
\be
\tee{AB}=\frac{c}{3}\log\left(\frac{\beta^2}{\pi^2\delta^2}\cosh(\frac{\pi}{\beta}(T_{A1}+T_{B1}))\cosh(\frac{\pi}{\beta}(T_{A2}+T_{B2}))\right).
\ee
In the disconnected phase we have $\tee{AB}=\tee{A}+\tee{B}$ where the paths are the same as in the single interval case with one on each of the two boundaries. One must take care that because time on the left boundary runs opposite to that on the right the timelike surfaces should contribute with \emph{opposite} sign. That is
\be\label{eq:BTZ_LR_TEE}
\begin{split}
\tee{A}=\frac{c}{3}\log\left(\frac{\beta}{\pi\delta}\sinh\left(\frac{\pi}{\beta}(T_{A2}-T_{A1})\right)\right)+\frac{c}{6}i\pi\\
\tee{B}=\frac{c}{3}\log\left(\frac{\beta}{\pi\delta}\sinh\left(\frac{\pi}{\beta}(T_{B2}-T_{B1})\right)\right)-\frac{c}{6}i\pi.
\end{split}
\ee
This can be more clearly demonstrated by using \eqref{eq:BTZ_across_geo} and the relation 
\be\label{eq:relate_L_R}
t_R=t_L+i\frac{\beta}{2}
\ee
between times on the left and right boundary. Note that here $t_L$ is defined with respect to the time on the left boundary which runs opposite of that on the right. By substituting such that both end points are on the same boundary the result is precisely \eqref{eq:BTZ_LR_TEE}.
As a consequence $\tee{AB}$ in the disconnected phase is entirely real. We define the timelike thermofield mutual information to be
\be
\jtee{AB}=\tee{A}+\tee{B}-\tee{AB}
\ee
with $A$ and $B$ on different boundaries. The timelike thermofield mutual information is then found to be
\be
\jtee{AB}=\max \begin{cases}
     \frac{c}{3}\log\left(\frac{\sinh\left(\frac{\pi}{\beta}(T_{A2}-T_{A1})\right)\sinh\left(\frac{\pi}{\beta}(T_{2B}-T_{1B})\right)}{\cosh(\frac{\pi}{\beta}(T_{A1}+T_{B1}))\cosh(\frac{\pi}{\beta}(T_{A2}+T_{B2}))}\right)\\
     0.
\end{cases}
\ee

\subsubsection{The necessity of timelike entanglement entropy in thermofield double state}

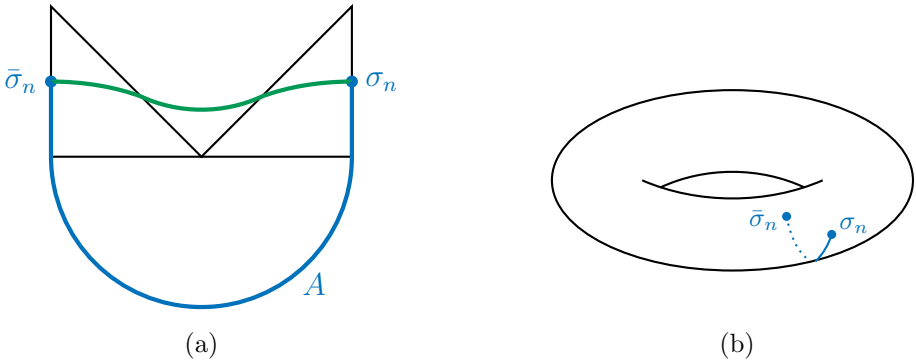
\begin{figure}[H]
    \begin{minipage}[b]{0.45\linewidth}
        \centering
        \begin{tikzpicture}[thick]
            \draw (-2,0) arc (-180:0:2 and 2);
            \draw (-2,0)--(-2,2)--(0,0)--(2,2)--(2,0)--cycle;
            \draw[fill=RoyalBlue,draw=none] (-2,1) circle (0.08);
            \draw[fill=RoyalBlue,draw=none] (2,1) circle (0.08);
            \begin{scope}[ForestGreen,ultra thick]
                \draw (-2,1) arc (90:53:2 and 1);
                \draw (2,1) arc (90:127:2 and 1);
                \draw (0.8,0.8) arc (-60:-120:1.6 and 1.3);
            \end{scope}
            \begin{scope}[RoyalBlue]
                \node at (2.4,1) {\large $\sigma_n$};
                \node at (-2.4,1) {\large $\bar{\sigma}_n$};
                \node at (1.5,-1.7) {\large $A$};
            \end{scope}
            \draw[RoyalBlue,ultra thick] (-2,1)--(-2,0) arc (-180:0:2 and 2)--(2,1);
        \end{tikzpicture}
        \subcaption{}
        \label{fig:HMa}
    \end{minipage}
    \begin{minipage}[b]{0.45\linewidth}
        \centering
        
        \begin{tikzpicture}[thick,scale=1.2]
            \draw (0,0) ellipse (2 and 1);
            \draw (-1,0) arc (-120:-60:2 and 1.5);
            \draw (-0.8,-0.08) arc (120:60:1.6 and 1.3);
            \draw[fill=RoyalBlue,draw=none] (0.6,-0.4) circle (0.05);
            \draw[fill=RoyalBlue,draw=none] (1.1,-0.6) circle (0.05);
            \begin{scope}[RoyalBlue]
                \draw (1.1,-0.6) arc (-20:-50:0.6 and 0.7);
                \draw[dotted] (0.6,-0.4) arc (200:250:0.4 and 0.8);
            \end{scope}
            \begin{scope}[RoyalBlue]
                \node at (1.35,-0.5) {\small $\sigma_n$};
                \node at (0.35,-0.45) {\small $\bar{\sigma}_n$};
            \end{scope}
        \end{tikzpicture}
        \vspace{0.5cm}
        \subcaption{}
        \label{fig:HMb}
    \end{minipage}
    \caption{(a) An interpretation of the configuration of $\sigma_n,\bar{\sigma}_n$ such that they are on the different sides of the black hole from each other. When the space is $\mathbb{S}^1$, the twist operators cannot be connected by a single spacelike region as depicted in (b). }
    \label{fig:HM}
\end{figure}

So far we have considered various choices of subsystems in two-sided BTZ black holes. Here we consider yet another choice of subsystem such that two twist operators are placed on the different sides of the black holes, depicted in figure \ref{fig:HMa} and figure \ref{fig:HMb}. As we will explain below, the corresponding entropy defined by the usual replica method, cannot be interpreted as the entanglement entropy, but as the timelike entanglement entropy. This provides an argument towards the necessity of introducing timelike entanglement entropy. Note that although this setup looks similar to \cite{Hartman:2013qma}, we consider only one twist operator in each AdS boundary, which is compactified on a circle.  

Let us consider the thermofield double state 
\begin{align}
    \ket{\text{TFD}}=\frac{1}{\sqrt{Z(\beta)}}\sum_n e^{-\beta E_n/2}\ket{n}_L\ket{n}_R,
\end{align}
where the Hilbert space of the CFT is doubled such that the left and right boundary of the BTZ are dual to CFT$_L$ and CFT$_R$.
This state is the zero eigenstate of the Hamiltonian 
\begin{align}\label{HTFD}
    H_{\text{TFD}}=\mathbf{1}\otimes H_R -H_L\otimes \mathbf{1},
\end{align}
i.e. the thermofield double state is invariant under the time evolution generated by $H_{\text{TFD}}$. In the bulk point of view, $\ket{\text{TFD}}$ is interpreted as the Euclidean construction of the Hartle-Hawking state in the BTZ black hole. The real time evolution generated by $H_{\text{TFD}}$ is realized as the gluing of the Euclidean geometry to the Lorentzian BTZ black hole geometry at $t=0$. In this geometry, time evolves forward in the right half and backward in the left half, the opposite direction of which corresponds to the relative sign in \eqref{HTFD}. 

As a model that has the non-trivial time evolution, we consider time evolution of the thermofield double state under another Hamiltonian
\begin{align}
    H=\mathbf{1}\otimes H_R+H_L\otimes \mathbf{1}.
\end{align}
As a result of flipping the relative sign from $H_{\text{TFD}}$, $H$ generates the forward time evolution in the both sides. In \cite{Hartman:2013qma}, a half plane $x<0$ in the planar CFT and a strip region $-L<x<L$ are considered as subsystems. In the both cases, geodesics connecting two twist operators on the left and right boundaries can be considered as candidate extreme surfaces. For example, if we consider a half plane as a subsystem, the entanglement entropy is calculated as the area of the extremal surface behaves as $S_A\propto t \ \ (t\gg \beta)$ \cite{Hartman:2013qma}.

Now, instead, let us consider the thermal state on a circle $\mathbb{S}^1$. We can again consider the configuration of two twist operators such that one is on the right boundary and the other is on the left. In the replica method we can calculate the corresponding entropy by the two-point function of the twist operators on the torus as before. In the AdS dual, this is a geodesic distance between the two points in the BTZ as sketched in figure \ref{fig:HMa}.
However, we cannot interpret this as EE because we cannot take the subsystem $A$ such that the twist operators are on $\partial A$. The Hilbert space is given by fields on two disconnected circles, situated in the left and right boundary and we only have a single twist operator in each of the circle, which makes the definition of the subsystem $A$ impossible.
However, we can understand this quantity as the timelike EE as depicted in figure \ref{fig:HMb}.

\subsection{BTZ with shock wave}\label{sec:shock}

Following \cite{2014JHEP...03..067S} we introduce a shock wave into the BTZ geometry. For early low energy perturbations along the $u=0$ horizon this takes the form of a simple shift $\alpha$ between two regions above and below such that $\tilde{v}=v+\alpha$. 

\begin{figure}[H]
    \centering
    \includegraphics[width=.75\textwidth,page=16]{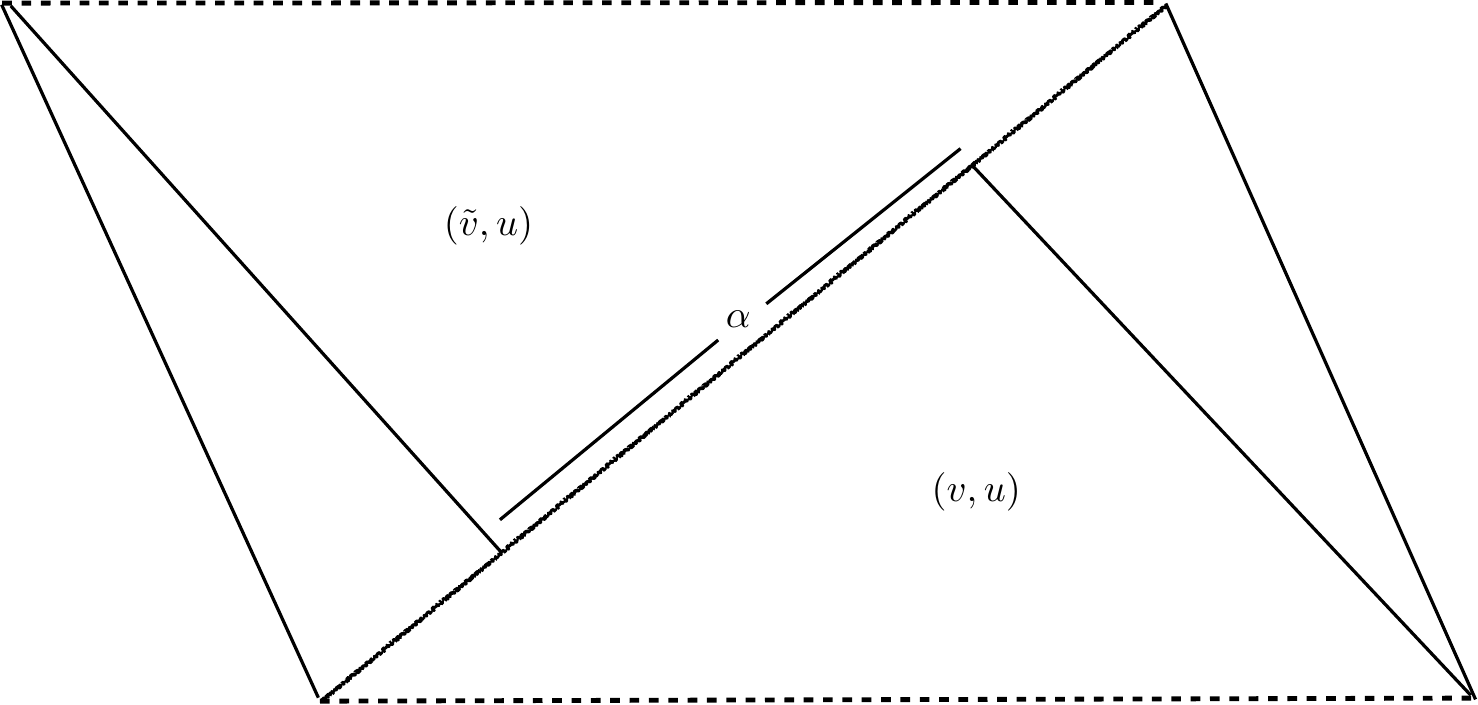}
    \caption{The BTZ geometry with a shock wave characterized by the parameter $\alpha$. The geometry above and below the $u=0$ horizon is BTZ but the two are shifted such that $\tilde{v}=v+\alpha$. Here $\tilde{v}$ is the shifted coordinate for $u>0$. }
    \label{fig:shock_geo}
\end{figure}

\noindent The metric has the form

\be
ds^2=-4\frac{dudv}{(1+u(v+\alpha\theta(u)))^2}+\frac{(1-u(v+\alpha\theta(u)))^2}{(1+u(v+\alpha\theta(u)))^2}r_+^2d\phi^2
\ee
and the geometry is shown in figure \ref{fig:shock_geo}.

Note that spacelike geodesics in the $t=0$ plane will not cross the horizon. As a result the entanglement entropies of spacelike regions will be unaffected by the shock wave and given by \ref{EES1}. If we consider the analytic continuation of the spacelike entanglement entropy we would incorrectly conclude that the timelike entanglement entropy should be similarly unaffected and given by \eqref{eq:BH_SI_tee} which is independent of $\alpha$. To see that this is incorrect we will carry out our optimization procedure and find in subsection \ref{sec:CFT_quench} that the result exactly matches the timelike entanglement entropy when computed from a careful Wick rotation of the boundary CFT.

We consider a single region $A=[T_1,T_2]$ on the right boundary where $T_1<T_2$ and use \eqref{eq:g_k_conv}.
\begin{figure}[H]
    \centering
    \includegraphics[width=.75\textwidth,page=18]{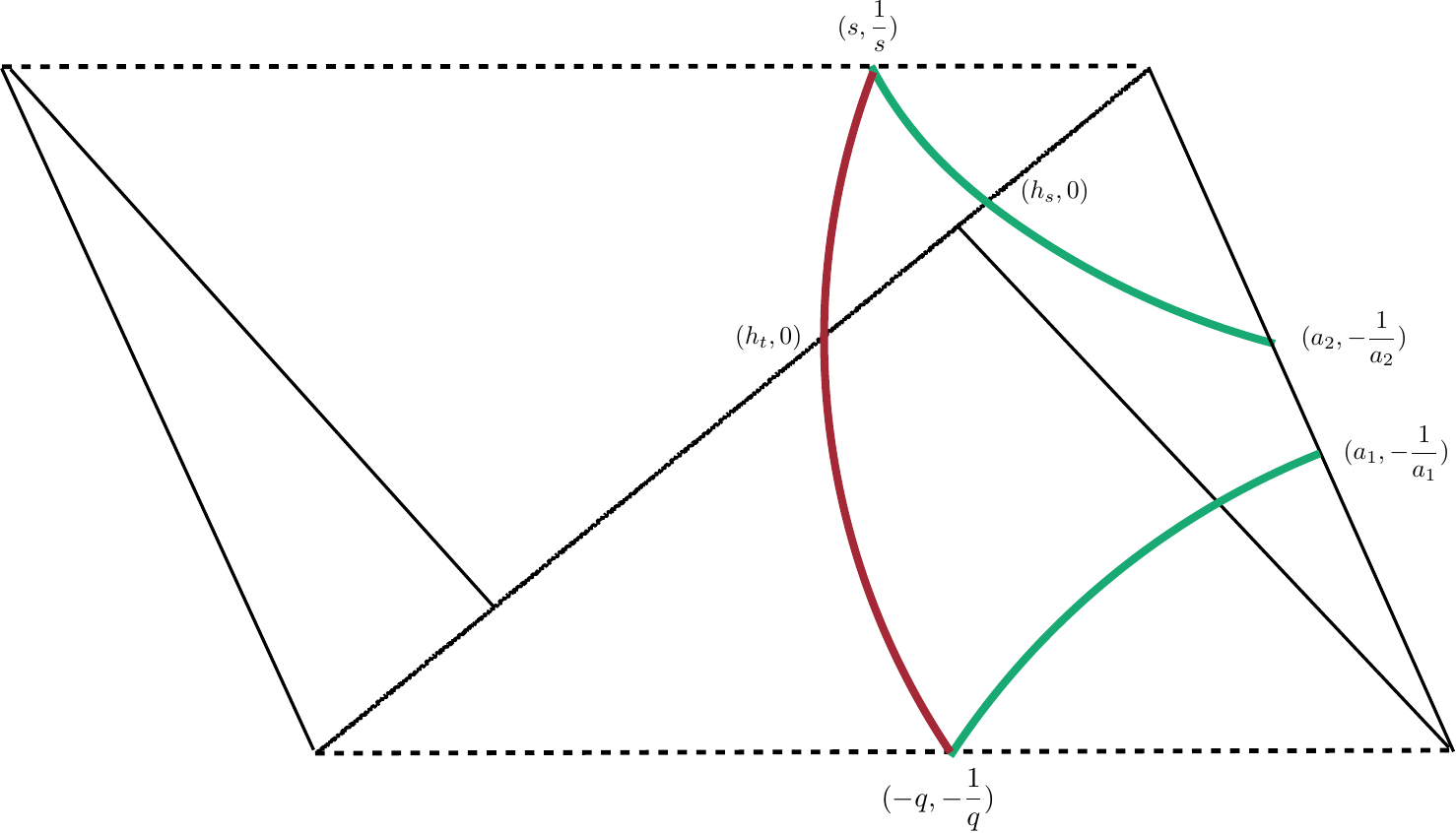}
    \caption{The five geodesics which make up the candidate paths $\Gamma_{A}.$ To account for the change of coordinates geodesics which cross the $u=0$ horizon where there is a shock wave are separated into two geodesics. The point which these geodesics cross the horizon as well as the location of the endpoints on the singularities are initially free parameters. These can all be fixed by requiring that $\Gamma_A$ is stationary. For $h_s,h_t$ this is equivalent to requiring that the geodesics are smoothly connected across the horizon.}
    \label{fig:shock_geo2}
\end{figure}
\noindent To define $\Gamma_A$ we use the collection of five geodesics shown in figure \ref{fig:shock_geo2}. This consists of three spacelike geodesics one from the boundary at $v=a_2$ to the horizon at $(h_s,0)$, from the horizon at $(h_s,0)$ to the future singularity at $v=s$ and one from the boundary at $v=a_1$  to the past singularity at $v=q$. We also include two timelike geodesic between the past singularity at $v=q$ and future singularity at $v=s$ which intersect the horizon at $(h_t,0)$. This forms a family of candidates $\Gamma_A$ where the values of $s,q,h_s,h_t$ are to be determined.

\paragraph{Upper spacelike geodesic}
We consider two geodesics: one below $u=0$ which we call $g_{u}$, and one above $u=0$ which we call $g_{\tilde{u}}$. These can be solved using \ref{eq:kruskal_geo} with the additional constraint that smoothness across the horizon at $u=0$ determines the point of intersection $h_s$
\be
\begin{split}
g_{u}:&\quad u(v)=\frac{v-h_s}{2h_sa_2-a_2^2-h_sv}\\
g_{\tilde{u}}:&\quad u(\tilde{v})=\frac{\tilde{v}-(h_s+\alpha)}{s^2-(h_s+\alpha)\tilde{v}}\\
&h_s=\frac{s^2+a_2^2-\alpha^2}{2(a_2+\alpha)}.
\end{split}
\ee
Together these have length
\be
\begin{split}
l_u+l_{\tilde{u}}&=\log\left(\frac{2}{\omega}\frac{a_2-h_s}{a_2}\right)+\log\left(\frac{\sqrt{(h_s+\alpha)^2-s^2}+h_s+\alpha-s}{\sqrt{(h_s+\alpha)^2-s^2}-(h_s+\alpha-s)}\right)\\
&=\log\left(\frac{1}{\omega}\frac{(a_2+\alpha)^2-s^2}{a_2s}\right)
\end{split}
\ee
Note that $h_s$ can also be determined directly from the geodesics length by considering $\partial_{h_s}(l_u+l_{\tilde{u}}=0$.
\paragraph{Lower spacelike geodesic} The lower spacelike geodesic $g_{l}$ remains in the lower half of the geometry and does not cross the shock wave the result is the same as the BTZ black hole
\be
g_l:\quad u(v)=\frac{2va_1-(a_1^2+q^2)}{2a_1q^2-(a_1^2+q^2)v}
\ee
\be
l_{l}=\log\left(\frac{1}{\omega}\frac{(q^2-a_1^2)}{a_1q}\right).
\ee
\paragraph{Timelike geodesic} The two timelike geodesics $g_t$ and $g_{\tilde{t}}$ are given by
\be
\begin{split}
g_t:&\quad u(v)=\frac{v-h_t}{q^2-h_tv}\\
g_{\tilde{t}}:& \quad u(\tilde{v})=\frac{\tilde{v}-(h_t+\alpha)}{s^2-(h_t+\alpha)\tilde{v}}
\end{split}
\ee
with smoothness across the horizon requiring
\be
h_t=\frac{s^2-q^2-\alpha^2}{2\alpha}.
\ee
The total length is
\be
\begin{split}
l_t+l_{\tilde{t}}=&2i\left[\arctan\left(\sqrt{\frac{q-h_t}{q+h_t}}\right)+\arctan\left(\sqrt{\frac{s-(h_t+\alpha)}{s+(h_t+\alpha)}}\right)\right]\\
=&2i\arctan\left(\frac{\sqrt{\frac{q-h_t}{q+h_t}}+\sqrt{\frac{s-(h_t+\alpha)}{s+(h_t+\alpha)}}}{1-\sqrt{\frac{q-h_t}{q+h_t}}\sqrt{\frac{s-(h_t+\alpha)}{s+(h_t+\alpha)}}}\right)
\end{split}
\ee
where $h_t$ can also be determined directly from the geodesics length by considering $\partial_{h_s}(l_t+l_{\tilde{t}})=0$. the timelike geodesics will be maximal with length $i\pi$ when the denominator is zero. As such this becomes the condition
\be
1-\frac{|\alpha-s+q|}{|\alpha+s-q|}=0 \longrightarrow s=q, \quad h_t=-\frac{\alpha}{2}.
\ee
Using that $s=q$ the spacelike geodesics have combined length
\be
l_u+l_{\tilde{u}}+l_l=\log\left(\frac{1}{\omega^2}\frac{((a_2+\alpha)^2-s^2)(s^2-a_1^2)}{a_1a_2s^2}\right)
\ee
which is stationary when $s^2=a_1(a_2+\alpha)$. Then
\be
l_u+l_{\tilde{u}}+l_l=\log\left(\frac{1}{\omega^2}\frac{(a_2-a_1+\alpha)^2}{a_1a_2}\right).
\ee
\begin{figure}[H]
    \centering
    \includegraphics[width=.75\textwidth,page=19]{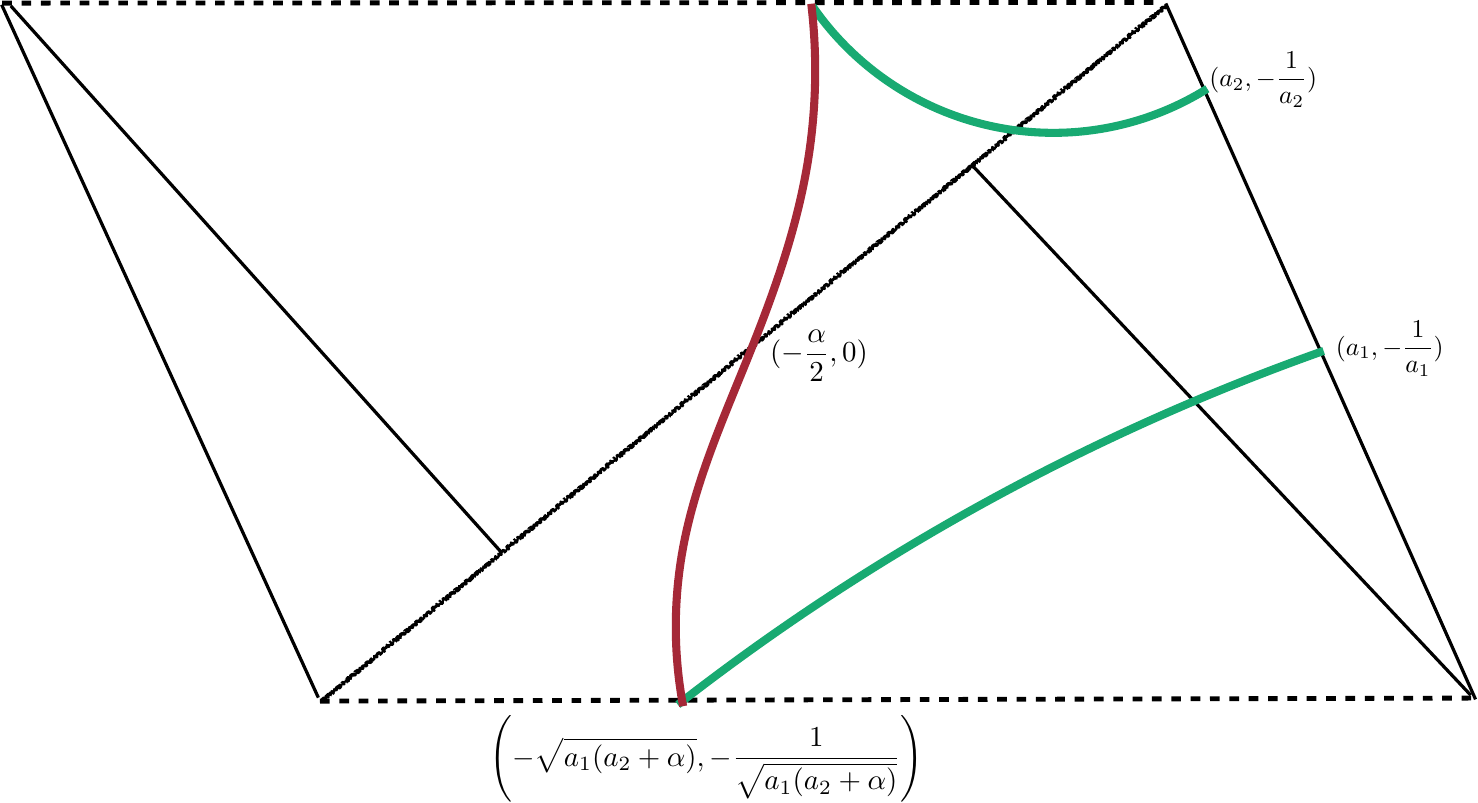}
    \caption{An example of a stationary path $\Gamma_A^*$ whose area calculates the the timelike entanglement entropy $\tee{A}$ in the BTZ geometry with a shock wave.}
    \label{fig:shock_geo_opt}
\end{figure}
\noindent Now with all the free parameters fixed it is found that the area of $\Gamma_A^*$ is given by
\be\label{eq:shock_gamma_area}
\frac{c}{3}\log\left(\frac{\beta}{\pi\delta}\left(\sinh(\frac{\pi}{\beta}(T_2-T_1))+\frac{\alpha}{2} e^{-\frac{\pi}{\beta}(T_2+T_1)}\right)\right)+\frac{c}{6}i\pi.
\ee
where there is a correction due to the shift $\alpha$. An example of a such path $\Gamma_A^*$ with this area is shown in figure \ref{fig:shock_geo_opt}. As a check the geodesic length between two points on opposite boundaries in the shock wave geometry was computed in \cite{2014JHEP...03..067S} to be:
\ba
\frac{c}{3}\log\left[\frac{\beta}{\pi\delta}\left(\cosh\frac{\pi}{\beta}(t_R-t_L)+\frac{\ap}{2}e^{-\frac{\pi}{\beta}(t_R+t_L)}\right)\right].
\ea
By continuing $t_R=T_2$ and $t_L=T_1-\frac{\beta}{2}i$ using \eqref{eq:relate_L_R}, we can reproduce (\ref{eq:shock_gamma_area}).

As a result we have shown that the two constructions --- the analytic continuation of the entanglement entropy given by \eqref{eq:BH_SI_tee}, which is independent of $\alpha$, and the optimization of extremal surfaces given by \eqref{eq:shock_gamma_area}, which depends on $\alpha$ --- are not the same. We return to this in section \ref{sec:CFT_quench} where we will show the correctness of  \eqref{eq:shock_gamma_area} by comparing this result with that of the CFT calculation.

\subsection{Timelike EE under local quench in thermofield double state from the CFT}\label{sec:CFT_quench}
\begin{figure}[H]
    \centering
\includegraphics[width=.7\textwidth]{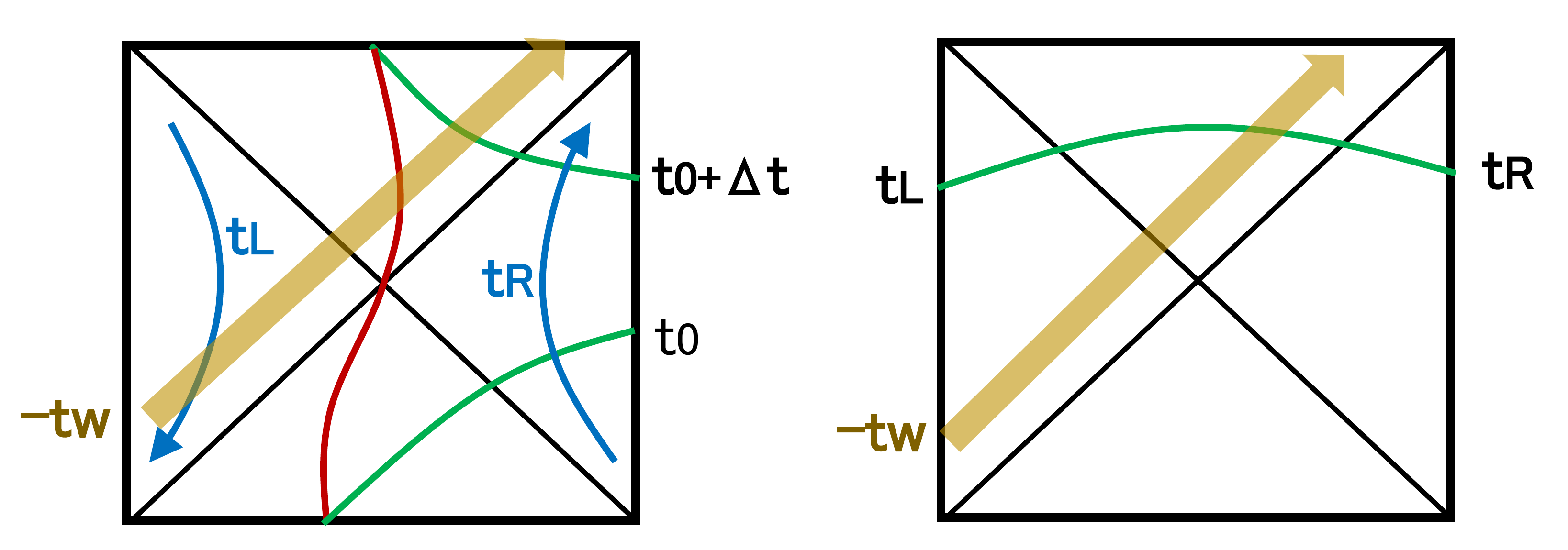}
    \caption{The sketches of geodesic in the gravity dual of local quench. The left picture describes the geodesics whose sum computes the timelike EE. The right panel shows the geodesic which connect a point in the left boundary and a point in the right panel. The brown arrows depict the localized shock wave created by the local operator. }
    \label{fig:TFDshock}
\end{figure}

In section \ref{sec:shock}, we have observed a surprising feature that the gravity dual calculation of timelike EE in the shock wave geometry is affected by the shock wave sent in the left boundary i.e. CFT$_L$, though the timelike EE is defined in the other boundary i.e CFT$_R$. In the gravity calculation this occurs because the geodesics whose sum computes the holographic timelike EE, extend to the region which is not causally related to the CFT$_R$ boundary.  This makes us wonder if the timelike EE in CFTs can have a similar behavior. To study this we consider the setup of a local quench in  a thermofield double state considered in \cite{Caputa:2015waa}. This is dual to a localized shock wave geometry, depicted in the left panel of figure \ref{fig:TFDshock}.

We insert the local primary operator $O(t_L,\phi_L)$ in CFT$_L$ at $t_L=-t_w$ and $\phi_L=0$ with a regularization parameter $\gamma$, which is infinitesimally small, such that $e^{-\gamma H_R}$ is multiplied on the local operator. The subsystem $A$ is defined as the interval between two points $(t_R,\phi_R)=(t_0,\phi_0)$ and $(t_R,\phi_R)=(t_0+\Delta t,\phi_0+\Delta\phi)$. In this setup we have
\ba
\mbox{Tr}[(\rho_A)^n]=\frac{\la O^n(y_1,\bar{y}_1)\sigma_n(y_2,\bar{y}_2)
\bar{\sigma}_n(y_3,\bar{y}_3)O^{n\dagger}(y_4,\bar{y}_4)\lb}{\la O(y_1,\bar{y}_1)O(y_4,\bar{y}_4)\lb^n},\label{fourpgh}
\ea
where we defined the coordinates on a cylinder such that 
\ba
&& y_1=-i\gamma+i\frac{\beta}{2},\ \  y_2=\phi_0-t_0-t_w,\ \ y_3=\phi_0+\Delta \phi-t_0-\Delta t-t_w,\ \ y_4=i\gamma+i\frac{\beta}{2},\no
&& \bar{y}_1=i\gamma-i\frac{\beta}{2},\ \  \bar{y}_2=\phi_0+t_0+t_w,\ \ \bar{y}_3=\phi_0+\Delta \phi+t_0+\Delta t+t_w,\ \ \bar{y}_4=-i\gamma-i\frac{\beta}{2}.\no
\ea
The coordinates $(t_R,\phi_R)$ are related to those on the complex plane via
\ba
(y,\bar{y})=\left(\phi_R-t_R,\phi_R+t_R\right),
\ea
and the time in the CFT$_L$ is related to $t_R$ via 
\ba
t_R=t_L+i\frac{\beta}{2}.
\ea
Now we perform the conformal transformation to a complex plane $w=e^{\frac{2\pi}{\beta}y}$ and introduce the cross ratio $z=\frac{w_{12}w_{34}}{w_{13}w_{24}}$.

Since $\gamma$ is infinitesimally small, we find 
\ba
&& z\simeq 1-\frac{2\pi i\gamma}{\beta}\cdot \frac{\sinh\left[\frac{\pi}{\beta}(\Delta\phi-\Delta t)\right]}{\cosh\left[\frac{\pi}{\beta}(\phi_0-t_0-t_w)\right]
\cosh\left[\frac{\pi}{\beta}(\phi_0+\Delta\phi-t_0-\Delta t-t_w)\right]},\no
&& \bar{z}\simeq 1+\frac{2\pi i\gamma}{\beta}\cdot \frac{\sinh\left[\frac{\pi}{\beta}(\Delta\phi+\Delta t)\right]}{\cosh\left[\frac{\pi}{\beta}(\phi_0+t_0+t_w)\right]
\cosh\left[\frac{\pi}{\beta}(\phi_0+\Delta\phi+t_0+\Delta t+t_w)\right]}.
\ea

This shows that when the subsystem $A$ is spacelike $\Delta\phi-\Delta t>0$, we have $(z,\bar{z})\to (1,1)$. Howeever when it is timelike  $\Delta\phi-\Delta t<0$, we find $(z,\bar{z})\to (e^{-2\pi i},1)$.
As in \cite{Asplund:2014coa}, the non-trivial monodromy around $z=1$ gives a non-zero entropy, assuming the HHLL conformal block approximation is justified in holographic CFTs.  Indeed the total entropy looks like ($\delta$ is the UV cut off):
\ba
S_A=\frac{c}{6}\log\left[\frac{\beta^2}{\pi^2\delta^2}\sinh\left(\frac{\pi}{\beta}(\Delta\phi-\Delta t)\right)\sinh\left(\frac{\pi}{\beta}(\Delta\phi+\Delta t)\right)\right]+S^{(\text{M})}_A,
\ea
where the monodromy contribution $S^{(\text{M})}_A$ is given by 
\ba
S^{(\text{M})}_A=\frac{c}{6}\log\left[\frac{z^{\frac{1}{2}(1-\ap_O)}
\bar{z}^{\frac{1}{2}(1-\ap_O)}(1-z^{\ap_O})(1-\bar{z}^{\ap_O})}
{\ap^2_O(1-z)(1-\bar{z})}\right].
\ea
The parameter $\ap_O$ is related to the (chiral) conformal dimension $h_O=\bar{h}_O$ of the primary operator $O$ via $\ap_O=\s{1-\frac{24h_O}{c}}$.

The EE for the spacelike interval does not depend on the local operator excitation because $S^{(\text{M})}_A=0$ 
and is identical to the standard result.
On the other hand, the timelike EE is computed as follows
(we set $\Delta\phi=0$):
\ba
\tee{A}&=&\frac{c}{3}\log\left[\frac{\beta}{\pi\delta}\sinh\left(\frac{\pi}{\beta}\Delta t\right)\right]
+\frac{c}{6}\log\left[\frac{\beta\sin(\pi\ap_O)}{\pi\ap_O\gamma}\right]\no
&&+\frac{c}{6}\log\left[\frac{\cosh\left(\frac{\pi}{\beta}(\phi_0-t_0-t_w)\right)\cosh\left(\frac{\pi}{\beta}(\phi_0-t_0-\Delta t-t_w)\right)}{\sinh\left(\frac{\pi}{\beta}\Delta t\right)}\right]+\frac{\pi}{6}ic.\no  \label{localpsge}
\ea
 The result above shows that the timelike EE for CFT$_R$ is influenced by the excitation in CFT$_L$ as in the shock wave analysis in section \ref{sec:shock}. Moreover, if we take the limit $t_w/\beta\to-\infty$, then we can expect that the setup gets closer to the gravity dual result for shock wave geometry in section \ref{sec:shock}, Indeed, in this limit we find from (\ref{localpsge}) the following behavior 
 \ba
\tee{A}\simeq &\frac{c}{3}\log\left[\frac{\beta^2}{\pi\delta\gamma}e^{-\frac{\pi}{\beta}(2t_w+t_0+t_0+\Delta t)}\right]+\frac{\pi}{6}ci.
 \ea
 If we identify $\ap=\frac{\beta}{\gamma}\cdot e^{\frac{2 \pi}{\beta}|t_w|}$ as in \cite{2014JHEP...03..067S}, then this agrees with (\ref{eq:shock_gamma_area}) by setting $t_0=T_1$ and $t_0+\Delta t=T_2$.
 
We can also show that the same result (\ref{localpsge}) can be obtained from the geodesic length in the gravity dual of the local quench. Consider the spacelike geodesic which connects the point $(t_R,\phi_R)=(t_R,\phi_0)$ on the right boundary and the point $(t_L,\phi_L)=(t_L,\phi_0)$, depicted in the right panel of figure \ref{fig:TFDshock}.
Its length divided by $4G_N$ was computed in \cite{Caputa:2015waa} as 
\ba
&&\frac{c}{6}\log\left[\frac{\beta^2}{\pi^2\delta^2}\cosh\left(\frac{\pi}{\beta}(t_R-t_L)\right)\right]
+\frac{c}{6}\log\left[\frac{\beta\sin(\pi\ap_O)}{\pi\ap_O\gamma}\right]\no
&&+\frac{c}{6}\log\left[\sinh\left(\frac{\pi}{\beta}(\phi_0-t_L-t_w)\right)\cosh\left(\frac{\pi}{\beta}(\phi_0-t_R-t_w)\right)\right].
\ea
By continuing the point in the left to the right we set
\ba
t_R=t_0+\Delta t,\ \ \ t_L=t_0-\frac{\beta}{2}i.
\ea
This gives the geodesic length between two timelike separated points, which computes the timelike EE, depicted in the left panel of figure \ref{fig:TFDshock}.
Indeed, this precisely reproduces the CFT result  (\ref{localpsge}). Notice that the standard EE for a spacelike interval in CFT$_R$ is static and indepenedent of the local operator excitation in CFT$_L$. This means that a naive analytic continuation from the spacelike subsystem to timelike one does not work because we need a careful analysis of conformal blocks in the holographic CFT. It is intriguing to note that the holographic analysis gives us a useful guide for the correct prescription.

\subsection{Rotating BTZ}

Finally we would like to study the rotating BTZ black hole. The metric looks like
\be
ds^2 = -\frac{(r^2-r_{+}^2)(r^2-r_{-}^2)}{r^2}dt^2+\frac{r^2}{(r^2-r_{+}^2)(r^2-r_{-}^2)}dr^2+r^2(dx+\frac{r_{+}r_{-}}{r^2}dt)^2
\ee
where we have compactified $x \sim x+L$ and imposed $r_{+}>r_{-}$. 

In order to consider the CFT dual to this system, we must perform a Wick rotation $t \rightarrow i\tau$ and compactify the imaginary time direction $\tau \sim \tau+\beta$, where $\beta$ is the inverse temperature. We obtain the following metric after the rotation.
\be
ds^2 = \frac{(r^2-r_{+}^2)(r^2-r_{-}^2)}{r^2}d\tau^2+\frac{r^2}{(r^2-r_{+}^2)(r^2-r_{-}^2)}dr^2+r^2(dx+i\frac{r_{+}r_{-}}{r^2}d\tau)^2
\ee

Near the outer horizon, i.e. $r=r_{+}+\eta$ where $\eta/r_{+} \ll 1$, the periodicity of time is determined as $\beta = 2\pi \frac{r_{+}}{r_{+}^2-r_{-}^2}$ by requiring that there is no conical singularity at the horizon. The cross term between $dx$ and $d\tau$ also implies we must further impose another compactification of the $x$ coordinate of the form, $x \sim x-i\beta\Omega$. Near the horizon, one can find that $\Omega=\frac{r_{-}}{r_{+}}$. These compactifications give rise to a density matrix of the form
\be
\rho = e^{-\beta H + \beta \Omega P}
\ee
where $H$ and $P$ are the Hamiltonian and the momentum of the CFT respectively. Taking a general subsystem interval, the holographic entanglement entropy is cited as \cite{Hubeny:2007xt}
\be
S_A = \frac{c}{6}\log\left(\frac{\beta_{+}\beta_{-}}{\pi^2\epsilon^2}\sinh\left(\frac{\pi(\Delta l+\Delta t)}{\beta_{+}}\right)\sinh\left(\frac{\pi(\Delta l-\Delta t)}{\beta_{-}}\right)\right)
\ee
where $\beta_{\pm}=\beta(1\pm\Omega)$, and $\Delta l$ and $\Delta t$ are the spatial and temporal lengths of the interval in the dual CFT. It becomes immediately clear that, in the limit where $\Delta l \rightarrow 0$, the term inside the logarithm becomes real and negative, leaving an imaginary contribution of $\frac{c}{6} \pi i$ to the entropy, which is a key characteristic of the timelike entanglement entropy.

We can then make the replacement $r_{-} \rightarrow ir_{-}$ and render the Euclidean theory real. However, this would in turn make the original Lorentzian metric complex-valued. Due to the dependence of $\Omega$ on $r_{-}$, both $\beta_\pm$ become complex (in fact, complex conjugates of each other), giving us complex-valued entanglement entropy, which we should interpret as pseudo entropy. The newly obtained pseudo entropy can be written as follows.
\be
\pe{A} = \frac{c}{6}\log\left(\frac{\beta^2(1+\Omega^2)}{\pi^2\epsilon^2}\sinh\left(\frac{\pi(\Delta l+\Delta t)}{\beta(1+i\Omega)}\right)\sinh\left(\frac{\pi(\Delta l-\Delta t)}{\beta(1-i\Omega)}\right)\right)
\ee
 In its most general form, the imaginary contribution can be expressed as
\begin{multline}
\operatorname{Im}(\pe{A}) = \frac{c}{6} \operatorname{Arg}\biggl(\left(\cosh{\frac{2\pi \Delta l}{\beta (1+\Omega^2)}}\cos{\frac{2\pi \Omega \Delta t}{\beta (1+\Omega^2)}} - i\sinh{\frac{2\pi \Delta l}{\beta (1+\Omega^2)}}\sin{\frac{2\pi \Omega \Delta t}{\beta (1+\Omega^2)}}\right) \\
- (\Delta l \leftrightarrow \Delta t)\biggr).
\end{multline}

The consideration of a system consisting of a rotating BTZ black hole has given us yet another example of pseudo entropy being manifest in a physical setting. The non-static nature of this system has allowed us to make a choice of equipping the Euclidean theory with a real metric while simultaneously equipping the corresponding Lorentzian theory with a complex-valued metric. The holographic entanglement entropy of the system is evaluated to be complex valued in general, even for spacelike intervals. This occurrence of the complex-valued  entanglement entropy should be correctly interpreted as pseudo entropy.

\section{Holographic timelike entanglement entropy in \texorpdfstring{AdS$_{d+1}$/CFT$_d$}{AdSd+1/CFTd}}\label{sec:4}

In the previous section, we only considered the holographic timelike EE in AdS$_3$/CFT$_2$. Since we have not used any specific features in three-dimensional gravity, it is natural to extend it to higher-dimensions. As generalizations of a temporal subsystem we have considered, here we consider two examples; a hyperbolic region $t^2-\mathbf{x}^2< T_0^2/4$ and a strip region $|t|<T_0/2$. For both cases, we first compute the timelike EE by performing the ordinary Wick rotation $t_{\text{E}}=it$, and then try to give geometric interpretations to the result in the similar way as we did in sections \ref{subsec:pureads} and \ref{subsec:extremalsurface}.

\subsection{Hyperbolic subsystem}

First let us consider a hyperbolic subsystem $A$ defined by $t^2-\mathbf{x}^2\le T_0^2/4$ as a generalization of a temporal interval in the $d=2$ case. After we obtain the EE for $A$ by applying the Wick rotation to the calculation for Euclidean AdS$_{d+1}$ (EAdS$_{d+1}$), we will find out the geometric configuration reproducing the result of the Wick rotation.

\subsubsection{Wick rotation}
First we review the ordinary computation of the EE for the Poincar\'{e} EAdS$_{d+1}$
\begin{align}
    ds^2=R_{\text{AdS}}^2\frac{dz^2+dy^2+dt_{\text{E}}^2+d\mathbf{x}^2}{z^2},
\end{align}
where $y$ is a direction that we regard as an alternative ``time'' direction and $\mathbf{x}\in\mathbb{R}^{d-2}$ are the remaining directions. 
We define a subsystem $A$ of the $y=0$ slice on the boundary as $t_{\text{E}}^2+\mathbf{x}^2\le T_{0}^2/4$, where we again adopt the Poincar\'{e} coordinates \eqref{PoincareEAdS}. Here we introduce a radial coordinate $\xi=\sqrt{t_{\text{E}}^2+\mathbf{x}^2}$, then the entanglement entropy is the extremal value of a functional 
\begin{align}
    \frac{R_{\text{AdS}}^{d-1}}{4G_N^{(d+1)}}\text{Vol}(\mathbb{S}^{d-2})\int dz \frac{\xi^{d-2}}{z^{d-1}}\sqrt{1+\xi'(z)^2},
\end{align}
where $\text{Vol}(\mathbb{S}^{d-2})=\frac{2\pi^{\frac{d-1}{2}}}{\Gamma(\frac{d-1}{2})}$ is the area of $\mathbb{S}^{d-2}$. By varying this functional, we find that the extremal surface is a hemisphere $z^2+\xi^2=T_{0}^2/4$. Therefore the entanglement entropy is 
\begin{align}
    S_A=\frac{R_{\text{AdS}}^{d-1}}{4G_N^{(d+1)}}\text{Vol}(\mathbb{S}^{d-2}) \int_{\frac{\epsilon}{T_{\text{E}}}}^{1}du \frac{(1-u^2)^{\frac{d-3}{2}}}{u^{d-1}},
\end{align}
where we have rescaled as $z=\frac{T_0}{2}u$ and introduced a UV cutoff $z=\epsilon$. For odd $d$,
\begin{align}
    S_A=\frac{R_{\text{AdS}}^{d-1}}{4G_N^{(d+1)}}\text{Vol}(\mathbb{S}^{d-2})\left[\sum_{k=0}^{\frac{d-3}{2}}\binom{\frac{d-3}{2}}{k}\frac{(-1)^k}{d-2k-2}\left(\frac{T_0}{2\epsilon}\right)^{d-2k-2}+\frac{(-1)^{\frac{d-1}{2}}\sqrt{\pi}\Gamma\left(\frac{d-1}{2}\right)}{2\Gamma\left(\frac{d}{2}\right)}\right]
\end{align}
and for even $d$,
\begin{align}
    S_A=\frac{R_{\text{AdS}}^{d-1}}{4G_N^{(d+1)}}\text{Vol}(\mathbb{S}^{d-2})\left[\sum_{k=0}^{\frac{d}{2}-2}\binom{\frac{d-3}{2}}{k}\frac{(-1)^k}{d-2k-2}\left(\frac{T_0}{2\epsilon}\right)^{d-2k-2}+\frac{(-1)^{\frac{d}{2}-1}\Gamma\left(\frac{d-1}{2}\right)}{\sqrt{\pi}\Gamma\left(\frac{d}{2}\right)}\log\frac{T_{0}}{\epsilon}\right].
\end{align}
Here we would like to perform the Wick rotation. As before, we replace $t_{\text{E}}\to it$ and $T_{0}\to iT_0$, then the extremal surface will be a hyperbolic surface
\begin{align}
    t^2-\mathbf{x}^2-z^2=\frac{T_0^2}{4}
\end{align}
which ends on $\partial A$ at the boundary. As before, we define a radial coordinate $\xi=\sqrt{t^2-\mathbf{x}^2}$ for the unit $\mathbb{H}^{d-2}$. Since the volume of the unit hyperboloid $\mathbb{H}^{d-2}$ diverges, in order to define the volume we regularize it as 
\begin{align}
    \text{Vol}(\mathbb{H}^{d-2})=\text{Vol}(\mathbb{S}^{d-3})\cdot 2\int_0^\Lambda dv\sinh^{d-3}v.
\end{align}
By taking $\Lambda\to i\frac{\pi}{2}$, the right-hand side becomes 
\begin{align}
    i^{d-2}\text{Vol}(\mathbb{S}^{d-3})\cdot 2\int_0^{\frac{\pi}{2}}d \theta\sin^{d-3}\theta=i^{d-2}\text{Vol}(\mathbb{S}^{d-2}).
\end{align}
Therefore $\text{Vol}(\mathbb{S}^{d-2})$ should be replaced with $(-i)^{d-2}\text{Vol}(\mathbb{H}^{d-2})$ under the Wick rotation. Note that the phase factor $(-i)^{d-2}$ is $i(-1)^{\frac{d-1}{2}}$ for odd $d$ and $(-1)^{\frac{d}{2}-1}$ for even $d$. Thus the timelike EE becomes 
\begin{align}\label{dEEwickodd}
    \tee{A}=\frac{R_{\text{AdS}}^{d-1}}{4G_N^{(d+1)}}\text{Vol}(\mathbb{H}^{d-2})\left[\sum_{k=0}^{\frac{d-3}{2}}\binom{\frac{d-3}{2}}{k}\frac{1}{d-2k-2}\left(\frac{T_0}{2\epsilon}\right)^{d-2k-2}+\frac{i\sqrt{\pi}\Gamma\left(\frac{d-1}{2}\right)}{2\Gamma\left(\frac{d}{2}\right)}\right].
\end{align}
for odd $d$ and 
\begin{align}\label{dEEwickeven}
    \tee{A}=\frac{R_{\text{AdS}}^{d-1}}{4G_N^{(d+1)}}\text{Vol}(\mathbb{H}^{d-2})\left[\sum_{k=0}^{\frac{d}{2}-2}\binom{\frac{d-3}{2}}{k}\frac{1}{d-2k-2}\left(\frac{T_0}{2\epsilon}\right)^{d-2k-2}+\frac{\Gamma\left(\frac{d-1}{2}\right)}{\sqrt{\pi}\Gamma\left(\frac{d}{2}\right)}\log\frac{T_0}{\epsilon}+\frac{i\sqrt{\pi}\Gamma\left(\frac{d-1}{2}\right)}{2\Gamma\left(\frac{d}{2}\right)}\right].
\end{align}
Note that all the divergent terms are real and the $\mathcal{O}(\epsilon^0)$ term is purely imaginary. It is remarkable that the form of the imaginary part does not depend on whether $d$ is odd or even.

\subsubsection{Extremal surface}

We would like to find the geometric configuration in the Lorentzian AdS$_{d+1}$ spacetime
\begin{align}
    ds^2=R_{\text{AdS}}^2\frac{dz^2+dy^2-dt^2+d\mathbf{x}^2}{z^2}.
\end{align}
which reproduces the result above.
In the Lorentzian case, we should extremize the functional 
\begin{align}
    \frac{R_{\text{AdS}}^{d-1}}{4G_N^{(d+1)}}\text{Vol}(\mathbb{H}^{d-2})\int dz \frac{\xi^{d-2}}{z^{d-1}}\sqrt{1-\xi'(z)^2}.
\end{align}
The solution extremizing this functional is 
\begin{align}\label{spextpo}
    \xi^2-z^2=T_0^2/4,
\end{align}
then the contribution to the entanglement entropy that comes from the two spacelike surfaces that end on the future and past part of $\partial A$ respectively is 
\begin{align}\nonumber
    S_A&=\frac{R_{\text{AdS}}^{d-1}}{4G_N^{(d+1)}}\text{Vol}(\mathbb{H}^{d-2})\int_{\frac{2\epsilon}{T_0}}^\infty du \frac{(1+u^2)^{\frac{d-3}{2}}}{u^{d-1}} \\
    &=\frac{R_{\text{AdS}}^{d-1}}{4G_N^{(d+1)}}\text{Vol}(\mathbb{H}^{d-2})
    \left\{\begin{aligned}\label{ddimreal}
    \quad\sum_{k=0}^{\frac{d-3}{2}}&\binom{\frac{d-3}{2}}{k}\frac{1}{d-2k-2}\left(\frac{T_0}{2\epsilon}\right)^{d-2k-2} & (d:\text{odd}) \\
    \sum_{k=0}^{\frac{d}{2}-2}&\binom{\frac{d-3}{2}}{k}\frac{1}{d-2k-2}\left(\frac{T_0}{2\epsilon}\right)^{d-2k-2}+\frac{\Gamma\left(\frac{d-1}{2}\right)}{\sqrt{\pi}\Gamma\left(\frac{d}{2}\right)}\log\frac{T_0}{\epsilon} & (d:\text{even})
    \end{aligned}\right.
\end{align}
We can see that this result reproduces the real part of \eqref{dEEwickodd} and \eqref{dEEwickeven}. In addition to the spacelike surfaces considered above, we also consider a timelike surface 
\begin{align}
    z^2-\xi^2=C^2,
\end{align}
where $C$ is an arbitrary positive constant. This is embedded into the global patch of AdS$_3$ as a geodesic at the center $\rho=0$. The area of this surface is 
\begin{align}\begin{aligned}
    \text{Vol}(\mathbb{H}^{d-2})\int_{C}^\infty dz \frac{\xi^{d-2}}{z^{d-1}}\sqrt{1-\xi'(z)^2}&=\text{Vol}(\mathbb{H}^{d-2})\cdot i\int_1^\infty du\frac{(u^2-1)^{\frac{d-3}{2}}}{u^{d-1}} \\
    &=\text{Vol}(\mathbb{H}^{d-2})\frac{i\sqrt{\pi}\Gamma\left(\frac{d-1}{2}\right)}{2\Gamma\left(\frac{d}{2}\right)}
\end{aligned}\end{align}
Thus the contribution of this timelike surface to the entanglement entropy is 
\begin{align}
    \frac{R_{\text{AdS}}^{d-1}}{4G_N^{(d+1)}}\text{Vol}(\mathbb{H}^{d-2})\frac{i\sqrt{\pi}\Gamma\left(\frac{d-1}{2}\right)}{2\Gamma\left(\frac{d}{2}\right)},
\end{align}
which is independent of $C$. This expression is identical to the imaginary part of \eqref{dEEwickodd} and \eqref{dEEwickeven}. In particular, this result reduces to $\frac{ic_{\text{dS}}}{6}$ when $d=2$. 

\subsubsection{Global coordinate}\label{subsec:highgl}
Next let us give an interpretation to the result above by considering the extremal surface in the global coordinate in the similar way to subsection \ref{subsec:extremalsurface}. 

We consider the AdS$_d$ slicing of AdS$_{d+1}$
\begin{align}
    ds_{\text{AdS}_{d+1}}^2=R_{\text{AdS}}^2(d\eta^2+\cosh^2\eta\, ds_{\text{AdS}_d}^2). 
\end{align}
Here we adopt the open slicing as the coordinate of AdS$_d$:
\begin{align}
    ds_{\text{AdS}_d}^2=-d\tau^2+\sin^2\tau\, ds_{\mathbb{H}^{d-1}}^2,
\end{align}
where $ds_{\mathbb{H}^{d-1}}^2$ is the metric of the unit hyperbolic space:
\begin{align}
    ds_{\mathbb{H}^{d-1}}^2=d\rho^2+\sinh^2\rho\, d\Omega_{d-2}^2.
\end{align}
As done in subsection \ref{subsec:extremalsurface}, we consider the gluing of this geometry at $\eta=0$ to the Wick-rotated geometry by $\eta\to i\tilde{\eta}$:
\begin{align}
    ds^2=R_{\text{AdS}}^2(-d\tilde{\eta}^2+\cos^2\tilde{\eta}\,ds_{\text{AdS}_d}^2).
\end{align}

We would like to evaluate the sum of the spacelike and timelike surfaces connecting $\tau=\tau_0/2$ and $\tau=-\tau_0/2$. Let us first consider the spacelike surface emanating from $\tau=\tau_0/2$ and reaching $\eta=0$. 
We regard a coordinate $\theta$ of $\mathbb{S}^{d-2}$ as a time and take a time slice $\theta=0$. Then the induced metric of the codimension-2 extremal surface $\tau=\tau(\eta)$ is 
\begin{align}
    ds^2=R_{\text{AdS}}^2\left[(1-\cosh^2\eta\,\tau'^2)d\eta^2+\cosh^2\eta\sin^2\tau\,ds_{\mathbb{H}^{d-2}}^2\right]
\end{align}
and the area is
\begin{align}
    R_{\text{AdS}}^{d-1}\text{Vol}(\mathbb{H}^{d-2})\int_0^{\eta_{\infty}}d\eta\sqrt{1-\cosh^2\eta\,\tau'^2}\cosh^{d-2}\eta\sin^{d-2}\tau
\end{align}
The solution of the equation of motion with respect to $\tau(\eta)$ reads
\begin{align}
    \tau''-(d-1)\sinh\eta\cosh\eta\,\tau'^3-(d-2)\cot\tau\,\tau'^2+d\tanh\eta\,\tau'+(d-2)\frac{\cot\tau}{\cosh^2\eta}=0.
\end{align}
The solution with a boundary condition $\tau(\infty)=\tau_0/2$ is given by
\begin{align}
    \cos\tau=\cos\frac{\tau_0}{2}\tanh\eta.
\end{align}
Note that this surface reaches $\tau=\pi/2$ at $\eta=0$. Also, we can check that this equation is equivalent to \eqref{spextpo} by a coordinate transformation to the Poincare coordinate. The area of this surface is
\begin{align}
    R_{\text{AdS}}^{d-1}\text{Vol}(\mathbb{H}^{d-2})&\int_0^{\eta_\infty}d\eta \sin\frac{\tau_0}{2}(1-\cos^2\frac{\tau_0}{2}\tanh^2\eta)^{\frac{d-3}{2}}\cosh^{d-2}\eta\\
    &=R_{\text{AdS}}^{d-1}\text{Vol}(\mathbb{H}^{d-2})\int_0^{\sin\frac{\tau_0}{2}\sinh\eta_\infty}du(1+u^2)^{\frac{d-3}{2}}.
\end{align}
Introducing the cutoff $\sinh\eta_\infty\simeq\frac{1}{2} e^{\eta_\infty}\equiv \frac{1}{\epsilon}$ and identifying as $\sin\frac{\tau_0}{2}=T_0$, we can see that this area reproduces the real part of the result above \eqref{ddimreal}. 

Let us move on to the timelike surface. The induced metric of a surface $\tau=\tau(\tilde{\eta})$ on a slice $\theta=0$ is given by 
\begin{align}
    ds^2=R_{\text{AdS}}^2\left[-(1+\cos^2\tilde{\eta}\,\tau'^2)d\tilde{\eta}^2+\cos^2\tilde{\eta}\sin^2\tau\,ds_{\mathbb{H}^{d-2}}^2\right],
\end{align}
and the area functional is
\begin{align}
    iR_{\text{AdS}}^{d-1}\text{Vol}(\mathbb{H}^{d-2})\int d\tilde{\eta}\sqrt{1+\cos^2\tilde{\eta}\,\tau'^2}\sin^{d-2}\tau\cos^{d-2}\tilde{\eta}.
\end{align}
In the same way as solving the spacelike one, we can find solutions to the equation of motion for this action:
\begin{align}
    \cos\tau=C\tan\tilde{\eta},
\end{align}
where $C$ is an arbitrary constant that cannot be determined by boundary conditions. From the similar calculation to the spacelike case, we find that the area is
\begin{align}
    iR_{\text{AdS}}^{d-1}\text{Vol}(\mathbb{H}^{d-2})\frac{\sqrt{\pi}\Gamma\left(\frac{d-1}{2}\right)}{2\Gamma\left(\frac{d}{2}\right)},
\end{align}
which does not depend on the constant $C$. Thus we have derived the same result as the one obtained by the Wick rotation.

\subsection{Strip subsystem}

Next, let us consider a strip subsystem. First, we compute the timelike entanglement entropy by Wick rotation. Consider again the Poincar\'{e} EAdS$_{d+1}$ with a metric 
\begin{align}
    ds^2=\frac{dz^2+dt_{\text{E}}^2+dy^2+d\mathbf{x}^2}{z^2}.
\end{align}
We take a subsystem $A$ as a strip $-T_0/2<t_{\text{E}}<T_0/2$ on the $y=0$ surface. Furthermore, we take an IR cutoff $L$ as the lengths of the remaining directions $\mathbf{x}$. In this setup, the entanglement entropy of the dual CFT can be obtained as the extremal value of a functional
\begin{align}
    \frac{R_{\text{AdS}}^{d-1}}{4G_N^{(d+1)}}L^{d-2}\int  dz\frac{\sqrt{1+t_{\text{E}}'(z)^2}}{z^{d-1}}.
\end{align}
The solution of the Euler-Lagrange equation with a boundary condition $t_{\text{E}}(0)=T_0/2$ is given by \cite{Ryu:2006bv}
\begin{align}
    \pm t_{\text{E}}(z)=\frac{z^d}{dz_*^{d-1}} {_2F_1}\left(\frac{1}{2},\frac{d}{2(d-1)},\frac{3d-2}{2d-2},\left(\frac{z}{z_*}\right)^{2d-2}\right)-\frac{\sqrt{\pi}}{d}\frac{\Gamma\left(\frac{3d-2}{2d-2}\right)}{\Gamma\left(\frac{2d-1}{2d-2}\right)},
\end{align}
where $z=z_*$ is the tip of the surface satisfying $t_{\text{E}}'(z_*)=0$ and given by
\begin{align}\label{zstar}
    z_*=\frac{T_0\Gamma\left(\frac{1}{2(d-1)}\right)}{2\sqrt{\pi}\Gamma\left(\frac{d}{2(d-1)}\right)}.
\end{align}
Then the resulting EE becomes 
\begin{align}
    S_A=\frac{R_{\text{AdS}}^{d-1}}{2(d-2)G_N^{(d+1)}}L^{d-2}\left[\frac{1}{\epsilon_{\text{AdS}}^{d-2}}-\frac{1}{2}\left(\frac{2\sqrt{\pi}\Gamma\left(\frac{d}{2(d-1)}\right)}{\Gamma\left(\frac{1}{2(d-1)}\right)}\right)^{d-1}\frac{1}{T_0^{d-2}}\right]\, .
\end{align} 
Let us perform the Wick rotation $t_{\text{E}}=it$. Simultaneously, we also replace with $T_0\to iT_0$, then the subsystem becomes $-T_0/2<t<T_0/2$. 
\begin{align}\label{LAdSdEE}
    \tee{A}=\frac{R_{\text{AdS}}^{d-1}}{2(d-2)G_N^{(d+1)}}L^{d-2}\left[\frac{1}{\epsilon_{\text{AdS}}^{d-2}}-\frac{1}{2}\left(\frac{2\sqrt{\pi}\Gamma\left(\frac{d}{2(d-1)}\right)}{\Gamma\left(\frac{1}{2(d-1)}\right)}\right)^{d-1}\frac{(-i)^{d-2}}{T_0^{d-2}}\right]\, .
\end{align}
Note that the imaginary factor $(-i)^{d-2}$ is sensitive to dimensions and the term depends on the width of strip $T_0$. These features are highly different from the spherical case and counterintuitive as we will see soon. 

Unfortunately, in the strip case, we do not have a clear interpretation as the bulk extremal surface in the Lorentzian Poincar\'{e} AdS$_{d+1}$
\begin{align}
    ds^2=R_{\text{AdS}}^2\frac{dz^2-dt^2+dy^2+d\mathbf{x}^2}{z^2},
\end{align}
for which the extremal surface are obtained by varying a functional
\begin{align}\label{intA}
    \frac{R_{\text{AdS}}^{d-1}}{4G_N^{(d+1)}}L^{d-2}\int_\epsilon  dz\frac{\sqrt{1-t'(z)^2}}{z^{d-1}}.
\end{align}
The Euler-Lagrange equation leads to 
\begin{align}\label{stripsf}
    \frac{t'(z)}{z^{d-1}\sqrt{1-t'(z)^2}}=C,
\end{align}
where $C$ is a real constant.
The leading area term proportional to $\epsilon_{\text{AdS}}^{-(d-2)}$ can be reproduced because it does not depend on the integral constant $C$, in other words it is determined only from the information around the boundary $z=0$. On the other hand, the sub-leading term of the area of \eqref{stripsf} depends on $C$.
As noted above, whether the sub-leading term is real or imaginary is sensitive to the dimension $d$, thus there seems to be no proper constant $C$ that reproduces the result by Wick rotation since $C$ should be real. 
One observation is that the imaginary part of \eqref{LAdSdEE} is reproduced by considering in the continued coordinate $z=i\tilde{z}$ and simultaneously performing $z_*=i\tilde{z}_*$. This is easily understood because this analytic continuation is equivalent to $T_0\to iT_0$ through \eqref{zstar}. This idea looks similar to the procedure in section \ref{subsec:extremalsurface}, but we do not have a clear construction of this surface in the complexified geometry as we had there. 

However, if we follow a higher dimensional extension of the  prescription shown in the end of section 
(\ref{subsec:extremalsurface}), we would obtain a different result because this assumes either timelike or spacelike extremal surfaces instead of the complexified surface. In this way, the comparison between the holographic timelike EE in higher dimensional CFTs and its field theoretic computation when the subsystem $A$ is a strip, look non-trivial. We will leave better understandings of this for a future problem.

\section{Holographic pseudo entropy in dS/CFT}\label{sec:5}
In this section, we study the holographic properties of quantum entanglement in de Sitter spacetime and its relation to the timelike EE in AdS/CFT. The dS$_{d+1}$/CFT$_d$ correspondence \cite{Strominger:2001pn} is a conjecture for holographic duality between the quantum gravity on $d+1$-dimensional de Sitter (dS$_{d+1}$) space 
\begin{align}\label{globaldS}
    ds^2=R_{\text{dS}}^2(-d\tau^2+\cosh^2\tau\,d\Omega_d^2),
\end{align}
where $d\Omega_d^2$ is the metric of the unit sphere $\mathbb{S}^d$ and a $d$-dimensional Euclidean CFT defined on the future boundary $(\simeq \mathbb{S}^d)$ of dS$_{d+1}$. 

The dS/CFT dictionary analogous to the GKPW relation \cite{Gubser:1998bc,Witten:1998qj} in AdS/CFT is given by 
\begin{align}\label{dSGKPW}
    \Psi_{\text{dS}}[\phi_0]=Z_{\text{CFT}}[\phi_0].
\end{align}
The left-hand side $\Psi_{\text{dS}}[\phi_0]$ is defined as a path integral over all fields $\phi$ on dS$_{d+1}$ with fixing the boundary condition $\phi_0$ at the future boundary:
\begin{align}
    \Psi_{\text{dS}}[\phi_0]=\int_{\phi|_{\tau=\tau_\infty}=\phi_0}\mathcal{D}\phi\, e^{iI_{\text{dS}}[\phi]}\Psi_{\text{in}},
\end{align}
where $\Psi_{\text{in}}$ denotes an initial state at $t=0$. Throughout this paper we only focus on the Hartle-Hawking initial state \cite{Hartle:1983ai}, which is prepared by the path integral on a half of the Euclidean dS$_{d+1} (\simeq \mathbb{B}^3)$ given by the Wick rotation $\tau_{\text{E}}=i\tau$: 
\begin{align}
    ds^2=R_{\text{dS}}^2(d\tau_{\text{E}}^2+\cos^2 \tau_{\text{E}}\, d\Omega_{d}^2).
\end{align}
Therefore the geometry can be seen as the gluing of the Lorentzian dS$_{d+1}$ and the Euclidean dS$_{d+1}$ along $\tau=\tau_{\text{E}}=0$ as depicted in figure \ref{fig:skt}. On the other hand, the right-hand side is a generating functional of correlation functions with $\phi_0$ being sources. It is defined by the Euclidean path integral with the measure $\int \mathcal{D}\Phi \,e^{-I_{\text{CFT}}[\Phi]}$, where $\Phi$ is a collection of fields of the CFT. Due to the relative $i$ factor in the exponential of the two path integrals in \eqref{dSGKPW}, we can predict that the dual CFT to dS may be non-unitary. Indeed, it is known that the central charge of the CFT$_{d}$ dual to dS$_{d+1}$ takes complex-valued \cite{Maldacena:2002vr}
\begin{align}
    c\sim (-i)^{d-1}\frac{R_{\text{dS}}^{d-1}}{G_N^{(d+1)}}.
\end{align}
This can be easily checked by using the relation of the AdS radius to the dS radius: $R_{\text{AdS}}=-iR_{\text{dS}}$. 
Therefore we can see that (at least when $d\neq1\ \text{mod}\ 4$) the dual CFT is a non-unitary theory. 
For example, when $d=2$ the Brown-Henneaux's formula \cite{Brown:1986nw} 
\begin{align}
    c_{\text{AdS}}=\frac{3R_{\text{AdS}}}{2G_N}
\end{align}
gives the central charge of the dual CFT to dS$_3$
\begin{align}
    c=-i\frac{3R_{\text{dS}}}{2G_N}=-ic_{\text{dS}},
\end{align}
where we have defined $c_{\text{dS}}\equiv 3R_{\text{dS}}/2G_N$.

We would like to define the entanglement entropy in the dual CFT, which is defined on $\mathbb{S}^{d}$ at the future $\tau=\tau_\infty\gg1$ with a large radius $\cosh \tau_\infty\simeq \frac{1}{2}e^{\tau_\infty}$. We choose an arbitrary direction in $\mathbb{S}^d$ as an imaginary time $t_{\text{E}}$ and take a subsystem $A$ on a time slice $t_{\text{E}}=0$. Note that the imaginary time $t_{\text{E}}$ is different from the time $\tau$ in the global dS with the metric \eqref{globaldS}. A state on the time slice is prepared by path integral as usual. However, since the CFT is non-unitary, a ket state $\ket{\Psi}$ and a bra state $\bra{\Psi}$, which are prepared by the path integrals of the past and future halves of $\mathbb{S}^d$ with the same insertions of operators, are different in general, i.e. $\bra{\Psi}\neq\ket{\Psi}^\dagger$. Therefore we should call the operator
\begin{align}
    \rho=\ket{\Psi}\bra{\Psi}
\end{align}
as a transition matrix instead of density matrix following \cite{Nakata:2021ubr}. This discussion is similar to that for timelike entanglement entropy in section \ref{subsec:formulation}. For this reason, henceforth we call the quantity 
\begin{align}
    \pe{A}=-\tr[\rho_A\log\rho_A],\qquad \rho_A=\tr_{\bar{A}}\rho
\end{align}
as pseudo entropy. Thus in dS/CFT, we expect that the holographic relation analogous to holographic entanglement entropy in AdS/CFT holds for pseudo entropy, which can be imaginary-valued, rather than entanglement entropy.

\subsection{Holographic pseudo entropy in \texorpdfstring{dS$_3$/CFT$_2$}{dS3/CFT2}}\label{sec:dscftpe}
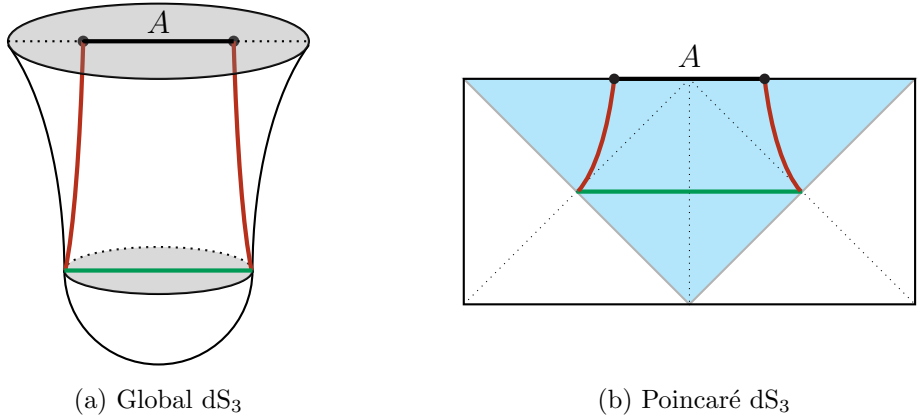
\begin{figure}[t]
    \begin{minipage}[b]{0.45\linewidth}
        \centering
        \begin{tikzpicture}[scale=0.5,thick]
            \begin{scope}
              \draw[yshift=-0.1cm] (4,0) arc (120:180:3 and 7);
              \draw[yshift=-0.1cm] (-4,0) arc (60:0:3 and 7);
            \end{scope}
              \draw (-2.5,-6.1) arc (180:360:2.5);
              \draw (-2.5,-6.1) arc (180:360:2.5 and 0.625);
              \draw[dotted] (2.5,-6.1) arc (0:180:2.5 and 0.625);
              \draw[draw=none,fill=gray,opacity=0.3] (0,-6.1) ellipse (2.5 and 0.625);
            \begin{scope}
                \draw[fill=Black,draw=none] (-2,0) circle (0.15);
                \draw[fill=Black,draw=none] (2,0) circle (0.15);
            \end{scope}
            \begin{scope}[ultra thick]
                \draw[BrickRed] (-2,0) arc (-20:-70:0.8 and 10.2);
                \draw[BrickRed] (2,0) arc (200:250:0.8 and 10.2);
                \draw[ForestGreen] (-2.5,-6.1)--(2.5,-6.1);
            \end{scope}
            \begin{scope}
              \draw (0,0) ellipse (4 and 1);
              \draw[draw=none,fill=gray,opacity=0.3] (0,0) ellipse (4 and 1);
              \draw[dotted] (-4,0)--(4,0);
              \draw[ultra thick] (-2,0)--(2,0);
            \end{scope}
            \begin{scope}
                \node at (0,0.5) {\large $A$};
            \end{scope}
        \end{tikzpicture}
        \subcaption{Global dS$_3$}
        \label{fig:dSgeodesicgl}
    \end{minipage}
    \begin{minipage}[b]{0.45\linewidth}
        \centering
        \begin{tikzpicture}
        \begin{scope}
            \begin{scope}[thick]
                    \draw (-3,1.5)--(3,1.5)--(3,-1.5)--(-3,-1.5)--cycle;
                    \draw[fill=Cyan,opacity=0.3] (-3,1.5)--(0,-1.5)--(3,1.5)--cycle;
            \end{scope}
            \begin{scope}
                    \draw[dotted] (-3,-1.5)--(0,1.5)--(3,-1.5);
                    \draw[dotted] (0,1.5)--(0,-1.5);
            \end{scope}
            \begin{scope}[ultra thick]
                \draw[ForestGreen] (-1.5,0)--(1.5,0);
                \draw[BrickRed] (-1,1.5) arc (-20:-70:0.8 and 2.5);
                \draw[BrickRed] (1,1.5) arc (200:250:0.8 and 2.5);
                \draw (-1,1.5)--(1,1.5);
            \end{scope}
            \node at (0,1.8) {\large $A$};
            \begin{scope}
                \draw[fill=Black,draw=none] (-1,1.5) circle (0.07);
                \draw[fill=Black,draw=none] (1,1.5) circle (0.07);
            \end{scope}
        \end{scope}
        \end{tikzpicture}
        \vspace{0.8cm}
        \subcaption{Poincar\'{e} dS$_3$}
        \label{fig:dSgeodesicpo}
    \end{minipage}
    \caption{The geodesic configuration whose length reproduces the pseudo entropy in the dual CFT. The red curve means the timelike geodesic, which contributes to the imaginary part of the pseudo entropy, and the green curve is the spacelike geodesic giving the real part, which goes through the boundary of the hemisphere $\mathbb{B}^3$. The left panel (a) shows the global dS$_3$ and the right panel (b) shows the Poincar\'{e} dS$_3$, which is depicted as the blue region, embedded into the Penrose diagram of the global dS$_3$.}
    \label{fig:dSgeodesic}
\end{figure}

In this subsection we restrict ourselves to the three-dimensional dS space given by 
\begin{align}\label{dSgl}
    ds^2=R_{\text{dS}}^2(-d\tau^2+\cosh^2\tau(dt_{\text{E}}^2+\cos^2t_{\text{E}} d\theta^2)),\qquad (\tau>0)
\end{align}
with the Euclidean part corresponding to the initial state
\begin{align}
    ds^2=R_{\text{dS}}^2(d\tau_{\text{E}}^2+\cos^2\tau_{\text{E}}(dt_{\text{E}}^2+\cos^2t_{\text{E}} d\theta^2)),\quad (0<\tau_{\text{E}}<\pi).
\end{align}
On the boundary $\tau=\tau_\infty$, we take a subsystem $A$ on a time slice $t_{\text{E}}=0$ as an arc of a region $-\phi_0/2<\phi<\phi_0/2$. In this case, the pseudo entropy of the CFT dual to dS$_3$ is expected to dual to the geodesic length. However, there seems to be a problem that we cannot connect the two edges $\partial A$ by a single geodesic because a geodesic emanating from an edge of $A$ should be timelike so that it cannot turn back to the other edge of $A$ in the global dS. Indeed, if we naively apply the geodesic formula, the length takes an imaginary value 
\begin{align}\begin{aligned}\label{dSgeodesic}
    D&=R_{\text{dS}}\cos^{-1}(\cosh^2\tau_\infty\cos\phi_0-\sinh^2\tau_\infty) \\
    &\simeq R_{\text{dS}}\cos^{-1}\left(-\frac{1}{2}e^{2\tau_\infty}\cos^2\frac{\phi_0}{2}\right)\\
    &=\pi R_{\text{dS}}-2iR_{\text{dS}}\log\left(e^{\tau_\infty}\cos\frac{\phi_0}{2}\right).
\end{aligned}\end{align}
Hence this is not a geodesic in the usual sense. Nevertheless we can give an interpretation to this result as follows. Remember that we are considering the Hartle-Hawking wave functional, i.e. the gluing of dS$_3$ and the Euclidean dS$_3$ at $\tau=\tau_{\text{E}}=0$. In this geometry, we have one possibility that a timelike geodesic emanating from an edge of $A$ goes through the Euclidean part of geometry and turns back to the other edge of $A$, as depicted in figure \ref{fig:dSgeodesicgl}. Let us assume this configuration and impose the ``extremality'' condition that both the real and imaginary parts of the length is extremized. 

We fix the points that the two timelike curves from $\phi_0/2$ and $-\phi_0/2$ will attach at $\tau=0$ as $\tilde{\phi}_1$ and $\tilde{\phi}_2$. First we consider the real part. The geodesic length of the Euclidean part is obtained by varying the length functional 
\begin{align}
    R_{\text{dS}}\int dt_{\text{E}}\sqrt{1+\cos^2t_{\text{E}}\left(\frac{d\phi(t_{\text{E}})}{dt_{\text{E}}}\right)^2}
\end{align}
with suitable boundary conditions.
This is the same problem with varying \eqref{TEEimarea} except for the absence of the overall $i$, so there is a condition 
\begin{align}\label{tildephicon}
    \tilde{\phi}_1-\tilde{\phi}_2=\pi
\end{align}
and the resulting length is 
\begin{align}\label{dSPEreal}
    \pi R_{\text{dS}}.
\end{align}
Therefore the geodesic takes the largest length in the hemisphere, depicted as a green line in figure \ref{fig:dSgeodesicgl}. 
Next we consider the imaginary part. The sum of geodesic length with fixing $\tilde{\phi}_1$ and $\tilde{\psi}_2$ is 
\begin{align*}
    R_{\text{dS}}\cos^{-1}&\left[\cosh\tau_\infty\cos\left(\frac{\phi_0}{2}-\tilde{\phi}_1\right)\right]+\cosh^{-1}\left[\cosh\tau_\infty\cos\left(\frac{\phi_0}{2}+\tilde{\phi}_2\right)\right] \\
    &\simeq iR_{\text{dS}}\log\left[e^{2\tau_\infty}\cos\left(\frac{\phi_0}{2}-\tilde{\phi}_1\right)\cos\left(\frac{\phi_0}{2}+\tilde{\phi}_2\right)\right]\\
    &=iR_{\text{dS}}\log\left[-\frac{1}{2}e^{2\tau_\infty}\left(\cos \phi_0+\cos 2\tilde{\phi}_1\right)\right],
\end{align*}
where in the second equality we have used the condition \eqref{tildephicon}. From the condition that the derivative with $\tilde{\phi}_1$ vanishes, we have $\sin 2\tilde{\phi}_1=0$, leading to 
\begin{align}
    \tilde{\phi}_1=\frac{\pi}{2},\qquad \tilde{\phi}_2=-\frac{\pi}{2}.
\end{align}
Then the contribution from the timelike geodesic reads 
\begin{align}\label{dSPEim}
    -2iR_{\text{dS}}\log\left(e^{\tau_\infty}\sin\frac{\phi_0}{2}\right).
\end{align}
Thus the sum of \eqref{dSPEreal} and \eqref{dSPEim} is identical to the naive calculation \eqref{dSgeodesic}. 

From the calculations above, the holographic pseudo entropy is 
\begin{align}\begin{aligned}\label{3ddSPE}
    \pe{A}&=\frac{1}{4G_N}\left[-2iR_{\text{dS}}\log\left(e^{\tau_\infty}\sin\frac{\phi_0}{2}\right)+\pi R_{\text{dS}}\right] \\
    &=-i\frac{c_{\text{dS}}}{3}\log\left(\frac{2\sin\frac{\phi_0}{2}}{\epsilon}\right)+\frac{\pi c_{\text{dS}}}{6},\quad (\epsilon\equiv 2e^{-\tau_\infty}),
\end{aligned}\end{align}
where $c_{\text{dS}}\equiv 3R_{\text{dS}}/2G_N$. Thus the pseudo entropy always takes a complex value.

Another observation from the result and the geodesic configuration in this subsection is the similarity with the timelike entanglement entropy in AdS/CFT we discussed in subsection \ref{subsec:extremalsurface}. In the next subsection, we will elaborate the relation between the two notions.

Finally let us describe an interpretation of the result in the Poincare dS$_3$
\begin{align}
    ds^2=R_{\text{dS}}^2\frac{-d\eta^2+dt_{\text{E}}^2+dx^2}{\eta^2}.
\end{align}
In the Poincar\'{e} dS$_3$, the geodesic configuration looks like the one depicted in figure \ref{fig:dSgeodesicpo}, where the blue region is the Poincar\'{e} patch embedded into the global dS$_3$. In that figure, the timelike geodesics (green ones) satisfy
\begin{align}
    -\eta^2+x^2=\frac{T_0^2}{4},
\end{align}
whose area is
\begin{align}
    -2iR_{\text{dS}}\log\frac{X_0}{\epsilon},
\end{align}
where we took a cutoff at $\eta=\epsilon$.
The red line in figure \ref{fig:dSgeodesicpo} is spacelike and the length is 
\begin{align}
    \pi R_{\text{dS}}.
\end{align}
Therefore the sum of the lengths matches the global dS$_3$ calculation by identifying $X_0=2\sin\frac{\phi_0}{2}$.

\subsection{Relation to timelike entanglement entropy}
In this subsection, we relate the pseudo entropy in dS/CFT to the timelike entanglement entropy in AdS/CFT. We will see that pseudo entropy for dS/CFT and timelike entanglement entropy in AdS/CFT can be regarded as the two different analytic continuations from the Euclidean AdS (EAdS). 

Let us first consider the analytic continuation from EAdS$_3$ to dS$_3$. For simplicity, we consider the Poincar\'{e} coordinates in EAdS$_3$
\begin{align}\label{PoincareEAdS}
    ds^2=R_{\text{AdS}}^2\frac{dz^2+dt_{\text{E}}^2+dx^2}{z^2}.
\end{align}
For later convenience, now we regard $x$ as the time direction and take a subsystem $A$ as an interval with length $T_0$ on the cutoff surface $z=\epsilon_{\text{AdS}}$. Of course, whether we take the subsystem as ``temporal'' or ``spatial'' is not essential in this phase because we are considering the Euclidean theory. The entanglement entropy for $A$ is 
\begin{align}\label{EAdSEE}
    S_A=\frac{c_{\text{AdS}}}{3}\log \left(\frac{T_0}{\epsilon_{\text{AdS}}}\right),
\end{align}
where $c_{\text{AdS}}$ is the central charge of the CFT dual to EAdS and in the bulk language $c_{\text{AdS}}=3R_{\text{AdS}}/2G_{N}$. The analytic continuation to dS is accomplished by Wick-rotating the radial coordinate $z$ as
\begin{align}
    z=-i\eta
\end{align}
and simultaneously the radius as\footnote{Note that this is a different convention from the one adopted in \cite{Hikida:2022ltr}.}
\begin{align}\label{radiusconti}
    R_{\text{AdS}}=-iR_{\text{dS}}.
\end{align}
Indeed, from these Wick rotations the metric \eqref{PoincareEAdS} is continued to the metric of dS in Poincar\'{e} coordinate
\begin{align}
    ds^2=R_{\text{dS}}^2\frac{-d\eta^2+dt_{\text{E}}^2+dx^2}{\eta^2}.
\end{align}
Let us take this analytic continuation for the entanglement entropy \eqref{EAdSEE}. The cutoff and the central charge are deformed as 
\begin{align}
    \epsilon_{\text{AdS}}&=-i\epsilon_{\text{dS}}, \label{cutoffconti} \\
    c_{\text{AdS}}&=\frac{3R_{\text{AdS}}}{2G_N}=-i\frac{3R_{\text{dS}}}{2G_N}\equiv -ic_{\text{dS}},
\end{align}
then the pseudo entropy becomes 
\begin{align}
    \pe{A}=-i\frac{c_{\text{dS}}}{3}\log\left(\frac{T_0}{\epsilon_{\text{dS}}}\right)+\frac{\pi c_{\text{dS}}}{6},
\end{align}
which is consistent with a result of \cite{Hikida:2022ltr}. Note that we choose the branch of the logarithmic function as taking the conventional principal value $-\pi<\arg z\le\pi$.

Next we consider another analytic continuation which takes EAdS to the Lorentzian AdS. We adopt the usual Wick rotation 
\begin{align}
    t_{\text{E}}=it
\end{align}
and simultaneously replacing the parameter $T_0$ with
\begin{align}
    T_0\to iT_0.
\end{align}
In this case, the cutoff and the central charge do not change. Therefore we have 
\begin{align}
    \tee{A}=\frac{c_{\text{AdS}}}{3}\log\left( \frac{T_0}{\epsilon_{\text{AdS}}}\right)+\frac{i\pi c_{\text{AdS}}}{6}.
\end{align}

\begin{figure}
    \centering
    \begin{tikzpicture}
        \draw[<->,very thick] (-1,0)--(1,0);
        \begin{scope}[xshift=-4cm,scale=0.8]
            \begin{scope}[thick]
                    \draw (-3,1.5)--(3,1.5)--(3,-1.5)--(-3,-1.5)--cycle;
                    \draw[fill=Cyan,opacity=0.3] (-3,1.5)--(0,-1.5)--(3,1.5)--cycle;
            \end{scope}
            \begin{scope}
                    \draw[dotted] (-3,-1.5)--(0,1.5)--(3,-1.5);
                    \draw[dotted] (0,1.5)--(0,-1.5);
            \end{scope}
            \begin{scope}[ultra thick]
                \draw[ForestGreen] (-1.5,0)--(1.5,0);
                \draw[BrickRed] (-1,1.5) arc (-20:-70:0.8 and 2.5);
                \draw[BrickRed] (1,1.5) arc (200:250:0.8 and 2.5);
                \draw (-1,1.5)--(1,1.5);
            \end{scope}
            \node at (0,1.8) {\large $A$};
            \begin{scope}
                \draw[fill=Black,draw=none] (-1,1.5) circle (0.07);
                \draw[fill=Black,draw=none] (1,1.5) circle (0.07);
            \end{scope}
            \node at (0,-2.5) {\large dS$_3$};
        \end{scope}
        \begin{scope}[xshift=3cm,rotate=-90,scale=0.8]
            \begin{scope}[thick]
                    \draw (-3,1.5)--(3,1.5)--(3,-1.5)--(-3,-1.5)--cycle;
                    \draw[fill=Cyan,opacity=0.3] (-3,1.5)--(0,-1.5)--(3,1.5)--cycle;
            \end{scope}
            \begin{scope}
                    \draw[dotted] (-3,-1.5)--(0,1.5)--(3,-1.5);
                    \draw[dotted] (0,1.5)--(0,-1.5);
            \end{scope}
            \begin{scope}[ultra thick]
                \draw[BrickRed] (-1.5,0)--(1.5,0);
                \draw[ForestGreen] (-1,1.5) arc (-20:-70:0.8 and 2.5);
                \draw[ForestGreen] (1,1.5) arc (200:250:0.8 and 2.5);
                \draw (-1,1.5)--(1,1.5);
            \end{scope}
            \node at (0,1.8) {\large $A$};
            \begin{scope}
                \draw[fill=Black,draw=none] (-1,1.5) circle (0.07);
                \draw[fill=Black,draw=none] (1,1.5) circle (0.07);
            \end{scope}
            \node at (4,0) {\large AdS$_3$};
        \end{scope}
    \end{tikzpicture}
    \caption{The analytic continuation of dS$_3$ to AdS$_3$. This is accomplished by the double Wick rotation \eqref{doubleWick}.}
    \label{fig:doubleWick}
\end{figure}
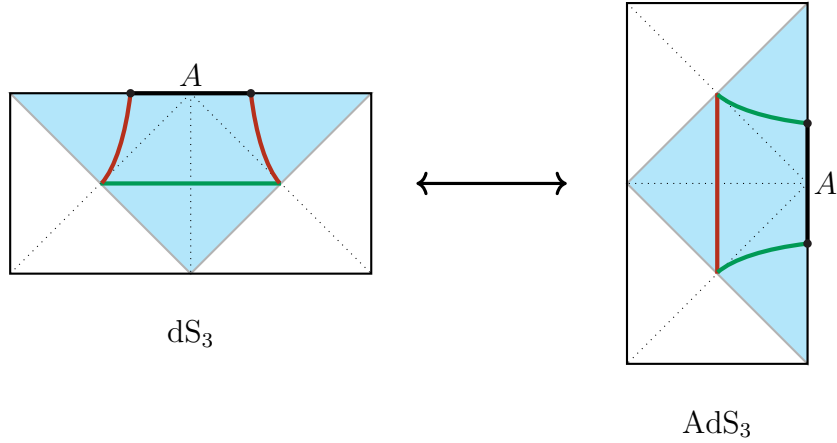

Intuitively, we can interpret the relation between the two notions as connected by the ``double Wick rotation''
\begin{align}\label{doubleWick}
    \eta\to z=-i\eta,\qquad t_{\text{E}}\to t=-it_{\text{E}}.
\end{align}
We can regard this as rotating the Penrose diagram by $90^{\circ}$ (see figure \ref{fig:doubleWick}) and the additional parts of the entropy as the effect of the boundary condition at the center of the diagram.

Next, we discuss in terms of the global coordinates. Since we are regarding $t_{\text{E}}$ as a spatial coordinate, Let us start with the global dS$_3$ \eqref{dSgl} with swapping $t_{\text{E}}$ and $\theta$:
\begin{align}
    ds^2=R_{\text{dS}}^2(-d\tau^2+\cosh^2\tau(d\theta^2+\cos^2\theta dt_{\text{E}}^2)).
\end{align}
As discussed above, the holographic pseudo entropy for a subregion of $-T_0/2<t_{\text{E}}<T_0/2$ takes 
\begin{align}
    \pe{A}=-i\frac{c_{\text{dS}}}{3}\log\left(\frac{2\sin\frac{T_0}{2}}{\epsilon_{\text{dS}}}\right)+\frac{\pi c_{\text{dS}}}{6},
\end{align}
where $\epsilon_{\text{dS}}\equiv 2e^{-\tau_\infty}$. We perform the following analytic continuation
\begin{align}\label{dSPEgl}
    \rho=i\theta,\quad R_{\text{dS}}=iR_{\text{AdS}}
\end{align}
and replace $\tau\to\eta$ and $t_{\text{E}}\to t$, then the metric becomes 
\begin{align}
    ds^2=R_{\text{AdS}}^2(d\eta^2+\cosh^2\eta(-\cosh^2\rho dt^2+d\rho^2)).
\end{align}
Correspondingly, we replace as $\epsilon_{\text{dS}}\to\epsilon_{\text{AdS}},c_{\text{dS}}\to ic_{\text{AdS}}$ in \eqref{dSPEgl}, then 
\begin{align}
    \tee{A}=\frac{c_{\text{AdS}}}{3}\log\left(\frac{2\sin\frac{T_0}{2}}{\epsilon_{\text{AdS}}}\right)+\frac{i\pi c_{\text{AdS}}}{6}
\end{align}

\subsection{Higher dimensions}
Let us move on to extension to higher dimensions. We consider the Poincar\'{e} dS$_{d+1}$
\begin{align}
    ds^2=R_{\text{dS}}^2\frac{-d\eta^2+dt_{\text{E}}^2+dy^2+d\mathbf{x}^2}{\eta^2}.
\end{align}
As in the previous subsection, we regard $t_{\text{E}}$ as a spatial direction for later convenience. We take a subsystem $A$ as a sphere with a radius $T_0/2$ on a slice $y=0$. It is useful to introduce a radial coordinate $\xi\equiv \sqrt{t_{\text{E}}^2+\mathbf{x}^2}$, then the subsystem is defined by $\xi<T_0/2$.

Analogous to the three-dimensional case, we consider the timelike surface
\begin{align}
    -\eta^2+\xi^2=\frac{T_0^2}{4}.
\end{align}
The area of this surface is 
\begin{align}
    R_{\text{dS}}^{d-1}\text{Vol}(\mathbb{S}^{d-2})\int_\epsilon^\infty d\eta\frac{\xi^{d-2}}{\eta^{d-1}}\sqrt{-1+\xi'(\eta)^2}&=i\frac{T_0}{2}R_{\text{dS}}^{d-1}\text{Vol}(\mathbb{S}^{d-2})\int_\epsilon^\infty d\eta \frac{(\eta^2+T_0^2/4)^{\frac{d-3}{2}}}{\eta^{d-1}}\\
    &=iR_{\text{dS}}^{d-1}\text{Vol}(\mathbb{S}^{d-2})\int_{\frac{2\epsilon}{T_0}}^\infty du \frac{(u^2+1)^{\frac{d-3}{2}}}{u^{d-1}}
\end{align}
The integral is identical to \eqref{ddimreal}. Similarly, the real part is given by a geodesic 
\begin{align}
    \eta^2-\xi^2=\frac{T_0^2}{4},
\end{align}
whose area is evaluated as 
\begin{align}
    R_{\text{dS}}^{d-1}\text{Vol}(\mathbb{S}^{d-2})\int_1^\infty d\eta\frac{(\eta^2-1)^{\frac{d-3}{2}}}{\eta^{d-1}}=R_{\text{dS}}^{d-1}\text{Vol}(\mathbb{S}^{d-2})\frac{\sqrt{\pi}\Gamma\left(\frac{d-1}{2}\right)}{2\Gamma\left(\frac{d}{2}\right)}
\end{align}
Therefore we have obtained the pseudo entropy 
\begin{align}\begin{aligned}\label{dSPEddim}
    \pe{A}&=\frac{R_{\text{dS}}^{d-1}}{4G_N^{(d+1)}}\text{Vol}(\mathbb{S}^{d-2})\frac{\sqrt{\pi}\Gamma\left(\frac{d-1}{2}\right)}{2\Gamma\left(\frac{d}{2}\right)}\\
    &+i\frac{R_{\text{dS}}^{d-1}}{4G_N^{(d+1)}}\text{Vol}(\mathbb{S}^{d-2})
    \left\{\begin{aligned}
    \quad\sum_{k=0}^{\frac{d-3}{2}}&\binom{\frac{d-3}{2}}{k}\frac{1}{d-2k-2}\left(\frac{T_0}{2\epsilon}\right)^{d-2k-2} & (d:\text{odd}) \\
    \sum_{k=0}^{\frac{d}{2}-2}&\binom{\frac{d-3}{2}}{k}\frac{1}{d-2k-2}\left(\frac{T_0}{2\epsilon}\right)^{d-2k-2}+\frac{\Gamma\left(\frac{d-1}{2}\right)}{\sqrt{\pi}\Gamma\left(\frac{d}{2}\right)}\log\frac{T_0}{2\epsilon} & (d:\text{even})
    \end{aligned}\right.
\end{aligned}\end{align}
We can easily check that this result is related to the timelike entanglement entropy \eqref{dEEwickodd} and \eqref{dEEwickeven} in AdS$_{d+1}$/CFT$_d$ by an analytic continuation
\begin{align}
    R_{\text{dS}}\to iR_{\text{AdS}},\quad \epsilon_{\text{dS}} \to i\epsilon_{\text{AdS}},\quad T_0\to iT_0,\quad \text{Vol}(\mathbb{S}^{d-2})\to (-i)^{d-2}\text{Vol}(\mathbb{H}^{d-2}).
\end{align}

Next, let us discuss the extremal surface in the global dS$_{d+1}$. The following discussion is parallel with that in section \ref{subsec:extremalsurface}. As a generalization of \eqref{dSgl}, we consider dS$_{d+1}$ with a metric
\begin{align}
    ds^2=R_{\text{dS}}^2\left(-d\tau^2+\cosh^2\tau\left(dt_{\text{E}}^2+\sin^2t_{\text{E}}\,d\Omega_{d-1}^2\right)\right),
\end{align}
where $d\Omega_{d-1}^2$ denotes the metric of the unit sphere $\mathbb{S}^{d-1}$. We regard a coordinate $\theta$ of $\mathbb{S}^{d-1}$ with $d\Omega_{d-1}^2=d\theta^2+\cos^2\theta\,d\Omega_{d-2}$ as a time direction. As described above, this is glued with the Euclidean geometry with the metric 
\begin{align}
    ds^2=R_{\text{dS}}^2\left(d\tau_{\text{E}}^2+\cos^2\tau_{\text{E}}(dt_{\text{E}}^2+\sin^2t_{\text{E}}\,d\Omega_{d-1}^2)\right)
\end{align}
at $\tau=0$.

First let us consider the Lorentzian part. We consider the codimension-2 surface given by $t_{\text{E}}=t_{\text{E}}(\tau)$ on $\theta=0$. The induced metric for the surface is
\begin{align}
    ds^2=R_{\text{dS}}^2\left[(-1+\cosh^2\tau\, t_{\text{E}}'^2)d\tau^2+\cosh^2\tau\sin^2 t_{\text{E}}\,d\Omega_{d-2}^2\right].
\end{align}
Then the pseudo entropy is evaluated as the extremal value of a functional
\begin{align}
    \frac{R_{\text{dS}}^{d-1}}{4G_N^{(d+1)}}\text{Vol}(\mathbb{S}^{d-2})\int d\tau\sqrt{-1+\cosh^2\tau\, t_{\text{E}}'(\tau)^2}\cosh^{d-2}\tau\sin^{d-2}t_{\text{E}}.
\end{align}
In the same manner as section \ref{subsec:highgl}, we can obtain the solution of the Euler-Lagrange equation with a boundary condition $t_{\text{E}}(\tau\to\infty)=T_0/2$:
\begin{align}
    \cos t_{\text{E}}=\cos \frac{T_0}{2}\tanh\tau.
\end{align}
Note that for $\tau=\frac{\pi}{2}$, the surface ends on $t_{\text{E}}=0$.
Inserting this solution, we can evaluate the contribution from the timelike surface 
\begin{align}
    i\frac{R_{\text{dS}}^{d-1}}{4G_N^{(d+1)}}&\text{Vol}(\mathbb{S}^{d-2})\int_0^{\sin\frac{T_0}{2}\sinh\tau_\infty}du(1+u^2)^{\frac{d-3}{2}}=i\frac{R_{\text{dS}}^{d-1}}{4G_N^{(d+1)}}\text{Vol}(\mathbb{S}^{d-2})\\
    &\times\left\{\begin{aligned}\label{ddimrealds}
    \quad\sum_{k=0}^{\frac{d-3}{2}}&\binom{\frac{d-3}{2}}{k}\frac{1}{d-2k-2}\left(\frac{\sin\frac{T_0}{2}}{2\epsilon}\right)^{d-2k-2} & (d:\text{odd}) \\
    \sum_{k=0}^{\frac{d}{2}-2}&\binom{\frac{d-3}{2}}{k}\frac{1}{d-2k-2}\left(\frac{\sin\frac{T_0}{2}}{2\epsilon}\right)^{d-2k-2}+\frac{\Gamma\left(\frac{d-1}{2}\right)}{\sqrt{\pi}\Gamma\left(\frac{d}{2}\right)}\log\left(\frac{\sin\frac{T_0}{2}}{\epsilon}\right) & (d:\text{even})
    \end{aligned}\right.
\end{align}

Next we move on to the Euclidean part. The induced metric for a surface $t_{\text{E}}=t_{\text{E}}(\tau_{\text{E}})$ defined on a $\theta=0$ slice is 
\begin{align}
    ds^2=R_{\text{dS}}^2\left[(1+\cos^2\tau_{\text{E}}\,t_{\text{E}}'^2)\,d\tau_{\text{E}}^2+\cos^2\tau_{\text{E}}\sin^2t_{\text{E}}\,d\Omega_{d-2}^2\right],
\end{align}
so we would like to evaluate the extremal value of
\begin{align}
    \frac{R_{\text{dS}}^{d-1}}{4G_N^{(d+1)}}\text{Vol}(\mathbb{S}^{d-2})\int d\tau_{\text{E}}\sqrt{1+\cos^2\tau_{\text{E}}\,t_{\text{E}}'^2}\cos^{d-2}\tau_{\text{E}}\sin^{d-2}t_{\text{E}}.
\end{align}
Varying this integral, we obtain the solutions
\begin{align}
    \cos t_{\text{E}}=C\tan\tau_{\text{E}},
\end{align}
where $C$ is an arbitrary constant. The contribution from these solutions is
\begin{align}
    \frac{R_{\text{dS}}^{d-1}}{4G_N^{(d+1)}}\text{Vol}(\mathbb{S}^{d-2})\frac{\sqrt{\pi}\Gamma\left(\frac{d-1}{2}\right)}{2\Gamma\left(\frac{d}{2}\right)}.
\end{align}

Thus the holographic pseudo entropy have been obtained by 
\begin{align}
    \pe{A}=&\frac{R_{\text{dS}}^{d-1}}{4G_N^{(d+1)}}\text{Vol}(\mathbb{S}^{d-2})\frac{\sqrt{\pi}\Gamma\left(\frac{d-1}{2}\right)}{2\Gamma\left(\frac{d}{2}\right)}+i\frac{R_{\text{dS}}^{d-1}}{4G_N^{(d+1)}}\text{Vol}(\mathbb{S}^{d-2})\\
    &\times\left\{\begin{aligned}
    \quad\sum_{k=0}^{\frac{d-3}{2}}&\binom{\frac{d-3}{2}}{k}\frac{1}{d-2k-2}\left(\frac{\sin\frac{T_0}{2}}{2\epsilon}\right)^{d-2k-2} & (d:\text{odd}) \\
    \sum_{k=0}^{\frac{d}{2}-2}&\binom{\frac{d-3}{2}}{k}\frac{1}{d-2k-2}\left(\frac{\sin\frac{T_0}{2}}{2\epsilon}\right)^{d-2k-2}+\frac{\Gamma\left(\frac{d-1}{2}\right)}{\sqrt{\pi}\Gamma\left(\frac{d}{2}\right)}\log\left(\frac{\sin\frac{T_0}{2}}{\epsilon}\right) & (d:\text{even})
    \end{aligned}\right.
\end{align}
which can be regarded as a generalization of the three-dimensional case \eqref{3ddSPE}.

Here we make a comment on a proposed holographic calculation of EE \cite{Narayan:2015vda,Sato:2015tta} in dS$_4$/CFT$_3$, in which a concrete example is known; a duality between a higher-spin gravity on dS$_4$ and large $N$ $Sp(N)$ vector model \cite{Anninos:2011ui}. The authors of \cite{Narayan:2015vda,Sato:2015tta} claimed that the leading term of the holographic entanglement entropy in dS$_{d+1}$/CFT$_d$ has an overall factor $(-i)^{d-1}$, so that the entanglement entropy takes real-valued when $d=3$. For example, the leading area term takes the form
\begin{align}\label{N-S}
    \frac{R_{\text{dS}}^2}{4G_N^{(4)}}\frac{\text{Vol}(\mathbb{S}^1)}{\epsilon}
\end{align}
One might think that this is not consistent with our calculation, in which the leading term always takes imaginary-valued. However, this is actually consistent with the imaginary part of our calculation. To see this, we have to be careful for definition of the cutoff $\epsilon$. As noted above, the cutoffs in AdS and dS are related by $\epsilon_{\text{AdS}}=-i\epsilon_{\text{dS}}$. In fact, the cutoff $\epsilon$ in \eqref{N-S} is the one defined in AdS, i.e. $\epsilon_{\text{AdS}}$. Therefore if we rewrite \eqref{N-S} by using $\epsilon_{\text{dS}}$, then the result is imaginary and consistent with the leading term of our calculation. 

On the other hand, the real part in \eqref{dSPEddim} is a new ingredient of our calculation. We can see that the real part 
\begin{align}
    \frac{R_{\text{dS}}^{d-1}}{4G_N^{(d+1)}}\text{Vol}(\mathbb{S}^{d-2})\frac{\sqrt{\pi}\Gamma\left(\frac{d-1}{2}\right)}{2\Gamma\left(\frac{d}{2}\right)}
\end{align}
is identical to a half of the de Sitter entropy in dS$_{d+1}$.

\section{Conclusions and discussions}\label{sec:6}

In this paper we introduced a new quantity called timelike entanglement entropy (EE) and studied its properties from both the field theoretic and the holographic viewpoint. The timelike EE $\tee{A}$ is defined by changing a spacelike subsystem $A$ into a timelike one. This is also equivalent to flipping the role of the time and space coordinates. We present analytical expressions of timelike EE in two dimensional CFTs based on both the replica method analysis and the numerical computation in free field theories. This quantity has a universal imaginary part and can be regarded as a special example of pseudo entropy. 

Next we considered a holographic calculation of timelike EE by extending the holographic entanglement entropy. Through calculations of holographic timelike EE in the basic examples of pure AdS$_3$ and BTZ, we propose the following definition for the holographic timelike entanglement entropy. Given a timelike boundary region $A$ we consider the union of timelike and spacelike extremal surfaces which form a single simply connected surface which is homologous to $A$. We call such a surface a ``path" and the collection of all such paths $\{\Gamma_A\}$. Over this space we vary the joining points and keep only those paths which are stationary with respect to this variation $\{\Gamma^s_A\}$. In general, these paths include  timelike geodesics which produce the imaginary part of timelike EE. This implies that the time coordinate emerges from the imaginary part of the timelike EE (or pseudo entropy)  generalizing the familiar idea that the space coordinate emerges from the entanglement entropy.  

In the examples of section \ref{sec:3} we did not construct all possible paths. Instead, only those paths were considered which allowed for the imaginary part to be maximal which matched the expectation coming from the boundary calculation via Wick rotation. This essentially forced the joining parts to lie as far to past or future as possible such that they occurred on horizons or singularities. In order for the answers to agree this was essentially necessary. It is possible that more care will be needed in defining the correct set of paths to optimize with respect to. It is possible that a generalization of the `Hartle-Hawking' like geometry shown in figure \ref{fig:global_HH}, which provided the specific locations of joining points of paths in global AdS$_3$, may provide additional insight.

Also in principle there may be multiple stationary paths with different complex valued areas in which case it is necessary to distinguish between them in order to correctly determine the path that corresponds to the timelike entanglement entropy. This already happened for the two party timelike entanglement entropies which exhibit a similar connected/disconnected transition compared with their spacelike counterparts. However, when considering the timelike mutual information for the examples presented we found it as defined to always be entirely real. As such for this specific situation it seems natural to select the saddle for which the timelike mutual information is maximized. This is however a prescription highly specific to the setups considered. One would desire a set of concrete rules to distinguish between stationary paths of different complex valued area. Presumably such a general prescription for multiple saddles will include some comparison of the relative magnitudes of the imaginary and real components of the areas of these saddles. At the present we are agnostic as to what the correct prescription is, additional explicit examples which include multiple saddles are needed to make further progress\footnote{See for example \cite{Li:2022tsv} which appeared while this paper was in preparation. Given multiple stationary paths the authors suggest that only those with the smallest imaginary component should be considered and among these the one of smallest real component selected. At present it is not clear to us that their or our examples can fully distinguish between this or other possible proposals.}. We leave such explorations to future work.

For the shock wave geometry and local operator quench we found the surprising result that the timelike entanglement entropy of a single region on one boundary is dependent on the shock wave and is capable of probing beyond the horizon. This conclusion was confirmed by explicit CFT calculations which implement the careful Wick rotation of the boundary CFT. This demonstrates that to determine the timelike entanglement entropy it is generally incorrect to simply analytically continue the entanglement entropy and more care must be taken. As a consequence the timelike entanglement entropy could play a larger role in our understanding of black hole information and the interior and the emergence of spacetime geometry from entanglement. 

We also studied a higher dimensional generalization of holographic timelike EE. This is straightforward for hyperbolic subsystems. However, the holographic calculation turned out to be very non-trivial for strip shaped subsystems, where a naive analytic continuation of the holographic result from the spacelike subsystem to timelike one is difficult to understand from the extremal surface viewpoint. We leave a better understanding of this issue for a future problem.

Finally we also found that if we apply the idea of holographic EE to the dS/CFT, we obtain a complex-valued entropy as the extremal surface consisting of both timelike and spacelike pieces. This looks analogous to the holographic timelike EE, though the imaginary part is universal in the latter, while it is not in dS/CFT. We argued that this complex-valued entropy can also be properly understood as the pseudo entropy. We also pointed out that the holographic pseudo entropy in dS/CFT is related to the holographic timelike EE in AdS/CFT via a double Wick rotation. Furthermore, we found that the real part of the holographic pseudo entropy can be interpreted as a half of the dS entropy in any dimensions. Recently, it was argued \cite{Chandrasekaran:2022cip} using algebraic considerations that the dS entropy has the maximal entropy. Although their definition of entropy is distinct from ours, relating these discussions to that of pseudo entropy is an interesting direction for possible future work.

\section*{Acknowledgements}

We are grateful to Xi Dong, Daniel Harlow, Patrick Hayden, Juan Maldacena and Jonathan Oppenheim for useful comments on pseudo entropy and timelike entanglement entropy. 
This work is supported by the Simons Foundation through the ``It from Qubit'' collaboration
and by MEXT KAKENHI Grant-in-Aid for Transformative Research Areas (A) through the ``Extreme Universe'' collaboration: Grant Number 21H05187.
This work is also supported by Inamori Research Institute for Science and by JSPS Grant-in-Aid for Scientific Research (A) No.~21H04469. Y.\,T. is supported by Grant-in-Aid for JSPS Fellows No.\,22J21950. TT would like to thank the 2022 Simons Collaboration on It from Qubit Annual Meeting where a part of the present work was presented. Some of the numerical calculations were carried out on Yukawa-21 at YITP in Kyoto University.

\appendix

\section{Calculation of thermofield mutual information for holographic CFTs}\label{apend:tmi}

In this appendix we derive the thermofield mutual information for a 2d holographic CFT. For a thermal state of inverse temperature $\beta$ the boundary manifold is a cylinder of radius $\beta$. We consider two intervals $A=[u_a,v_a]$ and $B=[u_b,v_b]$ with
\be
u_A=a, \quad v_A=b, \quad u_B=c+\frac{i\beta}{2}, \quad v_B=d+\frac{i\beta}{2}, \quad a<b, \quad c<d
\ee
this has the effect of placing the interval $B$ on the opposite side of the cylinder. Such a set up corresponds to placing one interval on each of the boundaries of the dual bulk geometry which is the BTZ black hole \cite{2013JHEP...07..081M}.

\begin{figure}[H]
    \centering
    \includegraphics[width=.5\textwidth,page=14]{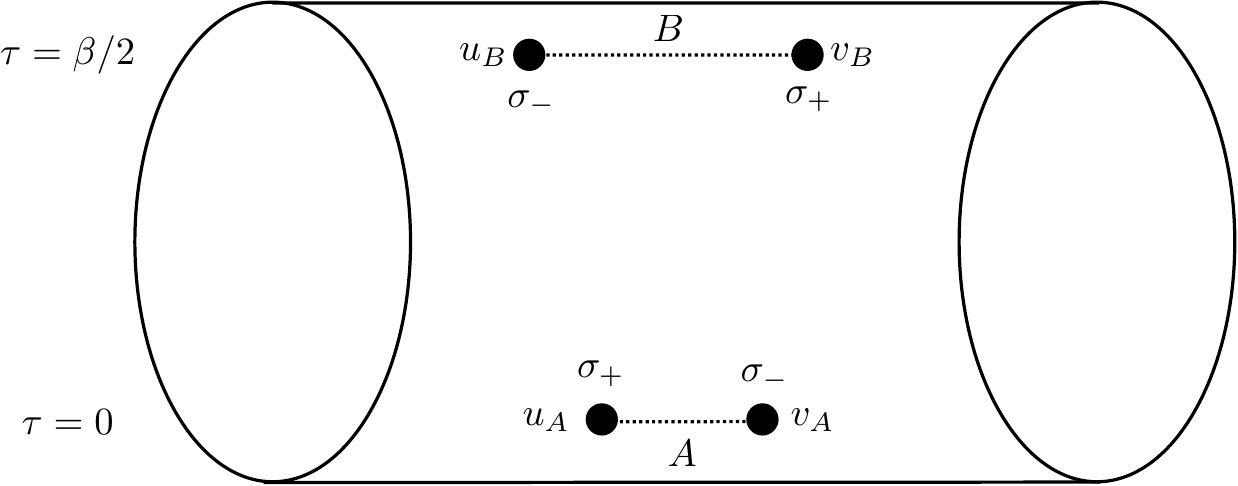}
    \caption{The boundary manifold with twist operator insertions for the calculation of $\tr(\rho_{AB}^n)$. The two intervals are placed a distance $\frac{\beta}{2}$ apart in euclidean time. This corresponds to one interval being located on each of the two boundaries of the BTZ black hole. Because the time runs in opposite directions on the two boundaries the correct prescription requires that the twist operators $\sigma_\pm$ be placed in the opposite orientation on one side. }
    \label{fig:PI_TMMI}
\end{figure}
\noindent The thermofield mutual information can be calculated using the replica method where twist operators $\sigma_{\pm}$ are inserted at the end points of boundary regions. Note the twist operators must be inserted with opposite ordering on one side due to the opposing time orientation of the two boundaries (see e.g. \cite{2013JHEP...07..081M,2015JHEP...08..011C}). The correct prescription is shown in figure \ref{fig:PI_TMMI}.

The thermofield mutual information is defined as the mutual information for these regions and is written in terms of the twist operators as
\be
\begin{split}
J_{AB}&=\lim_{n\rightarrow1}\frac{1}{1-n}\log\left[\frac{\tr(\rho_{A}^n)\tr(\rho_{B}^n)}{\tr(\rho_{AB}^n)}\right]\\
&=\lim_{n\rightarrow1}\frac{1}{1-n}\log\left[\frac{\langle\sigma^+_n(u_A)\sigma^-_n(v_A)\rangle_\beta\langle\sigma^+_n(u_B)\sigma^-_n(v_B)\rangle_\beta}{\langle\sigma^+_n(u_A)\sigma^-_n(v_A)\sigma^-_n(u_B)\sigma^+_n(v_B)\rangle_\beta}\right]
\end{split}
\ee
where $n$ is the number of replicas which we will analytically continued and then take to be one. To start we consider the two point functions which are entirely fixed by conformal symmetry and given by
\be
\langle\sigma^+_n(u_i)\sigma^-_n(v_i)\rangle_\beta=\left|\frac{\pi\epsilon}{\beta S_{u_iv_i}}\right|^{2\Delta_n}.
\ee
Here the conformal dimension $\Delta_n$ of the twist operators is given by
\be
\Delta_n=\frac{c}{12}\frac{(n-1)(n+1)}{n}
\ee
and for convenience we have defined
\be
S_{ij}=\sinh(\frac{\pi}{\beta}(i-j)), \quad C_{ij}=\cosh(\frac{\pi}{\beta}.(i-j)).
\ee
It is also necessary to consider the 4-point function
\be
\langle\sigma^+_n(u_A)\sigma^-_n(v_A)\sigma^-_n(u_B)\sigma^+_n(v_B)\rangle_\beta
\ee
We make use the conformal map $z=e^{\frac{2\pi w}{\beta}}$ which maps the cylinder to the plane
\be
\left|\left(\frac{2\pi\epsilon}{\beta}\right)^{4}e^{\frac{2\pi}{\beta}(u_A+v_A+u_B+v_B)}\right|^{\Delta_n}\langle\sigma^+_n(e^{\frac{2\pi}{\beta }u_A})\sigma^-_n(e^{\frac{2\pi}{\beta }v_A})\sigma^-_n(e^{\frac{2\pi}{\beta }u_B})\sigma^+_n(e^{\frac{2\pi}{\beta }v_B})\rangle
\ee
where for a function $f(\phi)$ correlation functions of primary operators transform as
\be
\langle\Phi(f(\phi_1))\cdots\Phi(f(\phi_n))\rangle=\prod_j^n\left|f'(\phi_j)\right|^{-\Delta_j}\langle\Phi(\phi_1)\cdots\Phi(\phi_n)\rangle.
\ee
In order to evaluate the 4-point function
\be
\langle\sigma^+_n(z_1)\sigma^-_n(z_2)\sigma^-_n(z_3)\sigma^+_n(z_4)\rangle
\ee
we need the twist operators to take the usual ordering. We make use of the Mobius transformation
\be
q(z)=\frac{(z-z_1)(z_4-z_3)}{(z-z_3)(z_4-z_1)}
\ee
which maps $z_1\rightarrow 0$, $z_3\rightarrow \infty$, $z_4\rightarrow 1$. $z_2$ determines the cross-ratio $x$ which is real with $0\leq x\leq1$ and given by
\be
x_s=x=\frac{S_{ba}S_{dc}}{C_{da}C_{cb}}, \quad x_t=1-x=\frac{C_{ca}C_{db}}{C_{da}C_{cb}},\quad \frac{x}{1-x}=\frac{S_{ba}S_{dc}}{C_{ca}C_{db}}.
\ee
Using conformal symmetry the 4-point functions are related by \cite{2018JHEP...01..115K}
\be
\langle\sigma^+_n(z_1)\sigma^-_n(z_2)\sigma^-_n(z_3)\sigma^+_n(z_4)\rangle=\left|\frac{x(x-1)}{z_{21}z_{43}z_{41}z_{32}z_{31}z_{42}}\right|^{\frac{4\Delta_n}{3}}\langle\sigma^+_n(q_1)\sigma^-_n(q_2)\sigma^+_n(q_3)\sigma^-_n(q_4)\rangle.
\ee
This 4-point function is now in the standard form and for the large $c$ holographic limit can be expanded in vacuum conformal blocks \cite{2013arXiv1303.6955H} $f_0(x)$
\be
\langle\sigma^+_n(q_1)\sigma^-_n(q_2)\sigma^+_n(q_3)\sigma^-_n(q_4)\rangle\sim c_n^2 e^{-\frac{c}{6}f_0(x)-\frac{c}{6}f_0(\bar{x})}
\ee
where $c^2_n\rightarrow1$ as $n\rightarrow1$.
There are two ways $f_0$ can be expanded corresponding to the $s$ and $t$ channels where to leading order $f_0$ is given by
\be
f_0(x)=\begin{cases}
(n-1)\log(x),\quad \text{$s$-channel}\\
(n-1)\log(1-x), \quad \text{$t$-channel}.
\end{cases}
\ee
Altogether the full 4-point function can be written as
\be
\langle\sigma^+_n(u_A)\sigma^-_n(v_A)\sigma^-_n(u_B)\sigma^+_n(v_B)\rangle_\beta =\left(\frac{\pi\epsilon}{\beta}\right)^{4\Delta_n}c_n^2\left|\frac{1}{C_{da}C_{cb}}\right|^{2\Delta_n}x_{s,t}^{-(n-1)\frac{c}{3}}.
\ee
We are now prepared to calculate the thermofield mutual information. Gathering the pieces we have

\be
\begin{split}
J_{AB}&=\lim_{n\rightarrow1}\frac{1}{1-n}\log\left[\frac{x_{s,t}^{(n-1)\frac{c}{3}}}{c_n^2}\left|\frac{C_{da}C_{cb}}{S_{ba}S_{dc}}\right|^{2\Delta_n}\right]\\
&=\frac{c}{3}\log\left[\frac{1}{x_{s,t}}\left(\frac{S_{ba}S_{dc}}{C_{da}C_{cb}}\right)\right]
\end{split}
\ee

so that
\be
J_{AB}=\frac{c}{3}\max \left(0, \log\left[\frac{S_{ba}S_{dc}}{C_{ca}C_{db}}\right]\right)
\ee
as required.

\bibliographystyle{JHEP}
\bibliography{dSPE}


\end{document}